\immediate\write18{makeindex \jobname.nlo -s acro.ist -o \jobname.nls}%try
\documentclass[12pt,oldfontcommands]{ociamthesis} 
\usepackage[a4paper,inner=35mm,outer=35mm,top=25mm,bottom=30mm]{geometry}
\usepackage[T1]{fontenc}
\usepackage{CJKutf8 }
\usepackage{kotex} %package for korean input 

\usepackage{amssymb}
\usepackage{cite}%ekledim
\usepackage{acro}% \usepackage{nomencl} trial
\usepackage{transparent}
\usepackage{eso-pic}
\usepackage{xfrac}    % for \xfrac macro
\usepackage{nicefrac}
\usepackage{subfig}
\usepackage{caption}
\usepackage{amsmath}
\usepackage{amsthm,amssymb,amsxtra}
\usepackage{mathtools}%mathtools
\usepackage{graphicx}
\usepackage{epsfig}
\usepackage{physics}  
\usepackage{fancyhdr}
\usepackage{appendix}
\usepackage{ifpdf}
\usepackage{url}
\usepackage{rotating}
\usepackage{times}

\usepackage{verse}  %package for poem in the quotation
\newcommand{\attrib}[1]{%
\nopagebreak{\raggedleft\footnotesize #1\par}}
 
\ifpdf
\usepackage[pdftex]{hyperref} 
\hypersetup{
colorlinks = true
} 
\fi

\ifpdf  % enable hyper links
\hypersetup{
 pdftitle={Thesis},
 pdfauthor={Geonmo Ryu},
 pdfkeywords={j/psi,charmed meson},
 citecolor=blue,%darkblue,
 filecolor=black,
 linkcolor=blue,
 urlcolor=blue,
 colorlinks=true}
\fi

\usepackage{siunitx} % use this package module for SI units 
\usepackage{enumitem} %deneme
\usepackage{pdfpages} %sondeneme

\DeclareAcronym{GP}{
  short=GP,
  long=Gross--Pitaevskii,
}
\DeclareAcronym{WC}{
  short=WC,
  long=weak chaos,
}
\DeclareAcronym{SC}{
  short=SC,
  long=strong chaos,
}
\DeclareAcronym{LP}{
  short=LP,
  long=Lifshits phase,
}
\DeclareAcronym{ST}{
  short=ST,
  long=Self-trapping,
}
\DeclareAcronym{AL}{
  short=AL,
  long=Anderson localization,
}
\DeclareAcronym{TIO}{
  short=TIO,
  long=Transfer Integral Operator,
}
\DeclareAcronym{BM}{
  short=BM,
  long=Bogoliubov mode,
}
\DeclareAcronym{BdG}{
  short=BdG,
  long=Bogoliubov-de Gennes,
}

\DeclareAcronym{LG}{
  short=LG,
  long=Lifshits glass,
}
\DeclareAcronym{LCE}{
  short=LCE,
  long=Lyapunov characteristic exponents,
} 
\DeclareAcronym{BEC}{
  short=BEC,
  long=Bose--Einstein condensate,
}
\DeclareAcronym{PDF}{
  short=PDF,
  long=Probability distribution function,
}
\DeclareAcronym{GS}{
  short=GS,
  long=ground state,
}

\begin{document}

%\maketitle                  % create a title page from the preamble info
%\include{dedication}        % include a dedication.tex file

\EnglishTitle{Equilibrium and Non-equilibrium Gross--Pitaevskii Lattice Dynamics: Interactions, Disorder, and Thermalization}
   \KoreanTitle{}
  \AuthorKoreanName{Yagmur Kati}
  \AuthorEnglishName{Yagmur Kati}
  \KoreanNameofDegree{물리학 박사}
  
   \EnglishNameofDegree{Doctor of Philosophy}
  \EnglishNameofPaper{Ph.D. Thesis}

  \GraduateDate{2021년 06월}
  \GraduateDateEnglish{June 2021}
  \SummittedDate{2021년 06월}
	\EnglishSummittedDate{June 2021}
  \EnglishDepartmentName{Basic Science}
  \KoreanDepartmentName{물리학과}
  \RefereeDate{2021년 06월}
  \RefereeChief{Hee Chul Park}
  \RefereeSecond{Sergej Flach}
  \RefereeThird{Alexei Andreanov}
	\RefereeFourth{Jung-Wan Ryu}
	\RefereeFifth{Boris Fine}
	\RefereeSixth{Andrey Miroshnichenko}
  \AdvisorKoreanName{Sergej Flach}
  \AdvisorEnglishName{Sergej Flach}
  \CoadvisorKoreanName{Alexei Andreanov}
  \CoadvisorEnglishName{Alexei Andreanov}
  \KeyWordsEnglish{disordered discrete nonlinear Schrodinger lattice, Gross-Pitaevskii, Lifshits glass, ergodicity, non-equilibrium dynamics}
  \KeyWordsKorean{DNLS, GPE}

%--------------------front cover-------------------
  \makefrontcover
  \cedp
%-----------------list of approval------------------
\includepdf[pages=-]{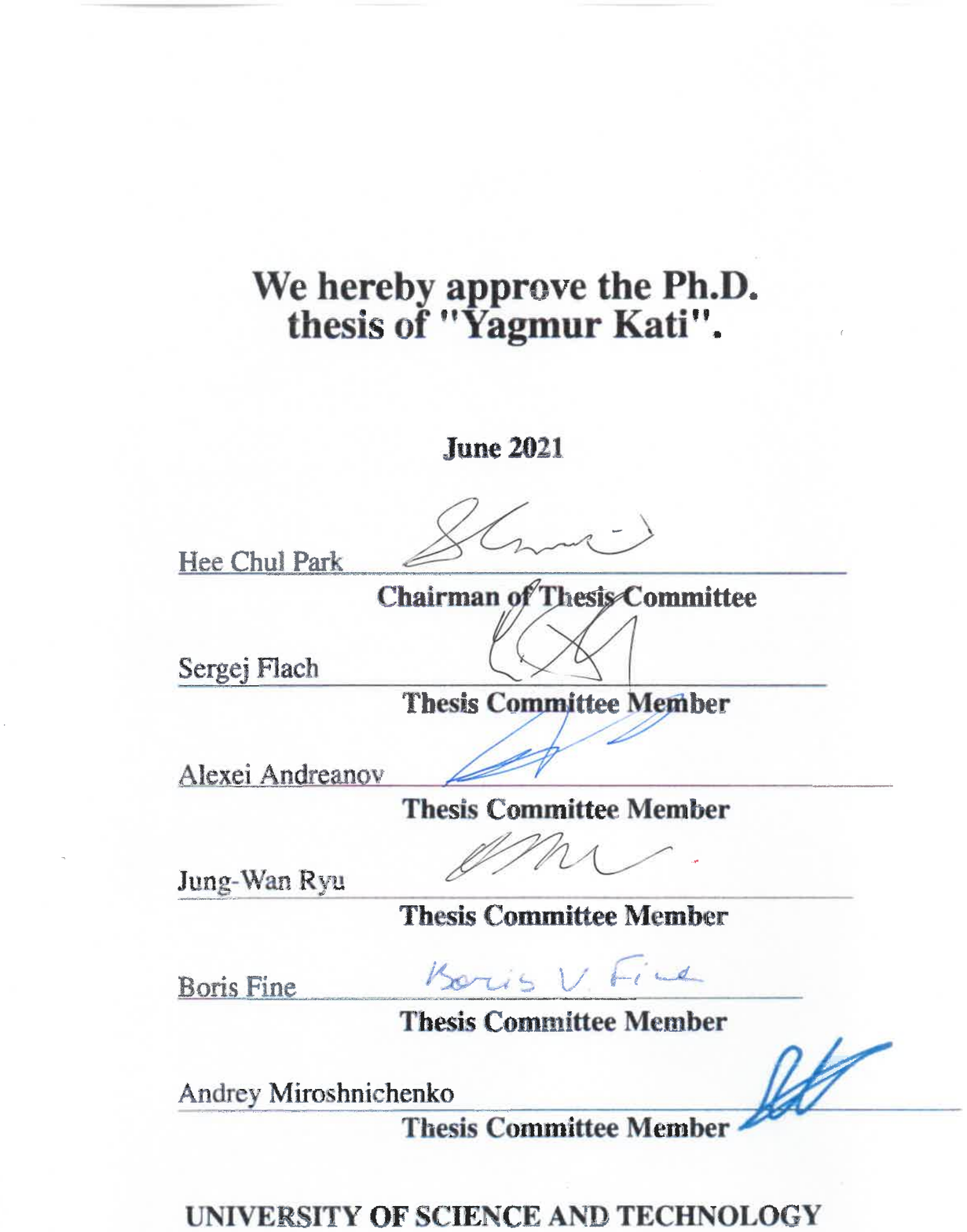}
%%  \makeenglishapproval %Yagmur put it back later
  \cedp 
%--------------------------------------------------------------
\pagenumbering{gobble}
\markboth{Quotation}{}
%\linespread{1.5}
\baselineskip=16pt
\begin{quotation}

\centering
%To all women in physics. 

%\vspace{0.9cm}

\textbf{State of Chaos}\\%State of Chaos
\vspace{0.35cm}
%%\baselineskip=13.6pt%14.5pt
%Interacting the waves of knowledge,\\
%she broke into a new chaotic world \\
%from her crystal perfect state.\\
%She was a true adventurer,\\
%dreaming to achieve things known as\\
%difficult…\\
%impossible…\\
%risky.
Interacting in the waves of knowledge,\\
she broke her crystal state \\
and opened the doors to a chaotic world.\\
As a true adventurer,\\
her trajectories dreamed of achieving everything\\
complex...\\ %difficult…\\
impossible...\\
risky.

%her region of happiness
Assertiveness heated the phase space \\
almost to an infinite temperature. \\
She was the reason, not the result. \\
She wanted to be the leader, not the follower. \\
She plotted a reliable and distinct scheme \\
to build her perception of the world. \\
Physics is so exciting, she thought one day.\\
%No one believed in her, except herself.\\
%Yet, she did not follow anyone,\\
%build her understanding of the world.\\
%Physics is so exciting, she said, one day.\\
I can solve the biggest mysteries of the universe.\\
I am curious, brave, and capable of anything.\\
I want to learn all the secrets before I die.\\%I want to learn all the secrets and not die before.\\
Slowly some disorder arose in her head,\\
trying to break her ergodic confidence,\\
by alluding many can thermalize better than her,\\
by hinting her frequency cannot create chaos,\\
by implying her ideas are glassy and transient,\\%her glassy ideas will not spread well,\\
over and over, every day.\\
She had to reach the ground state\\
to disentangle the branches of her thoughts down to the roots.\\
%to separate her islands of thoughts near to the root.\\
Then a perturbation distorted her equilibrium.\\
With all the elementary excitations, she said:\\
I know I am nonlinear.\\%smart.\\
I know I am dedicated.\\
I know I have potential.\\
She said, I can solve this problem,\\
also that problem,\\
and all the others.\\
I can...\\
I can...\\
I can.\\
%She read all the books she finds\\
%Challenging herself to be the best kid ever\\
%Always wanted to do things that people find\\
%difficult…\\
%impossible…\\
%risky.\\
%No one believed in her, except herself\\
%Yet, she did not follow anyone,\\
%build her understanding of the world\\
%Physics is so exciting, she said, one day\\%amazing
%I can solve the biggest mysteries of the universe\\
%I am curious, brave,  and capable of anything\\
%I want to learn all the secrets and do not die before %that\\
%Slowly some disorder arose in her head\\
% Trying to break her ergodic confidence\\
%with…\\ %ergodicity, equilibrium, GS, disorder, localized
%with…\\%correlation, nonlinearity, interaction, 
%with…\\
%over and over, everyday\\
%She had to reach to the ground state\\
%to separate her islands of thoughts near to the root\\
%then a perturbation distorted her equilibrium\\
%with all the elementary excitations, she said\\
%I know I am smart\\%intelligent
%I know I am dedicated\\
%I know I have potential\\
%She said I can solve this problem,\\
%also that problem,\\
%and all the others\\
%I can...\\
%I can...\\
%I can.\\
\attrib{---From Yagmur Kati\\
\hspace{1.5cm} To women in physics. } 
\end{quotation}
%\linespread{1.5}
\baselineskip=20pt
%, which is one of the fields in the world that the percentage of women is very small %this is temporary!!!CHANGE IT BACK
\cedp
\markboth{Aknowledgement}{}
\begin{Acknowledgement}

This thesis is the epitome of my journey in the understanding of the nonlinear and disordered dynamical systems during four years of my PhD Program at the Center for Theoretical Physics of Complex Systems. %in Institute for Basic Science.

In these four years, I have experienced and learned a wide variety of new things. Not only I have endeavored to grasp the fundamental physics at the core of our research, but also I have discovered new countries, new places, and new people. %, and new food. 
Eventually, I became more confident, more social and open, and more equipped.  
%not only in physics, but also discovered new countries, new places, new people, and new food. I became more confident, more social and open, and more equipped.  %It was a complete adventure from the beginning. 
Now I have the chance to thank all the people who had an impact during my PhD adventure.

%Dr. Anton Desyatnikov has been a terrific supervisor for many years. I am grateful for all the times he has provided new ideas when I am stuck, and enthusiasm when I am discouraged. He persistently and patiently questions me until I really understand new concepts. 

My thanks to my advisor Prof. Sergej Flach must be the first to reach because no need to tell that there would not be this thesis or even current myself without him. %He has been an extraordinary advisor for many years. 
He has been an extraordinary advisor and a lead of our projects who can interpret any scientific result at a glance, offer creative ideas, ask great questions, and easily see the core of the problems we have been working, and their probable future. He has also spent his time teaching me how to scientifically discuss and present \& deliver our results to the audience. I feel extremely lucky that he believed in me, helped me in all the difficult steps through my PhD, and guided me to grow as a scientist.

%I would like to acknowledge Prof. Andrey Miroshnichenko, Prof. Boris Fine,  Prof. Jung-Wan Ryu, and Prof. Hee Chul Park for accepting to be thesis referees. Their comments on this document and questions  put forward during the oral defense improved my thesis in a . 
%Prof. Boris Fine and Prof. Andrey Miroshnichenko kindly accepted to be thesis referees. I would like to thank them for their interest in our work, and comments which improved the thesis. 
%I would like to thank Hee Chul Park and Jung-Wan Ryu for having accepted to be members of the thesis committee, and for the questions and comments put forward during the oral defense.

I have been fortunate to have had the opportunity to collaborate with a number of great people during our projects. I am grateful to Prof. Mikhail Fistul who has taught me the origin of the statistical physics problems we have been working. I have evolved my analytical skills in statistical mechanics by working with him and Sergej. Here, I want to thank Prof. Alexander Yu Cherny who provided valuable insight and ideas during the time we have worked. He has helped to elucidate the crucial points of our research. Prof. Alexei Andreanov gave inspiration and ideas from the beginning of my PhD. His suggestions have always been valuable, and I am honored that he is my co-advisor. 
%%Prof. Andrey Miroshnichenko and Prof. Boris Fine kindly accepted to be referees  
I would like to thank Dr. Tilen Cadez, who is an expert, for teaching me the transfer matrix method in detail. ‪Ihor Vakulchyk has to be mentioned at this point. I am glad that such an expert in numerics has been in our center over these years. I would like to thank Prof. Juzar Thingna for our lively discussions, and Prof. Ivan Savenko for introducing me hybrid Bose-Fermi mixtures. 
%I also  owe many thanks to Dr. Carlo Danieli, and all of my other  collaborators due to their valuable contributions to our projects.
I also owe many thanks to my collaborators Dr. Carlo Danieli, Prof. David Kelly Campbell, Dr. Mithun Thudiyangal, and Dr. Xiaoquan Yu due to their valuable contributions to our projects. %I remember in the beginning of our Josephson-junction project, Carlo introduced me our novel method on board, and I could capture it in ten minutes absolutely without any question left. 
%%%I learnt from Carlo to share knowledge openly, clearly and fearlessly to others, since this is a profile of a researcher that anyone wish to work together. %%I would like to appreciate Carlo's ability to share knowledge in a right way. 
I feel indebted to Minyoung Lee, and Jaehee Kwon who have been always kind and helpful to me. 
%I remember in the beginning of our Josephson-junction project, Carlo introduced me our novel method  on board, and I captured it in ten minutes absolutely without any question left. I would like to appreciate his ability to share knowledge in a right way. 
%teaching abilities, and always sharing  his knowledge in a right way. 
%This is definitely a quality of a researcher that anyone wish to work together. 
%and want to acknowledge them for being always kind and helpful to me. %am also grateful to
%%I want to express my thanks to my friend Dr. Stephen Angus for the chats on string theory, and for telling me about Homestuck. 
%I want to extend my thanks to Dr. Stephen Angus for introducing me string theory, and for the friendly conversations about Homestuck.
Hereby, I would very much like to acknowledge my thesis referees Prof. Andrey Miroshnichenko, Prof. Boris Fine,  Prof. Jung-Wan Ryu, and Prof. Hee Chul Park for their fruitful comments on this document and enlightening questions put forward during the oral defense. 
%There are several special people to thank, and here
%unforgettable

%I want to use the chance to properly thank all the special people who made my PhD years more enjoyable.

Here I would like to thank the special people %who are all special for me, and 
who have made my PhD years more awesome. I feel very lucky to be friends with two great professors Sarika Jalan, and Sandra Maluckov. I will always remember our hiking trips and our chats on life and science. 
I want to express my thanks to my friend Dr. Stephen Angus for the chats on string theory, and for telling me about Homestuck. 
I am grateful to my coolest friends who organize several fun events: Letizia, Pramod, and Diana; and my unique friends who company: Henry, Jolin, Kabya, Nelli,  Bagrat, Ilias, Lauri,  Meng, Merab, Kodo, Ibrahim, Kristian, Niladri, Dominik, Taufiq and Nana. 
%Letizia Diamante, Pramod Padmanabhan, and Thomas Engl; and my unique friends who company: Hongyu Liu, Jolin Wei, Stephen Angus, Nelli Boichenko, Kabyashree Sonowal, Diana Thongjaomayum, Bagrat Mailyan, Ilias Amanatidis, Lauri Toikka,  Meng Sun, Merab Malishava, Wulayimu Maimaiti, Kristian Villegas, Niladri Gomes, and Dominik Šafránek. 
I am happy for the times we discovered new places in Korea, tried adventurous food together, played board games, go festivals,  camping, sightseeing, hiking, swimming, skiing, and even for the times we just met at the gym or kitchen of our center. %I am grateful to have known such wonderful people. %coffee, icecream, gym, shopping, swimming
I also have never forgotten the nicest people on earth: Brian Capper, and my officemates at Massey University; Sophie Shamailov, Andrew Punnet, and Peter Jeszenszki who live in the most beautiful country: New Zealand. I wish to meet you again.

%I also have never forgotten the nicest people on earth: Brian Capper, and my officemates at NZIAS Sophie Shamailov, Andy Punnet, and Peter Jeszenszki who live in the most beautiful country: New Zealand. I wish to meet with you again.  %I am especially grateful to Brian Capper, and my officemates Sophie, Andy, and Peter. %I am glad that I have met with these nicest people on earth. %, the nicest people I have ever met

%Peter Jeszenszki 

In my MS thesis, I wrote
"I am hugely indebted and thoroughly grateful to Sinan Gundogdu for his precious help and support during the time it has taken me to finalize this thesis". After many years, I would like to repeat the same statement for Sinan today, who is now my life partner. I cannot find the words to tell how perfect and intelligent he is, and how I am sure that I cannot find someone like him in the world. %Thank you Sinan.

%Lastly, I would like to thank my family and my best friend Meryem for their infinite love and support despite the large geographical distances. 

Finally I would like to extend my deepest gratitude to my parents Nur\c{s}en and Adil Kat{\i} %, and my family %and my best friend Meryem 
for their infinite love and support despite the large geographical distances.

\end{Acknowledgement}   % include an acknowledgements.tex file
\cedp
%>>>>>>>>>>>>>>>>>>English Abstract<<<<<<<<<<<<<<<<<<<<<

\markboth{Abstract}{}
\begin{EnglishAbstract}

The interplay of fluctuations, ergodicity, and disorder in many-body interacting systems has been striking attention for half a century, pivoted on two celebrated phenomena: Anderson localization predicted in disordered media \cite{Anderson58}, and Fermi--Pasta--Ulam--Tsingou (FPUT) recurrence observed in a nonlinear system. The destruction of Anderson localization by nonlinearity \cite{Pikovsky08} and the recovery of ergodicity after long enough computational times lead to more questions. This thesis is devoted to contributing to the insight of the nonlinear system dynamics in and out of equilibrium. 
Focusing mainly on the \ac{GP} lattice, we investigated elementary fluctuations close to zero temperature, localization properties, the chaotic subdiffusive regimes, and the non-equipartition of energy in non-Gibbs regime. 

Initially, we probe equilibrium dynamics in the ordered GP lattice and report a weakly non-ergodic dynamics, and an ergodic part in the non-Gibbs phase that implies the Gibbs distribution should be modified. Next, we include disorder in GP lattice, and build analytical expressions for the thermodynamic properties of the ground state, and identify a Lifshits glass regime where disorder dominates over the interactions. In the opposite strong interaction regime,  we investigate the elementary excitations above the ground state and found a dramatic increase of the localization length of Bogoliubov modes (BM) with increasing particle density. Finally, we study non-equilibrium dynamics with disordered GP lattice by performing novel energy and norm density resolved wave packet spreading. In particular, we observed strong chaos spreading over several decades, and identified a Lifshits phase which shows a significant slowing down of sub-diffusive spreading.
\end{EnglishAbstract}

\cedp
% -*- mode : latex; coding: utf-8 -*-
%%%%%%%%%%%%%%%%%%%%%%%%%%%%%%%%%%%%%%%%%%%%%%%%%%%%%%%%%%%%%%
%% 국문초록
%%%%%%%%%%%%%%%%%%%%%%%%%%%%%%%%%%%%%%%%%%%%%%%%%%%%%%%%%%%%%%
\begin{KoreanAbstract}

\begin{CJK}{UTF8}{mj}
다체 상호 작용 시스템에서 요동, 에르고딕 성질 및 무질서의 상호 작용은 반세기 동안 두 가지 유명한 현상을 중심으로 주목을 끌어왔다: 무질서한 매질에서 예측된 앤더슨 국소화 및 비선형 계에서 관측된 FPUT (Fermi--Pasta--Ulam--Tsingou) 되풀이. 비선형에 의한 앤더슨 국소화의 파괴와 충분한 계산 시간 후에 나타나는 에르고딕 성질의 회복은 더 많은 질문으로 이어진다. 이 논문은 평형과 비평형의 비선형 계 동역학에 대한 통찰에 기여하는 데 전념한다. 주로 Gross--Pitaevskii (GP) 격자에 초점을 맞춰 영도에 가까운 기본 요동, 국소화 특성, 혼돈스러운 부확산 영역 및 비-Gibbs 영역에서 에너지의 비균등 분할을 조사하였다.

처음에는 정렬된 GP 격자에서 평형 동역학을 조사하고 약한 비에르고딕 동역학과 Gibbs 분포가 수정되어야 함을 암시하는 비-Gibbs 상태의 에르고딕 영역을 보고한다. 다음으로 GP 격자에 무질서를 포함하고, 바닥 상태의 열역학적 특성에 대한 해석적 표현을 구축하며, 무질서가 상호 작용보다 지배적인 Lifshits 유리 영역을 확인한다. 강한 상호 작용 영역에서 우리는 바닥상태 위의 기본 들뜸을 조사하고 입자 밀도가 증가함에 따라 Bogoliubov 모드 (BM)의 국소화 길이가 급격히 증가하는 것을 발견하였다. 마지막으로, 우리는 에너지 및 표준 밀도를 이용한 새로운 파동 묶음 확산을 수행하여 무질서한 GP 격자에서 비평형 동역학을 연구한다. 특히, 우리는 디케이드에 걸쳐 퍼져 나가는 강한 혼돈을 관찰하였고, 부확산의 현저한 감속을 보이는 Lifshits 상태를 확인하였다.
%평평한 띠(FBs)는 .....
\end{CJK}

\end{KoreanAbstract}
          % include the abstract
\cedp

\clearpage %lets try without it
\begin{romanpages}          % start roman page numbering
\tableofcontents            % generate and include a table of contents

\listoffigures              % generate and include a list of figures
\listoftables                % generate and include a list of tables

\printacronyms

%%\printnomenclature             % generate nomenclature
\cedp
\end{romanpages}            % end roman page numbering 

%this baselineskip gives sufficient line spacing for an examiner to easily
%markup the thesis with comments
\baselineskip=20pt%15pt%15pt%20pt plus1pt %16pt%

\pagenumbering{arabic}

%set the number of sectioning levels that get number and appear in the contents
\setcounter{secnumdepth}{3}
\setcounter{tocdepth}{3}

  \pagestyle{fancy}%
%  \lhead{\changefont \leftmark}%
%  \rhead{}%
%  \chead{}%
%  \cfoot{\thepage}%
\fancyhead[LE]{\slshape \nouppercase{\rightmark}} %section
\fancyhead[RE]{}
\fancyhead[RO]{\slshape \nouppercase{\leftmark}} % chapter
\fancyhead[LO]{}
  \renewcommand{\headrulewidth}{1pt}

%now include the files of latex for each of the chapters etc
\chapter{Motivation}
\label{Chapter1}
\ifpdf
    \graphicspath{{Chapter1/Figs/Raster/}{Chapter1/Figs/PDF/}{Chapter1/Figs/}}
\else
    \graphicspath{{Chapter1/Figs/Vector/}{Chapter1/Figs/}}
\fi

Although disordered and complex systems are ubiquitous in nature, physicists tend to prefer “simple” systems which obey simple laws, and which can be represented by simple mathematical equations \cite{Goldenfeld87}. The reductionist hypothesis was highly effective in physics until the first half of the 20th century. In 1972, P. W. Anderson criticized the reductionism approach by stating that we can start from fundamental laws and reconstruct the universe instead of reducing everything to simple laws since each level of complexity brings new properties and so is equally important \cite{moreisdifferent}. He elucidated his point in 2011 as

\vspace{0.4cm}

\textit{I argue against, not the reductionist program itself but
the rationale and programmatic which is often associated with it, which gives the "Theory of Everything" the status of a "God Equation" from which all knowledge follows. Rather, I see the structure of the world as a hierarchy with the levels separated by stages of emergence, which means that each is intellectually independent from its substrate. Reduction has real value in terms of unifying the sciences intellectually and strengthening their underpinnings, but not as a program for comprehending the world completely.}

\vspace{0.4cm}
and epitomized it briefly:
\vspace{0.4cm}

\textit{"have no fear: More is Different."} \cite{Anderson2011}

\section{Anderson localization}

\rightskip=2.95cm
\textit{"Very few believed (localization) at the time, and even fewer saw its importance; among those who failed to fully understand it at first was certainly its author. It has yet to receive adequate mathematical treatment, and one has to resort to the indignity of numerical simulations to settle even the simplest questions about it."}

\rightskip=0pt
\hspace*{2.6cm} —Philip W. Anderson, Nobel lecture, 1977\\

One of the key ingredients of the indispensable imperfectness in natural phenomena is disorder. Comprehension of its role has been one of the most enigmatic inquiries in physical sciences which has gained more attention in the last few decades. The existence of disorder is usually disregarded in the initial theoretical approaches to the portrayal of physical systems, although even a tiny amount of it is capable of yielding major differences. Given its natural inevitability on all scales due to the impurities or defects induced by external fields, it is important to understand how it fundamentally distorts our theoretical predictions on clean and idealized models. 

\begin{figure}[hbt!] 
\centering
\includegraphics[width=0.8\textwidth]{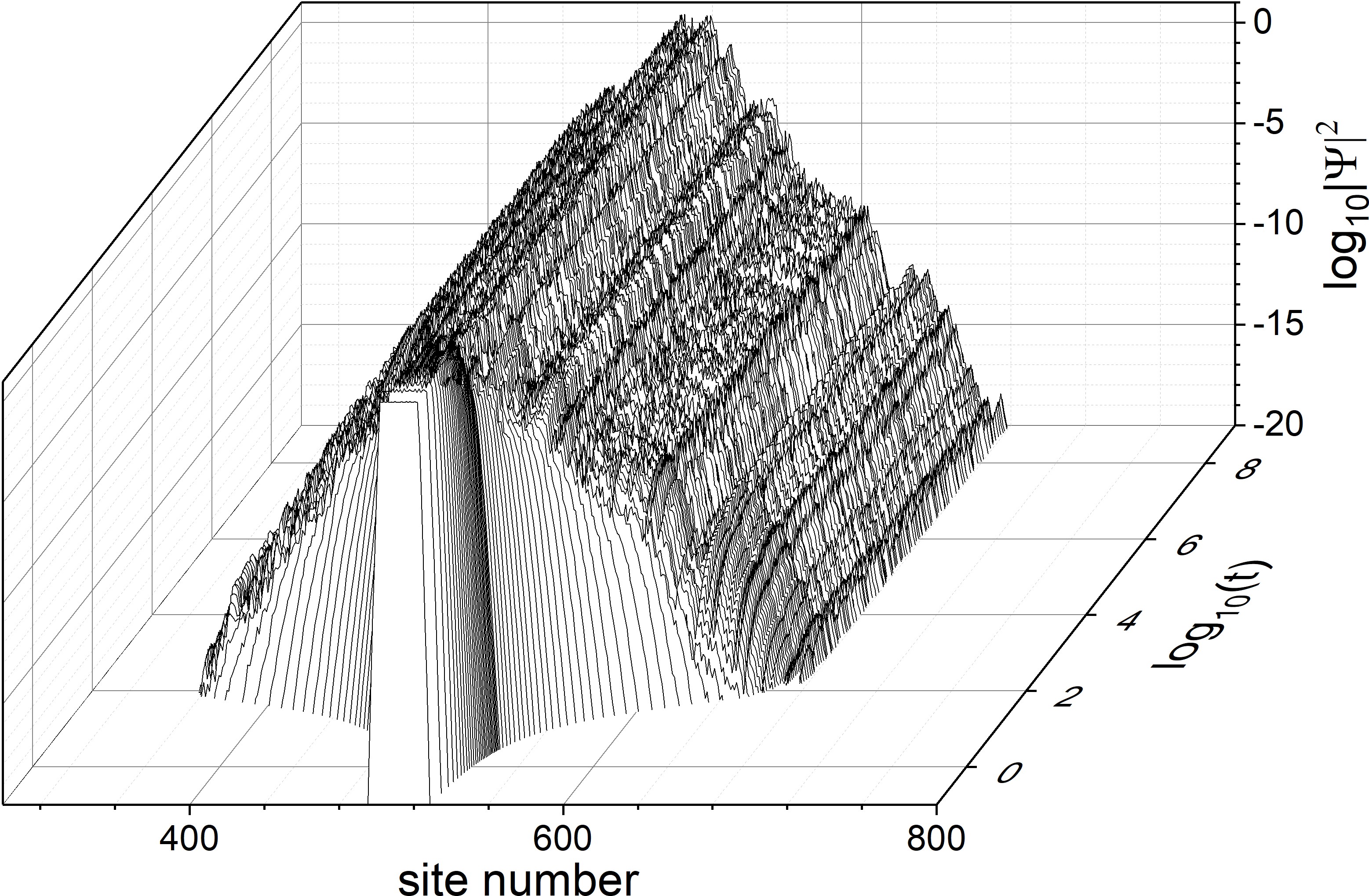}%anderson4.png}%
\caption[Anderson localization of a wave packet]{Anderson localization of a  wave packet in a tight-binding chain (Eq. \ref{eq:tbc}) with disorder strength $W=4$. The intensity of local particle densities $\log_{10}{|\psi_\ell|^2}$ is plotted in log time for each site number $\ell$. The initial width of the wave is $L_0=21$ which is equal to the total norm $\mathcal{A}$, where $\mathcal{H}=0$. Total system size is $N=2^{10}$.}%The random potential varies from -2 to 2.
\label{fig:ALfigure}
\end{figure}

Disorder is well-known to be the lead of the remarkable phenomena of localization of classical waves and non-interacting quantum mechanical particles. In condensed matter physics itself, this paradigm was brought to the forefront by the seminal work of P. W. Anderson (1958) \cite{Anderson58}, where he theoretically proposed that in disordered crystals there is a subtle quantum interference effect of noninteracting electron scatterings by defects of random potential that localizes single-particle wave functions in space with an upper bound on the localization length. The inability to carry currents over macroscopic length scales due to the existence of localized states had  substantial effects on the transport properties of materials. The presence of disorder is so critical that even a weak disorder may be sufficient to inhibit any transport and turn a conductor into an insulator. While the key ingredients of \ac{AL} are coherence and disorder, all kinds of waves and quantum particles may be affected by this phenomenon. Therefore, AL has  been embraced by not only solid-state physics \cite{Zehender,Choi:15,Mookherjea,PhysRevLett.70.2932} but also a variety of fields including optics \cite{Fallert,ALlight,Billy:2008aa,matterwave,PhysRevLett.94.073901,PhysRevLett.98.143901,Stano,PhysRevA.88.023426,PhysRevA.89.063812,Sapienza1352}, acoustics \cite{WEAVER1990129}, biological systems \cite{ALbiology,Vattay2014} and atomic physics \cite{2009Aspect}. An example of Anderson localization is exhibited in Fig. \ref{fig:ALfigure} for a one-dimensional tight-binding chain with Hamiltonian %GP lattice (see Chapter \ref{chapter5} for more details).
\begin{equation}\label{eq:tbc}
\mathcal{H}=  \sum_{{\ell=1}}^N \epsilon_\ell |\psi_\ell|^2 +{\psi_\ell}^* \psi_{\ell+1} +\psi_\ell \psi_{\ell+1}^*,
\end{equation}
where $\psi_\ell$ are complex variables, $N$ is total number of sites, $\epsilon_\ell$ is the onsite potential defined randomly in $[-W/2,W/2]$ where $W$ is the disorder strength. 

With $\dot{\psi_\ell}=\partial\mathcal{H}/\partial i {{\psi_\ell}^*}$ we obtain the dynamics of equations of motion
\begin{equation}
    i\dot{\psi_\ell}=\sum_{{\ell=1}}^N \epsilon_\ell\psi_\ell-\psi_{\ell+1}-\psi_{\ell-1}
\end{equation}
which conserves the total energy $\mathcal{H}$ and total norm $\mathcal{A}=\sum_\ell |\psi_\ell|^2$. Fig. \ref{fig:ALfigure} shows the time evolution of a wave packet of size $L_0$ defined in the middle of the chain with $\psi_\ell=1$, while rest of the lattice sites own $\psi_\ell=0$. %, where $N=2^10$. 

%The initial states are described as an excited region ($\psi_\ell>0$) in the middle of the lattice which occupies a size $L_0\ll N$. We leave the rest of the lattice sites unexcited ($\psi_\ell=0$) so that the spreading process begins with a wave packet length $L_0$. The initial wave packet is defined with two densities: norm density $a=\mathcal{A}/L_0$, and energy density $h=\mathcal{H}/L_0$.

%The energy $\mathcal{H}$ and total norm $\mathcal{A}=\sum_\ell |\psi_\ell|^2$ are conserved.

%Anderson localization of a wave packet]{Anderson localization of a  wave packet in a tight-binding chain with an uncorrelated random potential defined in $[-w,w]$ where $w=2$. The intensity of local particle densities $\log_{10}{|\psi_\ell|^2}$ is plotted in log time for each site number $\ell$. The initial width of the wave is 21 which is equal to the total norm. 

Originally the general notion of Anderson localization concerns non-interacting particles. Given the fact that nonlinearity is also an inevitable ingredient in condensed matter physics, understanding how disorder combines with interactions has perplexed physicists since Anderson's celebrated article on the localization phenomenon \cite{Anderson58}. The prominent challenge here comes from the existence of only a few theoretical tools which allow the treatment of disorder and interactions on an equal footing. %There are several reasons why we use the Gross-Pitaevskii model in this thesis, yet one main reason is this model tune the interaction strength and the disorder amplitude in a fair manner.

\section{Non-equilibrium phenomena}

Thermalization, ergodicity, equipartition, and chaos are fundamental concepts in the context of many-body interacting systems. %Statistical mechanics, which is mainly built by Gibbs, Boltzmann, and Maxwell, 
Since Gibbs coined the name "statistical mechanics" crediting Boltzmann, and Maxwell in 1884 \cite{Klein1990}, there has been search for the patterns in the chaotic behavior of many-body systems, considering both collective and individual components. The general statistical laws have limitations close to the integrable limits of a dynamical system which is characterized by a countable set of preserved actions, for example, harmonic lattice vibrations in crystals. Near to the integrable (exactly solvable) limit, the time required for equipartition may increase immensely. This violation of the equilibrium statistical mechanics is mathematically explained by Kolmogorov-Arnold-Moser (KAM) theorem (see Sec. \ref{sec:kam}). 

%The ergodic hypothesis is first introduced by Boltzmann in 1898 \cite{LBoltzman}, and proven by Birkhoff \cite{Birkhoff656},  John von Neumann \cite{Neumann}, and Khinchin \cite{Khinchin:232978}  that the time averages of an ergodic system must coincide with the space averages. There exists a relaxation time $\tau$ for every significant physical observable $f$ such that the time-average coincides with the phase average for all times larger than it \cite{Neumann,Neumann2} .

In the real world, not all systems are ergodic and thermalized, for example, glasses and gels possess very long metastable states. Moreover, it is notable that Anderson localization is in fact a paramount example of nonergodicity, due to the existence of exponentially localized states led by the interplay between disorder and interference. To assess the transport features and thermalization processes, the ergodicity properties of particles are crucial. As knowing that a disordered GP lattice has high complexity due to possessing both short and long-range networks, in this thesis, we will discuss the weak ergodicity breaking of GP lattice without the disorder.

\section{Outline}
 This thesis  is dedicated to the understanding and characterization of the phenomena of Anderson localization and non-equipartition, approaching it from complementary facets.  We identify new pathways to probe the equilibrium and non-equilibrium dynamics of interacting many-body systems. We primarily address the interplay between disorder and interactions and the fate of ergodicity in Gross--Pitaevskii lattices.

The thesis is organized as follows. In the second chapter, we introduce the concepts to bring a clear perspective to the report. This includes the Gross--Pitaevskii model definition, its equilibrium dynamics, the ergodic hypothesis, non-equipartition phenomena, and a brief discussion of discrete breathers.  While this chapter includes some of our results, it will be useful as a background in all the rest of the thesis. 
In Chapter 3, we present  our results of equipartition and non-equipartition of energy in GP lattice, including the numerical methods to test ergodicity. We examine the equilibrium fluctuation of the Gross--Pitaevskii equation and the corresponding impact of localized excitations without disorder. From Chapter 4, the disorder and thus its relation with interactions are introduced to the thesis. In chapter 4, we present ground state statistics and elementary excitations in disordered GP lattice.  In Chapter 5, we turn our attention to the density resolved GP wave packet spreading. We exhibit and discuss our results of different subdiffusive regimes in which the wave packets spread, such as strong chaos, weak chaos, self-trapping, Lifshits, and strongly interacting regimes. 
%Finally, in Chapter \ref{chapter6} we briefly comment on our results and the relation between our current results and our future projects.
Finally, in Chapter \ref{chapter6} we briefly comment on our research outcome and the connection between our current results and the research directions for the future. 

%We will introduce a systematic way to characterize the fluctuations of such systems in equilibrium, as well as the relaxation dynamics from extreme non-equilibrium states. This includes the long-time computational evolution of the systems and the statistics of fluctuations. Our goal is to analyze the impact of the existence of coherent states on the statistics of so-called extreme events, i.e. long-time fluctuations off equilibrium. We will also use our methods to characterize out-of-equilibrium dynamics of many-body systems, where the Gibbs distribution fails to provide statistical grounds for computing averages.

During my Ph.D., I also worked on the non-equipartition phenomena in the Josephson junction and the Klein--Gordon chains; details can be found in \cite{mithun2019,danieli2019}.
%in two different nonlinear systems:
\cedp
\chapter{Introduction} %background
\label{chapter2}
\ifpdf
\graphicspath{{Chapter2/Figs/}{Chapter2/Figs/PDF/}{Chapter2/Figs/}}
\else
\graphicspath{{Chapter2/Figs/}{Chapter2/Figs/}}
\fi

\textit{The perfect square has no corners \\
Great talents ripen late \\
The highest notes are hard to hear \\
The greatest form has no shape \\
}
\hspace*{3cm} —Lao Tzu, 6th-century BC\\

As we may presume several behaviors of cosmos by simple rules, the most surprising and extraordinary things that happened in its history are coming from the complex and chaotic behavior. 

The intriguing phenomena of Anderson localization, non-equilibrium nonlinear dynamics, and ergodicity breaking have been principally %substantially
tested in many-body systems.
Classical nonlinear wave equations are mostly studied as mean-field approximations to interacting many-body systems, and as for proper tools to describe photonic networks. The Gross--Pitaevskii lattice is one of the famous models which can be used for this purpose.

In this chapter, we first present the complex many-body system that we mainly tested and discovered its chaotic dynamics during my Ph.D.: \textit{The Gross--Pitaevskii equation}. 
In Sec. \ref{sec:gpl}, the Gross--Pitaevskii lattice model is defined in detail. Next, we describe the GP lattice thermodynamics with Transfer Integral Operator method. 
In section \ref{sec:ergodicity}, we introduce one of the major concepts in statistical mechanics and nonlinear dynamics: \textit{ergodicity}. We give its background information that can be useful to gain an insight into the results discussed in Chapter \ref{chapter3}.

\section{Gross--Pitaevskii equation}\label{sec:gpl}

The Gross--Pitaevskii equation is a model which can describe Bose--Einstein condensates of ultracold atoms within the Hartree--Fock approximation. It is integrable in one dimension, and it can be used in many areas in solid-state physics, such as nonlinear optics \cite{spreading}.

We are using a discrete form named Gross--Pitaevskii (GP) lattice, aka discrete nonlinear Schr\"{o}dinger equation, which is a semiclassical reference model of the equation of motion with breather solutions, providing the opportunity to study the many-body problem. The model, also labeled as semiclassical Bose--Hubbard, can be used to explore the wave packet localization, and to explain the non-equipartition phenomena in nonlinear systems which may or may not include disorder.  The simple and rich mathematical structure of GP lattice provides an approximate description of diverse physical situations, e.g.,  the two-body interactions in dilute Bose--Einstein condensates trapped in external periodic lattices \cite{Franzosi:2011}, the dynamics of high-frequency Bloch waves \cite{Kosevich:2002}, the discrete breathers in the networks of various interacting optical waveguides \cite{Aceves:1996}, and electronic transport in biomolecules \cite{scott}. A statistical explanation of its underlying physics in the GP lattice model has been unearthed by Rasmussen \textit{et al} in \cite{Rasmussen:2000}. The Hamiltonian of the ordered one-dimensional  GP lattice

\begin{equation}\label{eq:Hamiltonian}
H=\sum_{\ell=1}^N \left[\frac{g}{2} |\psi_\ell|^4 - J(\psi_\ell \psi_{\ell+1}^* + {\psi_\ell}^* \psi_{\ell+1}) \right],
\end{equation}
where $\ell$ is the site index of the lattice with size $N$, and $i {\psi_\ell}^*, \psi_\ell$ form canonically conjugate pairs of variables. The Hamiltonian consists of a quadratic nonlinear term with the tunable nonlinearity parameter $g>0$, and the nearest neighbor hopping terms with the tunneling amplitude $J$. 
%The sign of the hopping term does not matter due to   the symmetry held by the gauge transformation whereas 
The nonlinear term $g$ stands for the two-body interactions in Bose--Einstein condensates, and has a positive sign due to the repulsiveness of the cold atoms \cite{Iubini:2012}. The Hamiltonian is considered in dimensionless unit, while measuring the energy in units of $J$. Moreover, the norm $|\psi_\ell|^2$  is uniformly rescaled to tune the nonlinear parameter $g$. Hence,    $J\equiv g\equiv 1$ in all of our computations, unless stated otherwise. 
The equations of motion are generated by $\dot{\psi_\ell} = \frac{\partial H}{\partial i \psi_\ell^* }$:
\begin{equation}\label{eq:eqsofmotion}
i \dot{\psi_\ell} = g |\psi_\ell|^2 \psi_\ell - (\psi_{\ell+1} + \psi_{\ell-1})
\end{equation}
where the overdot represents the time derivative. Eq. (\ref{eq:eqsofmotion}) conserves the total norm $\mathcal{A}$ and the total energy $\mathcal{H}$. The total norm, which is analogous to the total number of particles in the system, can be written as

\begin{equation}\label{eq:totalnorm}
 \mathcal{A}= \sum_\ell  |\psi_\ell|^2.   
\end{equation}
  
The GP lattice has rich statistical  properties, by virtue of the conservation of two significant quantities ($\mathcal{H},\mathcal{A}$),  with the discreteness-induced bounded kinetic energy part in Eq. (\ref{eq:Hamiltonian}). On account of two integrals of motion,  ensembles of initial conditions with given values of norm and energy will create a two-dimensional microcanonical equilibrium phase diagram (Fig. \ref{fig:simplephase}) which is strictly equivalent to a grand canonical description of the same initial state by the temperature $T$ and the chemical potential $\mu  $\cite{Iubini:2013}. The detailed statistical analysis based on the grand-canonical partition function explored the phase diagram of GP lattice for the re-scaled parameter space ($g a$, $g h$), where $a = \langle\mathcal{A}\rangle/N$ is the norm density, $h = \langle\mathcal{H}\rangle/N$ is the energy density, and $g=1$.

In all simulations shown in this thesis, Eq. (\ref{eq:eqsofmotion}) is integrated by using the symplectic procedure $SBAB_2$ described in Appendix \ref{sec:symp} implemented with time step $d t$. The time step has to be chosen smaller for large energies and large norms. Therefore, we chose $d t=0.01 - 0.02$ in Chapter \ref{chapter3} for fully excited lattice dynamics, and $d t=0.05-0.1$ in Chapter \ref{chapter5} for the spatiotemporal evolution of a wave packet. The periodic boundary condition $\psi_1=\psi_{N+1}$ is used for all the results. 

\subsection{Gibbs and non-Gibbs states}\label{sec:nongibbs}%Gibbs Distribution  \& 

The disorder-free translationally invariant GP lattice model shows a non-Gibbs phase \cite{Rasmussen:2000,Johansson:2004}, albeit its space-continuous equation does not. 
%At variance to its space-continuous counterparts, the disorder-free translationally invariant GP lattice model shows a non-Gibbs phase \cite{Rasmussen:2000,Johansson:2004}. 
Hence, the phase diagram of GP lattice can be separated into two regions, one called {\it Gibbs} regime where grand-canonical partition function based statistical explanation is applicable to the thermalization ($T\ge0$), and a second one called {\it non-Gibbs} regime where the former partition function does not apply and no clear concept of temperature $T$ seems to be well-defined \cite{Lebowitz:1988,Rasmussen:2000,Johansson:2004}.

The temperature in the non-Gibbs regime is sometimes referred to as negative temperatures since this region corresponds to the undefined $T<0$ field of Gibbs thermodynamics, causing a divergent partition function. 
%formally described by Gibbs distributions with negative temperature. %reached only when $T<0$ is substituted into the formula of Gibbs thermodynamics. 
The existence of negative temperatures was first proposed in \cite{Onsager}, later its concept was introduced via the nuclear-spin systems experiments \cite{Ramsey1951278,PhysRev.81.279,PhysRev.81.156}, and more recently is discovered in a physical system of Bose--Einstein condensate \cite{Braun52}. The discussions on the physical meaning of negative temperatures, and the correct definition of entropy has been puzzling scientists for the last century \cite{einstein,Pearson19853030,Jaynes,Berdichevsky19912050,CAMPISI2005275,Adib2004581,LAVIS2005245,dunkel,PhysRevE.90.062116,sokolov,PhysRevE.91.052147,SWENDSEN201624,Ramsey195620,baldovin2021statistical,iubini2021chaos}. 

\subsection{Gibbs distribution}
In the Gibbs regime, any pair of realizable densities $\{a, h\}$ can be defined by a Gibbs distribution
\begin{equation}
\rho=\frac{1}{\mathcal{Z}} \exp(-\beta (\mathcal{H}+\mu \mathcal{A})),
\end{equation}
where $\beta$ is the inverse temperature, $\mu$ is the chemical potential, and $\mathcal{Z}$ is the grand canonical partition function. 
%%Their relation will be depicted in the next section. But, as noted earlier, for the non-Gibbs regime, there has to be a modification in the Gibbs distribution definition.
The Hamiltonian (\ref{eq:Hamiltonian}) possesses two integrable limits:

\begin{enumerate}[label=(\roman*)]
\setlength\itemsep{0em}
\item $g a\rightarrow 0$ in which the nonlinear part of $\mathcal{H}$ is negligible,
\item $g a\rightarrow\infty$ in which the nearest neighbor interactions  get negligible. 
\end{enumerate}
The limit (i) produces the linear GP lattice which will be introduced in the next section, and examined in Appendices \ref{ap:betamu_a_h}, and \ref{ap:temperature_mu_a_h_g}. The limit (ii) is briefly discussed in Chapter \ref{chapter3}. Both integrable limits are also valid in the presence of disorder, which is studied in Chapter \ref{chapter4}.

\subsection{Statistical mechanics for linear lattice}%Statistical Mechanics for $g=0$}

\begin{figure}[hbt!] 
    \centering
    \includegraphics[width=0.75\textwidth]{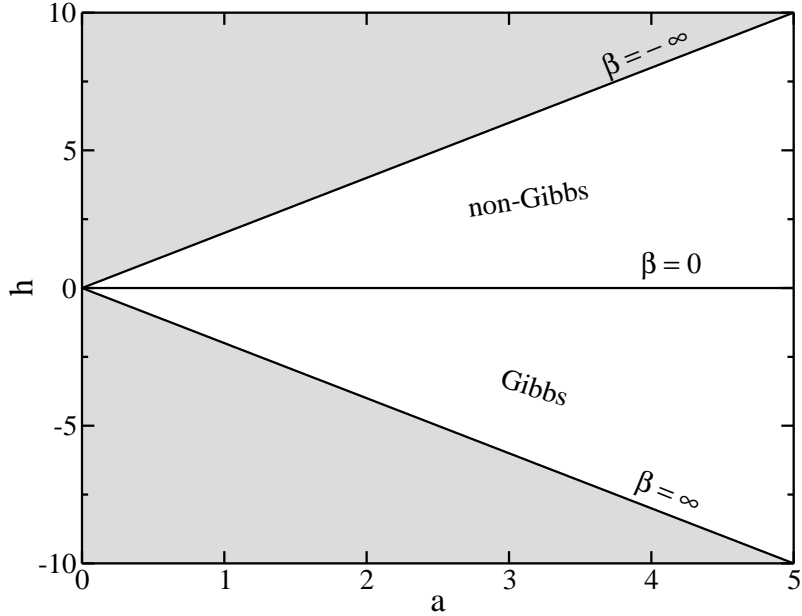}
    \caption[The phase diagram of ordered \& linear GP lattice]{The phase diagram of ordered linear GP lattice ($g=0$), where the shaded areas are inaccessible. The thick lines are the inverse temperature lines $\beta=-\infty$, $\beta=0$  and $\beta=\infty$, as defining the Gibbs and non-Gibbs regimes. }
    \label{fig:linearphasediagram}
\end{figure}

Following Eq. (\ref{eq:Hamiltonian}), the dynamics of a simple lattice without nonlinearity ($g=0$) is governed by the Hamiltonian
\begin{equation}\label{eq:Hamiltonian_g0}
\mathcal{H}=\sum_{\ell=1}^N  - J(\psi_\ell \psi_{\ell+1}^* + {\psi_\ell}^* \psi_{\ell+1}).
\end{equation}
Using $\psi_\ell = \sqrt{a_\ell} e^{i \phi_\ell}$ in Eq. (\ref{eq:Hamiltonian_g0}) where we define the local norm per site as $a_\ell \ge 0$ and the local phase $ | \phi_\ell| \le \pi$. Hence, the Hamiltonian transforms into 
\begin{equation}\label{eq:hamiltonian_cosg0}
\mathcal{H}=\sum_\ell  - 2\sqrt{a_\ell a_{\ell+1}} \cos(\phi_\ell - \phi_{\ell+1}). 
\end{equation} 
The Hamiltonian in Eq. (\ref{eq:hamiltonian_cosg0}) can be minimized when there is no phase difference between all sites $\phi_\ell=\phi_{\ell+1}$ that results in $\mathcal{H}= -2a N$, and maximized when it is the largest, i.e., $|\phi_\ell-\phi_{\ell+1}|=\pi$ that gives $\mathcal{H}=2aN$;  all the amplitudes are held as $\psi_\ell=\sqrt{a}$ to optimize.  Hence,  $h=2a$ is the maximum reachable energy density, and the ground state has the relation $h= -2a$. These limits produces the phase diagram exhibited in Fig. \ref{fig:linearphasediagram}, with two inaccessible parts: $h>2a$ and $h<-2a$. The temperature $\beta$ and chemical potential $\mu$ can be defined analytically in terms of $a$ and $h$, derived in Eq. (\ref{eq:betamu}), as 
\begin{equation}\label{eq:betamuc}
    \beta=\frac{-2h}{4a^2-h^2}, \quad \mu=\frac{4a^2+h^2}{-2ah}.
\end{equation}

This scheme of linear GP lattice, which is described in detail in Appendix \ref{ap:betamu_a_h}, is useful to analyze how the nonlinear GP lattice dynamics alters in the $g a\rightarrow0$ integrable limit.

\subsection{Statistical mechanics for nonlinear lattice}\label{sec:zerotemperatureorder} %$g\neq 0$}

The generic GP lattice is nonlinear and is defined with a nonzero $g>0$. Hereby, we will display how the temperature lines $\beta=-\infty$, $\beta=0$, and $\beta=\infty$ in Fig. \ref{fig:linearphasediagram} change in the presence of nonlinearity.

We apply the canonical transformation $\psi_\ell = \sqrt{a_\ell} e^{i \phi_\ell}$ to Eq. (\ref{eq:Hamiltonian}), thus the Hamiltonian transforms into

\begin{equation}\label{eq:hamiltonian_cos}
    \mathcal{H}=\sum_\ell  - 2\sqrt{a_\ell a_{\ell+1}} \cos(\phi_\ell - \phi_{\ell+1})+\frac{g}{2}a_\ell^2,
    \end{equation} 
with the norm $\mathcal{A}=\sum_\ell a_\ell$. %As derived also in \cite{Rasmussen:2000}, 
The Hamiltonian function (\ref{eq:hamiltonian_cos}) reaches its minimum value when all the phases are the same $\phi_\ell=\phi_{\ell+1}$, and the amplitudes of each site are equal to each other $\psi_\ell=\sqrt{a}$, which leads to 
$
\mathcal{H}= \frac{g a^2 N}{2}-2a N    
$.
Hence, the energy density $h$ at the ground state is
\begin{equation}\label{eq:groundstatezerodisorder}
h= \frac{g }{2}a^2-2a.   
\end{equation}

Eq. (\ref{eq:hamiltonian_cos}) does not hold any upper limit of energy density. When nonlinearity $g>0$ is introduced to the Hamiltonian, it removes the upper bound so that all states with $h>2a$ become available and shifts the lower bound, aka the ground state, to $h<-2a+ga^2/2$.

\begin{figure}[hbt!] 
    \centering
    \includegraphics[width=0.75\textwidth]{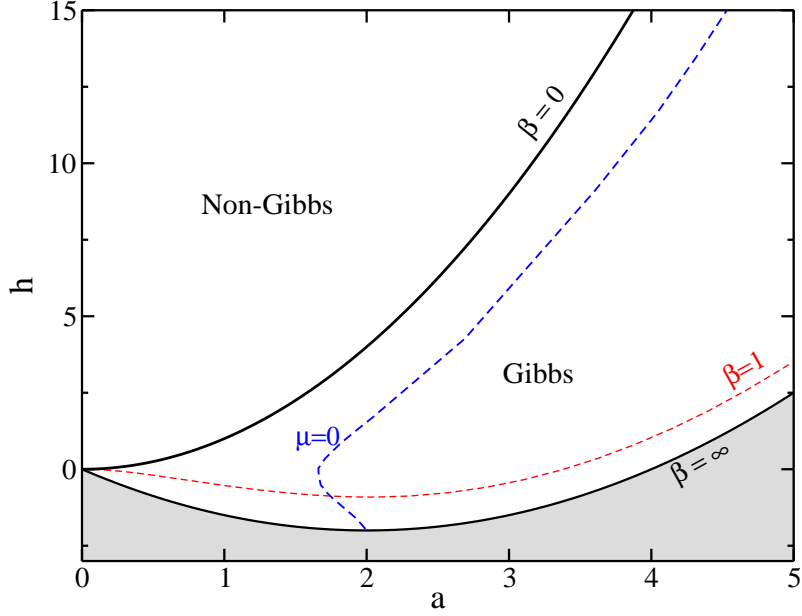}
    \caption[The phase diagram of ordered GP lattice]{The phase diagram of ordered GP lattice ($g=1$), where the shaded area is inaccessible. The thick lines are the inverse temperature lines $\beta=0$  and $\beta=\infty$, as defining the finite temperature Gibbs regime. We show  $\beta=1$ (red) and $\mu=0$ (blue) dashed lines for reference.}
    \label{fig:simplephase}
\end{figure}

\section{Gross--Pitaevskii lattice thermodynamics} %Equilibrium Dynamics
\label{sec:betamu_xy}

The GP lattice possesses both Gibbs and non-Gibbs distribution regimes that are separated by the infinite temperature line, shown in Fig. \ref{fig:simplephase}. For the equilibrium case, the analytic relations of average energy and norm densities at zero and infinite temperatures (derived in Appendix) specifies the Gibbs regime.

%As explained in the previous subsection, in the non-Gibbs regime, the average densities 
%Although it has been found that there is a part in non-Gibbs phase that shows ergodic behavior \cite{PhysRevLett.120.184101},

While one searches the equilibrium dynamics in the non-Gibbs regime, employing negative temperatures into the formula of Gibbs thermodynamics results in a divergence of the partition function. Thus the "negative temperature" description is indeed not reasonable. What truly happens in the non-Gibbs regime is that the complex field of $\psi_\ell$ dynamics separates into two components: high density localized spots and delocalized wave excitations with presumably infinite temperature \cite{Rasmussen:2000,magnus,Rumpf:2007,Rumpf:2008,rumpf3}.
% While  average densities  in  the non-Gibbs phase can be formally described by Gibbs distributions with negative temperature, in truth the dynamics show a separation of the complex field $\psi_\ell$ into a two component one - a first component of high density localized spots and a second component of delocalized wave excitations with infinite temperature \cite{Rasmussen:2000,magnus,rumpf1,rumpf2,rumpf3}. 
There is still not a clear mathematical way to separate these two parts. Yet there has been a lot of research on the observation of high density localized spots, which has a similar concept to self-trapping and discrete breathers \cite{sergej1} that will be explained in Sec. \ref{discretebreathers}.

First, let me elucidate how we computed the Gibbs average of physical observables, e.g., participation ratio, when the densities $\{a,h\}$ are realizable.

\subsection{Transfer integral operator method}\label{sec:tim}

The classical grand-canonical partition function is
\begin{equation}\label{eq:Z}
    \mathcal{Z}=\int \exp(-\beta (\mathcal{H}+\mu \mathcal{A})) d\Gamma =\int_0^{\infty} \int_0^{2\pi} \prod_m d\phi_m dA_m \exp{[-\beta (\mathcal{H}+\mu \mathcal{A})]},
\end{equation} 
where $\mu$ is analogous to the chemical potential which is introduced in order for the conservation of $\mathcal{A}$. 
Inserting Eq. (\ref{eq:hamiltonian_cos}) in Eq. (\ref{eq:Z}), we obtain 
\begin{align}\label{eq:Zbessel}
    \mathcal{Z}=(2\pi)^N \int_0^\infty & \prod_m  dA_m  I_0(2\beta \sqrt{A_m A_{m+1}}) \nonumber \\
    &\times \exp{-\beta \sum_m \frac{g}{4}(A_m^2+A_{m+1}^2)+\frac{\mu}{2}(A_m+A_{m+1})}
\end{align}
Assuming the thermodynamic limit, $N\rightarrow \infty$, we can calculate the integral by using the transfer integral operator (TIO) method:

\begin{equation}
    \int_0^\infty dA_m \kappa(A_m,A_{m+1})y(A_m)=\lambda y(A_{m+1}),
\end{equation}
where the kernel $\kappa$ is
\begin{equation}\label{eq:kernelintro}
    \kappa(x,z)=I_0(2\beta \sqrt{x z}) \exp{-\beta \left( \frac{g}{4}(x^2+z^2)+\frac{\mu}{2}(x+z)\right)}
\end{equation}
This integral equation (\ref{eq:Zbessel}) corresponds to an eigenvalue problem that can be solved numerically. In the thermodynamic limit, we obtain $Z \approx (2\pi\lambda_0)^N$, where $\lambda_0$ is the largest eigenvalue of the operator. With this technique, the transfer integral operator (TIO) method \cite{Rasmussen:2000,aubry}, we can find the average energy density and average excitation norm:
\begin{equation}
    a=-\frac{1}{\beta \lambda_0}\frac{\partial \lambda_0}{\partial \mu}, \qquad h=-\frac{1}{\lambda_0}\frac{\partial \lambda_0}{\partial \beta}-\mu a.
\end{equation}

It is significant to note the followings found by the transfer integral operator method:

\begin{enumerate}[label=(\roman*)]
  \setlength\itemsep{0em}
\item For zero nonlinearity, the Kernel defined in Eq. (\ref{eq:kernelintro}) diverges as $\mu \rightarrow 2$.
\item The right hand side of $\mu=0$ line in Fig. \ref{fig:simplephase} corresponds to negative $\mu$ values, while its left hand side is positive. Hence, near to the origin, the value of $\mu$ is always positive.
\item As $\beta \rightarrow \infty$, $c \rightarrow -2$ which corresponds to $\mu \rightarrow 2$. In this case $\beta \times \mu=\infty$
\item As $\beta \rightarrow 0$, $c \rightarrow 0$, and $\mu \rightarrow \infty$. In this case $\beta \times \mu=\text{const}$. %, so that we can compute it safely.
\end{enumerate}

GP lattice is a simple physical model which allows studying thermalization and temperature concepts and can approximate many other nonlinear systems. The significance of the above notes stems from the fact that they provide the mapping between the lattice parameters $(a,h)$ and the thermodynamic parameters $(\beta,\mu)$. In Appendix \ref{ap:temperature_mu_a_h_g}, you can find how the temperature and chemical potential change as $(a,h)$ approaches the origin of the phase diagram from different directions. It is calculated by the TIO method and compared with  Eq. (\ref{eq:betamuc}) in the small norm density limit.

\subsection{Inverse participation ratio}\label{sec:participationnumber}
Inverse participation ratio is a numerical tool that gives the total number of strongly excited sites in the GP lattice. It is mainly used to approximate the localization length or the extent of a wave or to spot if there are inhomogeneities such as breather formations in the distribution of an equilibrium system. The inverse of the participation ratio in real space is 
%Participation number, aka participation ratio, is a numerical tool which gives the total number of highly excited sites in the GP lattice. It is mainly used to approximate the localization length or the extent of a wave, or to spot if there are inhomogeneities such as breather formations in the distribution of an equilibrium system. The inverse of the participation number in real space is
\begin{equation}
P^{-1}=\frac{\sum_\ell{{a_\ell}^2}}{\mathcal{A}^2}= \frac{1}{\mathcal{A}^2} \sum_{\ell=1} ^N |\Psi_\ell|^4. 
\end{equation}
The mean value of the inverse participation ratio can be defined by using the Gibbs distribution as
\begin{equation}\label{eq:pminus}
\begin{split}
\langle P^{-1} \rangle= \int P^{-1} \rho d \Gamma= \frac{1}{Z} \int P^{-1} \exp(\beta (\mathcal{H}+\mu \mathcal{A})) d \Gamma,
\end{split}
\end{equation}
where $d \Gamma=\prod_\ell d \mathcal{A}_\ell d \phi_\ell$. With
\begin{equation}
\frac{\partial \mathcal{H}}{\partial g} = \frac{1}{2} \sum_\ell |\Psi_\ell|^4 =  \frac{\mathcal{A}^2}{2} P^{-1}
\end{equation}
it follows
\begin{equation}\label{eq:delz}
\begin{split}
\frac{\partial \mathcal{Z}}{\partial g} &= \int d\Gamma \frac{\partial}{\partial g}\left(\exp(-\beta (\mathcal{H}+\mu \mathcal{A}))\right) \\
 & = -\frac{ \beta \mathcal{A}^2}{2} \int d\Gamma P^{-1} \exp(-\beta (\mathcal{H}+\mu \mathcal{A})). 
\end{split}
\end{equation}
%-\frac{\beta}{2}  \int d \Gamma \left ( A^2 P^{-1} \right) \exp(-\beta (H+\mu A))
Inserting Eq. (\ref{eq:pminus}) into Eq. (\ref{eq:delz}) gives
\begin{equation}
\frac{\partial \mathcal{Z}}{\partial g} = \left(-\frac{\beta}{2}\right) \mathcal{A}^2 \langle P^{-1} \rangle \mathcal{Z}.    
\end{equation}
Hence, we define the Gibbs average of inverse participation ratio as
\begin{equation}
{\langle P^{-1} \rangle}_G  = -\frac{2}{\beta \mathcal{A}^2 \mathcal{Z}} \frac{\partial \mathcal{Z}}{\partial g}.  
\end{equation}
According to the relation $\frac{1}{\textit{Z}}\frac{\partial \textit{Z}}{\partial g}=\frac{\partial \log{Z}}{\partial g}$ obtained from Eq. (\ref{eq:Z}), we can write 

\begin{equation}
{\langle P^{-1} \rangle}_G  = -\frac{2}{\beta \mathcal{A}^2} \frac{  \partial (\log \mathcal{Z})}{\partial g}. 
\end{equation}
By the approximation $ \mathcal{Z}\simeq (2\pi\lambda_0)^N$ from transfer integral operator method, we can simplify it further as
 \begin{equation}
{\langle P^{-1} \rangle}_G     =-\frac{2 N}{\beta  \mathcal{A}^2 \lambda_0}\frac{\partial\lambda_0}{\partial g} =- \frac{2}{\beta a^2 N \lambda_0} \frac{\partial \lambda_0}{\partial g},
  \label{eq:pgibbs}
\end{equation}
where we used the relation $a=<\mathcal{A}> / N$.

\section{Ergodicity and thermalization}\label{sec:ergodicity}

Ergodicity arose from the problems on the statistical properties of dynamical systems. It offers mathematical methods to study the long-term average behavior of complex systems and constitutes key aspects of the phenomenon of thermalization. 
To exemplify the ergodic theory, let me give an example from the book of mathematician Steven Kalikow \cite{kalikow2010outline}: 

\hspace{0.1cm}

\textit{Imagine a potentially oddly shaped billiard table having no pockets and a frictionless surface. Part of the table is painted white and part of the table is painted black. A billiard ball is placed in a random spot and shot along a trajectory with a random velocity. You meanwhile are blindfolded and don’t know the shape of the table. However, as the billiard ball careens around, you receive constant updates on when it’s in the black part of the table, and when it’s in the white part of the table. From this information you are to deduce as much as you can about the entire setup: for example, whether or not it is possible that the table is in the shape of a rectangle.}%\\

\hspace{0.1cm}

The subject matter of these situations spans the ergodic theory, which models them under the abstraction of measure-preserving transformations. Here, let me define measures and transformations. A measure is a concept of size that expresses the probability of an event such that the total measure of a probability space is equal to 1. 
A transformation, on the other hand, is a tool to map a space to itself, assigning one point to another and mostly indicates the evolution in time due to many modeling applications. For example, it maps the density space of GP lattice in time: $(a({t_1}),h({t_1}))$ to  $(a({t_2}),h({t_2}))$ while $t_1\neq t_2$. 
We can interpret that the expected frequencies of certain events must be time invariant under the measure-preserving transformations.

Ergodic theory has the inquiry on the behavior of time averages of various functions along trajectories of dynamical systems. In other words, it studies the long-term behavior of systems preserving a definite form of energy \cite{Coudene2016}. 
From the point of view of modern mathematics, the ergodic theory and dynamical systems are originated at the beginning of the twentieth century, led by the famous polymath Henri Poincar$\acute{e}$ %(1854–1912)
\cite{Coudene2016}. Poincar$\acute{e}$'s recurrence theorem has an important role in statistical physics which states that certain systems will return to their initial state after a long enough time. 
Having said that, the ergodic hypothesis is first introduced by Boltzmann in 1898 \cite{LBoltzman} and proven by Birkhoff \cite{Birkhoff656},  John von Neumann \cite{Neumann}, and Khinchin \cite{Khinchin:232978}  that the time averages of an ergodic system must coincide with the space averages. John von Neumann \cite{Neumann,Neumann2} stated there exists a relaxation time $\tau$ for every significant physical observable $f$ such that the time average coincides with the phase average for all times larger than it.

 Ergodic hypothesis, in a nutshell, consists of two main postulates as declared in \cite{Attard:1540150}, as we consider a phase space of microstates with the same energy, in an isolated system:  1) a single trajectory connects all the points with the same energy 2) the points in phase space of equal energy are equally likely. 
These principles of ergodicity characterize the temporal evolution of a dynamical system.
In this illustrated scheme, the time a system spends in some region (an arbitrary set) of phase space is proportional to the volume of the region (ensemble weight or mathematically called as the size of the set) such that the trajectories (solutions) have to visit almost all states of the available phase space during the motion after long, yet a finite period of time. In other words, for an ergodic system,  the memory of the observable is lost such that  its initial state  cannot be anticipated. 
Expectantly, the trajectories will spend most of their time in regions where the macroscopic variables take their equilibrium values since those regions occupy most of the phase space.  
Consequently, the infinite time average of any observable is equivalent to its phase space average. The system is counted as thermalized when the average of an observable $f$ reaches its ensemble average: $\langle f \rangle_\tau =\langle f\rangle_\Gamma$: 

\begin{equation}
\langle f \rangle = \int_\Gamma f(z) d\Gamma(z)= \lim_{\tau \rightarrow \infty} \frac{1}{\tau} \int_0^{\tau} f(t) dt, 
\end{equation} 
where $\tau$ is the time for thermalization, and $\Gamma$ represents the volume of the phase space.

\section{Ergodicity breaking}

\rightskip=2.95cm
\textit{I had my own contribution to the mix of simple physical models showing complex behavior, namely the spin glass, which initially was a model proposed for an observed phase transition in certain dilute magnetic alloys,  but soon, reappeared as a model for evolutionary landscapes, for neural networks, as a new way of thinking about classifying computational complexity. It adds two new words to the physics glossary: \textbf{frustration} and \textbf{non-ergodicity}.}

\rightskip=0pt
\hspace*{2.2cm} —Philip W. Anderson, More and Different \cite{Anderson2011}\\

Although thermalization, ergodicity, and chaos have been foremost notions of statistical mechanics, in nature there are systems in which these processes do not always emerge, or they even do not take place at all.
Spin glass is one of the complex disordered systems which shows a complicated form of ergodicity breaking where the thermalization properties are hard to foretell. %predict
As a well-known example, conventional glasses also violate ergodicity in a complicated way such that the diffusion process possibly leading to an ergodic behavior is extremely slow. Hence, on very long time scales these "glassy" systems may behave as liquids, but on sufficiently short time scales as solids \cite{Palmer}. 
On the other hand, for the FPUT-like systems (see Sec. \ref{sec:fput}), the diffusion process becomes extremely slow, when the specific energy tends to zero. In this case, the orbits may get trapped in some regions in the phase space within a quite long time scale. If we define the freedom of orbits to visit the whole energy surface as a \textit{liquid} phase, the emergence of trapping may lead to a kind of phase transition from a \textit{liquid} phase to a \textit{solid} one \cite{fputproblem}.

\subsection{FPUT experiment}\label{sec:fput}

In 1923, Fermi introduced his theorem to solve the ergodic problem such that all generic and non-integrable Hamiltonian systems are ergodic. Later, he wanted to test his argument numerically. 
The first computer simulation performed by Fermi, Pasta, Ulam, and Tsingou (FPUT) is published in the 1950s, pretty much the same time as the discovery made by Kolmogorov (see Sec. \ref{sec:kam}). The FPUT experiment started a new approach of studying the issue of ergodicity in classical and quantum many-body interacting systems and yielded a vast number of publications in the last decades including the discovery of solitons \cite{zabusky,ZABUSKY1967126}, and the striking progress in the physics of chaos in Hamiltonian systems \cite{CHIRIKOV197311,CHIRIKOV1979263}.

It is known that without interaction between normal modes of a Hamiltonian system, the equilibrium would never be reached. Hence,  FPUT ran a numerical simulation of the dynamics of a classical chain of 64 particles with a nonlinearity term \cite{Fermi:1955}, violating integrability, and anticipating to reach the thermalization of all the normal modes of the system due to the generation of mode-mode couplings. Yet, they observed an unexpected lack of equipartition for a long time scale due to the suppression of energy exchange between modes despite the presence of nonlinearity. 

They initiated the numerical experiment on a string with a quadratic force and a sine wave. The first few modes $k=1,2,\dots,7$ were successively excited, reaching a state close to equipartition, however, after remaining in the near equipartition state for a while, they had then departed from it. They run the computations for a long time assuming the system would be thermalized. What they expected was an ergodic behavior s.t. all the traces of the initial modes of vibration would fade away while all modes become excited nearly at the same level.
Instead, the classical chain system displayed a very complicated quasiperiodic behavior. In contrast to reaching or staying near to the equilibrium, the trajectory of the observable started to visit the non-equilibrium states again for a while and came back near to the thermalized state.  
These results were quite surprising in the physical community since it was against the main postulate of the Boltzmann-Gibbs statistical mechanics. 
This experiment yielded a lot of questions,  and since then the statistical understanding of this non-equipartition phenomena have been a big puzzle \cite{Ford:1992, Porter:2009}.

\subsection{KAM theorem}\label{sec:kam}

In integrable systems, the action variables are conserved.
The original problem of the KAM theorem is that whether a lasting quasiperiodic orbit occurs when an integrable dynamical system is slightly disturbed. 
The KAM theorem, initially proven by Kolmogorov in 1954, is meticulously  proven and extended by Moser in 1962 for twist maps and by Arnold in 1963 for analytic Hamiltonian systems.
According to the theorem, if a sufficiently small perturbation is applied to an integrable system, many of the solutions of the perturbed system will be confined close to the unperturbed system and will stay stable for an infinite time. Thus, there is only a small change in action for many solutions of the perturbed conservative dynamical system.

In the beginning, the KAM theorem seemed to be the explanation of the delay of ergodicity in the FPUT model. 
However, the theorem states, for a Hamiltonian system of $N\gg 1$, if $\varepsilon\ll \varepsilon_c$ in the Hamiltonian
\begin{equation}\label{eq:h1}
H(I_k,\theta_k,\varepsilon)=H_0(I_k)+\varepsilon H_1(I_k,\theta_k),    
\end{equation}
then the KAM tori can survive on the constant energy surface. Here $I_k$ are countable set of actions, and $\theta_k$ are canonically conjugated angles, where $k={1,2,\dots, N}$. Since the critical perturbation value $\varepsilon_c$ is expected to diminish quickly with growing system size, the FPUT simulations are possibly corresponding to $\varepsilon>\varepsilon_c$.  The finite time lack of ergodicity was explained by a large but finite-time scale beyond which thermalization was observed.

\section{Chaotic discrete breathers}%chaotic breathers
\label{discretebreathers}

As soon as it is understood that the KAM theorem fails to fully explain the FPUT phenomenon, there has been an ongoing debate to unfold the reason for the delay of thermalization in FPUT and related oscillator chains. More recent developments pointed out the existence of isolated structures that the energy is localized on the chain with high-frequency modes \cite{Mackay:1994,sergej1,Campbell2004,sergej3,IVANCHENKO2004120}. 
These exponentially localized  compact structures -known as breathers- are responsible for the delay of the thermalization process of such nonintegrable Hamiltonian lattice models. 
Originally, breathers are discovered as periodic solutions to partial differential equations, and they are analogous to solitons at low frequencies. However, in nonintegrable Hamiltonian lattices, these structures with a high-frequency mode located on the chain are not entirely stable. Although they have  long-time stability as being generic solutions of nonlinear Hamiltonian chains,  they can get dissipated eventually because of chaos. The discovery of these "chaotic breathers" significantly contributes to the understanding of the deferral of ergodization in FPUT-like systems \cite{Rasmussen:2000,Eleftheriou2005,PhysRevLett.77.5225,ELEFTHERIOU200320,PhysRevLett.95.264302,PhysRevE.92.022917}. The lifetimes of these chaotic breathers are so-called \textit{long-time  excursions of a trajectory out of equilibrium}  in Chapter \ref{chapter3}.

In order to have spatially localized modes in a linear differential equation, an "impurity" term has to be introduced to the system so that we can break the discrete translational symmetry of the discretized equations. However, it is also possible to conserve the discrete symmetry of the system and obtain localized excitations at the same time via adding nonlinear terms to the linear differential equations.  Nonlinear terms are present in many applications of many-particle dynamics. Due to nonlinearities, higher harmonics of the excitation frequency are generated and have to be considered too. Here, it is advantageous to use a discrete system, since the finite upper bound of the linear spectrum still allows for frequencies whose whole higher harmonics may lie outside of the linear spectrum. Hence, our target is to investigate the equilibrium dynamics impact of solutions of discrete nonlinear lattices, namely GP, which are spatially localized and periodic in time. These objects act like an \textit{impurity} in Hamiltonian lattices in such a way that the continuous ergodicity of the system is interrupted for a short or long time. 
These high excitations cannot move freely on nonlinear lattices and can exist when a constructive interaction occurs between discreteness and nonlinearity \cite{RevModPhys.82.2257}.

It has been asserted that  if a system has a disorder, discrete breathers survive \cite{sergej1}. However, while adding disorder, the nature of the normal modes of the linear wave equation will change drastically from extended to localized. Normal modes can survive as q-breathers in a weekly nonlinear regime \cite{Flach2008,PhysRevE.73.036618,PhysRevLett.95.064102,PhysRevLett.102.175507}. Therefore, when we add disorder to the GP lattice, the normal modes get localized in the linear regime \cite{sergej2,sergej3,sergejspreading,sergej4,sergej5,Eleftheriou2005}. In this sense, we expect to observe q-breathers in normal mode space and discrete breathers in real space dynamics.

The existence of breathers is further manifested by experiments in different fields of solid-state physics, e.g., superfluids \cite{Ganshin_2009}, semiconducting lasers \cite{PhysRevLett.107.053901}, arrays of waveguides \cite{Eisenberg,LEDERER20081}, optical fibers \cite{Ropers2007}, and microwave cavities \cite{PhysRevLett.104.093901}. The exploration of discrete breathers both answered and brought questions, and this lead to a growing interest in how to properly describe their statistical mechanics.

\cedp
\chapter{Order: Ergodic and nonergodic GP lattice dynamics}
\label{chapter3}
\ifpdf
    \graphicspath{{Chapter3/Figs/}{Chapter5/Figs/PDF/}{Chapter3/Figs/}}
\else
    \graphicspath{{Chapter3/Figs/}{Chapter3/Figs/}}
\fi

\textit{I specifically remember discussions among ourselves and with visitors about what is now known as nonlinear mathematics—truly a strange expression, for it is like saying “I will discuss nonelephant animals”.}
\hspace*{3cm} — S. Ulam (1909–1984) 

\vspace*{1cm}

In the 1950s, the nonintegrable Hamiltonian systems and their dynamics were not well-perceived, as 
Stanislaw Ulam stated above. After their celebrated FPUT experiment and the mathematical description of slightly nonintegrable systems by KAM, a vast amount of studies have been directed to nonlinear systems. Now, nearly integrable dynamical systems \cite{KOMADA198514,yagasaki2021nonintegrability} are considered as a principal class of models in mathematics, statistical and condensed matter physics. 

If an integrable system  with a set of preserved actions is subjected to a nonlinear  weak perturbation (e.g., Eq. (\ref{eq:h1})) it causes a short or long-ranged coupling network in action space. The main purpose of our studies presented in this chapter is basically to analyze the dynamics of observables which become the conserved actions in the integrable limit. %An integrable system is linear, predictable, and exactly solvable with a finite number of nonchaotic solutions. 
To satisfy the rule of ergodicity, which says the trajectory of an observable has to visit everywhere in its phase space, chaos has to be introduced. Hence, an integrable system is always nonergodic by its definition. On the other hand, a nonlinear Hamiltonian system is expected to show ergodic properties due to non-integrability and chaos. This may imply that the dynamics of a perturbed nonlinear system is going to switch from ergodic to nonergodic in the integrable limit. A lot of questions can be raised: Is there a sudden change in the dynamics while crossing over from the ergodic to non-ergodic phase? How does this process occur, and how close a system has to be to the integrable limit to reveal a non-ergodic behavior? Can discrete breathers be excited in a lattice system at thermal equilibrium? 
When the chaotic discrete breathers form, how do they affect the dynamical properties of a thermalized many-body system? In order to address these inquiries, a clear method to quantitatively assess and systematically probe the ergodic to non-ergodic transition near the integrable limits is necessary.

In this chapter, you will find a reliable and distinct scheme to study the gradual loss of ergodicity in many-body interacting systems. The chapter is organized as follows. In the next section,  the ergodicity problem of the discrete GP model is briefly introduced, and in the following section,  the initial states of our computations are defined.  In Sec. \ref{lce}, the reputed Lyapunov characteristic exponents are explained along with our results. Then, the main test with a macroscopic observable to confirm ergodicity at Gibbs temperatures is presented. In Sec. \ref{sec:pdf_method}, we introduce you to the novel method that we tested ergodicity: \textit{statistics of fluctuations}. We explicitly disclose its application to GP lattice and interpret the possible outcomes of the method. Next section \ref{sec:results3n6}, we will share and discuss our results found by the statistics of fluctuations method. In the concluding section, we will recap the main results and comment on their effect on future studies.

%sergej asked me to erase "also called aging and glassy phenomena"
%%In this chapter, you will find a reliable and distinct scheme to study weak nonergodicity in many-body interacting systems. The chapter is organized as follows. In the next section, I briefly introduce the ergodicity problem of the discrete GP model, and in the following section, I define the initial states of our computations.  In Sec.\ref{lce}, I will explain the reputed Lyapunov characteristic exponents, and share our results. I then present the main test with a macroscopic observable to confirm ergodicity at Gibbs temperatures. In Sec.\ref{sec:pdf_method}, I will introduce you to the novel method that we tested ergodicity: \textit{statistics of fluctuations}. In this section, I explicitly disclose its application to GP lattice and interpret the possible outcomes of the method. Next section, I will share and discuss our results found by the statistics of fluctuations method. In the concluding section, I will recap the main results and comment on their effect on future studies. %overall effect in the literature. %A series of Appendices provides technical details.

\section{Introduction }\label{sec:intro}

The classical perturbed Hamiltonian systems with a weak nonlinear component are known as suitable to study ergodicity. The computer experiment by FPUT (see Sec. \ref{sec:fput}) and the celebrated KAM theorem (see Sec. \ref{sec:kam}) initiated the discussion on the ergodicity of weakly nonintegrable Hamiltonian systems. It is later found that, close to their integrable limit, the chaotic discrete breathers (see Sec. \ref{discretebreathers}) can exist and confine the system from thermalization for divergently long time periods \cite{PhysRevX.4.011054,PhysRevLett.98.047202,Flach:1998}. %, as famously studied for FPUT chain (see Sec.\ref{sec:fput}).  
Hence, in their equilibrium dynamics, we expect to observe formations of chaotic breathers in consequence of the nonlinear localization of energy as a signature of a probable nonergodic behavior.

Among all FPUT-like nonlinear Hamiltonian systems, %Among all FPUT-like Hamiltonian systems such as Klein--Gordon and Josephson junction chains, 
we choose the nonlinear GP lattice as a proper candidate to study ergodicity with two constants of motion. % which is a proper candidate to study ergodicity due to possessing two constants of motion. 
The model is defined in detail in Sec. \ref{sec:gpl}. Our study considers %several unresolved problems: the fundamental phenomena of localization, a proper examination of nonequilibrium steady states, and 
a possible connection between the Gibbs to non-Gibbs transition with the phenomenon of ergodicity breaking. The thermodynamics of the Gibbs regime, as explained in Sec. \ref{sec:betamu_xy} with the Hamiltonian Eq. (\ref{eq:Hamiltonian2}), obeys statistical mechanics based on the grand-canonical ensemble. Hence, we expected thermalization and ergodic dynamics for the Gibbs regime of GP lattice. Additionally, we subjected the Gibbs dynamics to a preliminary ergodicity test as in Sec. \ref{sec:ergodicGibbs} to ensure the system is convenient,  before computing statistics of fluctuations in Sec. \ref{sec:pdf_method}. We discover how the system transforms its dynamics from ergodic to nonergodic by  gradually increasing the energy/norm density until a nearly integrable state, where the long-time excitations of chaotic discrete breathers are standstill.

\section{Initial state}\label{sec:initialstate}

\begin{figure}[hbt!] 
    \centering
    \includegraphics[width=0.75\columnwidth]{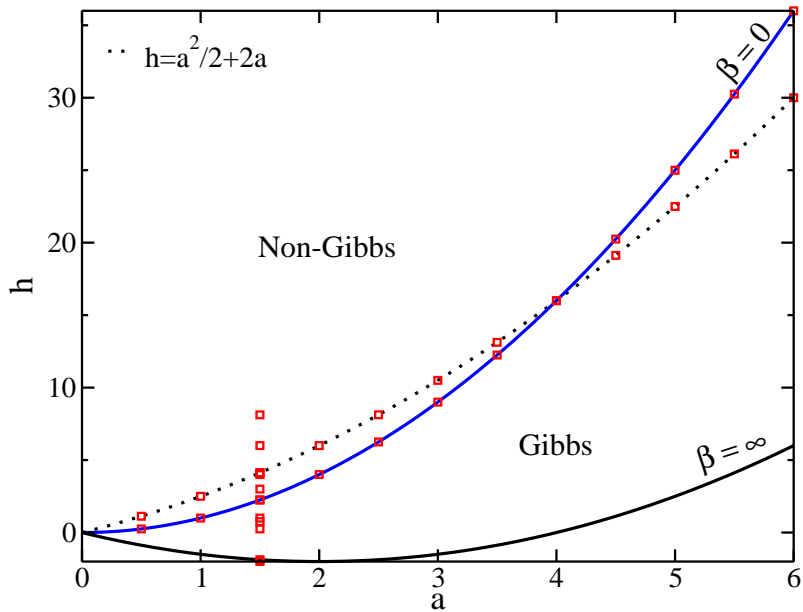}
    \caption[The parameter space diagram with the initial states tested for ergodicity]{{ The phase diagram for the parameters ($a,h$). The red squares are the initial states with $a\leq 6$, whose ergodic test results are exhibited in this chapter. The Gibbs regime is defined in between the solid black and blue lines:  $h=-2a+\frac{a^2}{2}$ ($\beta=\infty$) and  $h=a^2$ ($\beta=0$). Inhomogeneity in the norm distribution is required to reach above the black dotted line: $h=2a+\frac{a^2}{2}$.} }
    \label{fig:initialstates}
\end{figure}

Let me recall the Hamiltonian of discrete GP equation (\ref{eq:Hamiltonian}):
\begin{equation}\label{eq:Hamiltonian2}
H=\sum_{\ell=1}^N \left[\frac{1}{2} |\psi_\ell|^4 - (\psi_\ell \psi_{\ell+1}^* + {\psi_\ell}^* \psi_{\ell+1}) \right],
\end{equation}%measuring the energy in units of $J$ and varying $a$ which is rescaled to tune $g$ 
in which we considered nonlinearity $g=1$, and the hopping strength $J=1$, since the energy is measured in units of $J$, and average norm density $a$ is rescaled to tune $g$. The initial states are chosen from the Gibbs, and non-Gibbs regions of the parameter space in Fig. \ref{fig:initialstates} to record their dynamics and to detect if there is any non-ergodic transition. %, shown as a red dashed line in Fig.\ref{fig:phasediagram_alpha}. 

The initial states can be characterized by two densities $(a,h)$, and the ones considered in this chapter are exhibited by red squares in Fig. \ref{fig:initialstates}. The states with the densities relation $h\leq a^2/2+2a$  can be 
%The initial states $(a,h)$ defined for $h\leq a^2/2+2a$, corresponds to the area below the black dotted line in Fig.\ref{fig:initialstates}. These states 
easily described by $\psi_\ell = \sqrt{a} e^{i \ell \Delta \phi}$ where  the phase  difference of each site is fixed as 
\begin{equation}
\Delta\phi=\arccos{(-\frac{h}{2a}+\frac{a}{4})}. %\Delta\phi=\arccos{(-\frac{h}{2a}+\frac{ga}{4})}. 
\end{equation}

Adding perturbation to an ordered system is necessary to break its exact solution, and may lead to the generation of long-lived localized excitations of unstable modes \cite{Cretegny:1998,Dauxois:1993}. To break the homogeneous distribution of norm and phase, we weakly perturb both of them, i.e., $\psi_\ell=[\sqrt{a}+ \eta_\ell]e^{i [\ell \Delta \phi+\nu_\ell]}$, where $\eta_\ell$ and $\nu_\ell$ are two random realizations distributed over $[-10^{-4},10^{-4}]$. For an initial state above the black dotted line in Fig. \ref{fig:initialstates}, we define $\psi_\ell=\sqrt{a_\ell} e^{i [\ell \pi+\nu_\ell]}$ where $a_\ell=a_1,\dots, a_N$ is a set of positive numbers with a random realization distributed over $[0,2a]$, where $a$ is the targeted average norm density. We observed no noticeable difference in the statistics of fluctuations (explained in Sec. \ref{sec:pdf_method}) of initial conditions with strong or weak perturbations, corresponding to the same thermodynamic state in the densities space. %The time step $\Delta t= 0.02$, unless stated otherwise. %is mainly used and 
If the parameters are not stated otherwise,  the integration was performed with discrete time steps $d t= 0.02$ until $t=10^8$ by using the $SBAB_2$ integration scheme (explained in Appendix \ref{sec:symp}) with $N=2^{10}$, which approximately takes a two weeks CPU time. We examined the equilibrium fluctuations for different system sizes from $N=2^7$ to $N=2^{12}$ \cite{sergej} and  found no influence of system size in the dynamics.

\section{Lyapunov characteristic exponents}\label{lce}
  
\ac{LCE} first introduced by Lyapunov in 1992 \cite{Lyapunov:1992} in the context of the stability of nonstationary solutions of ordinary differential equations. Later, it has been applied to many dynamical systems and is known as a successful method of deterministic chaos. Lyapunov exponent basically quantifies the average exponential rate of divergence of infinitesimally close state-space trajectories of a dynamical system in time \cite{Wolf:1985}. 

Let me define the initial separation between two trajectories of an observable living in the phase space: $\delta_0$. We characterize the divergence of their separation $\delta(t)$ in time with the exponent $\lambda$ as $|\delta(t)|\approx e^{\lambda t} |\delta_0|$. A spectrum of Lyapunov exponents arises when different initial conditions are used, where the number of exponents is equal to the phase space dimension. The maximal LCE (mLCE) is an important parameter which can be approximately found after a long-time observation of the exponent $\lambda$, which is defined as $\lambda_\text{max}=\lambda(t\rightarrow\infty)$. mLCE predicts the degree of nonintegrability and the amount of chaos in a system. The method to extract maximal Lyapunov exponents for GP lattice is demonstrated in \cite{borisfine}.  %predictability,indicates
The necessary time to extract mLCE from a system’s dynamics is called \textit{ergodization time} \cite{Tarkhov_2018}:

\begin{equation}
 \tau_\text{erg}= \frac{1}{\langle\delta\lambda^2(t)\rangle}  {\int_0}^\infty  \langle \delta\lambda(t)\delta\lambda(0) \rangle dt,
\end{equation}
where $\delta\lambda(t)=\lambda(t)-\lambda_\text{max}$. 
%The maximal LCE (mLCE) gives the future of dynamical systems by predicting the nature of the orbits. 
%A positive nonzero value of mLCE indicates the chaotic motion. %The maximal LCE determines the notion of predictability, the degree of nonintegrability, and the amount of chaos in a system. 
If mLCE is positive, the dynamics of the considered system is labeled as chaotic. For the non-chaotic or regular motion, LCE approaches zero with the power-law behavior $\Lambda(t) \propto t^{-1}$ hence mLCE is zero. We apply this method to different initial states of our system with densities $(a,h)$ in order to detect if there is any reasonable difference in chaosity in ergodic/non-ergodic and Gibbs/non-Gibbs phases. 

We first add a weak random perturbation $\delta_\ell(1:N)$ to all the equilibrium states $\psi_\ell=\sqrt{a}e^{i\phi_\ell}$ of each initial state and find their equations of motion by linearizing Eq. (\ref{eq:eqsofmotion}) to the first order in $\delta_\ell$:
\begin{equation}\label{eq:DNLSlce}
i\dot{\delta}_\ell=-(\delta_{\ell+1}+\delta_{\ell-1})+2|\psi_\ell|^2\delta_\ell+\psi_\ell^2\delta_\ell^{\ast}. 
\end{equation}
We numerically solve this equation (see the details in appendix \ref{ap:lce}) to find out the discrete-time dynamics of the perturbation and calculate the LCE for the trajectory $\psi_\ell$ as
\begin{equation}
\Lambda(t)=\lim_{t\to\infty} \frac{1}{t}\log\frac{||\delta(t)||}{||\delta(0)||},    
\end{equation}
 where $||\delta(t)|| =\sqrt{\sum_{\ell=1}^N|\delta_\ell(t)|^2}$  \cite{Johansson:2010}.
%%$\Lambda(t)=\lim_{t\to\infty} \frac{1}{t}\log\frac{||\delta(t)||}{||\delta(0)||},$ where $||\delta(t)|| =\sqrt{\sum_{\ell=1}^N|\delta_\ell(t)|^2}$  \cite{Johansson:2010}. %and $\delta_\ell$ represents the deviation from the trajectory $\psi_\ell$

\begin{figure}[pht] 
\includegraphics[width=7.6cm,height=5.6cm]{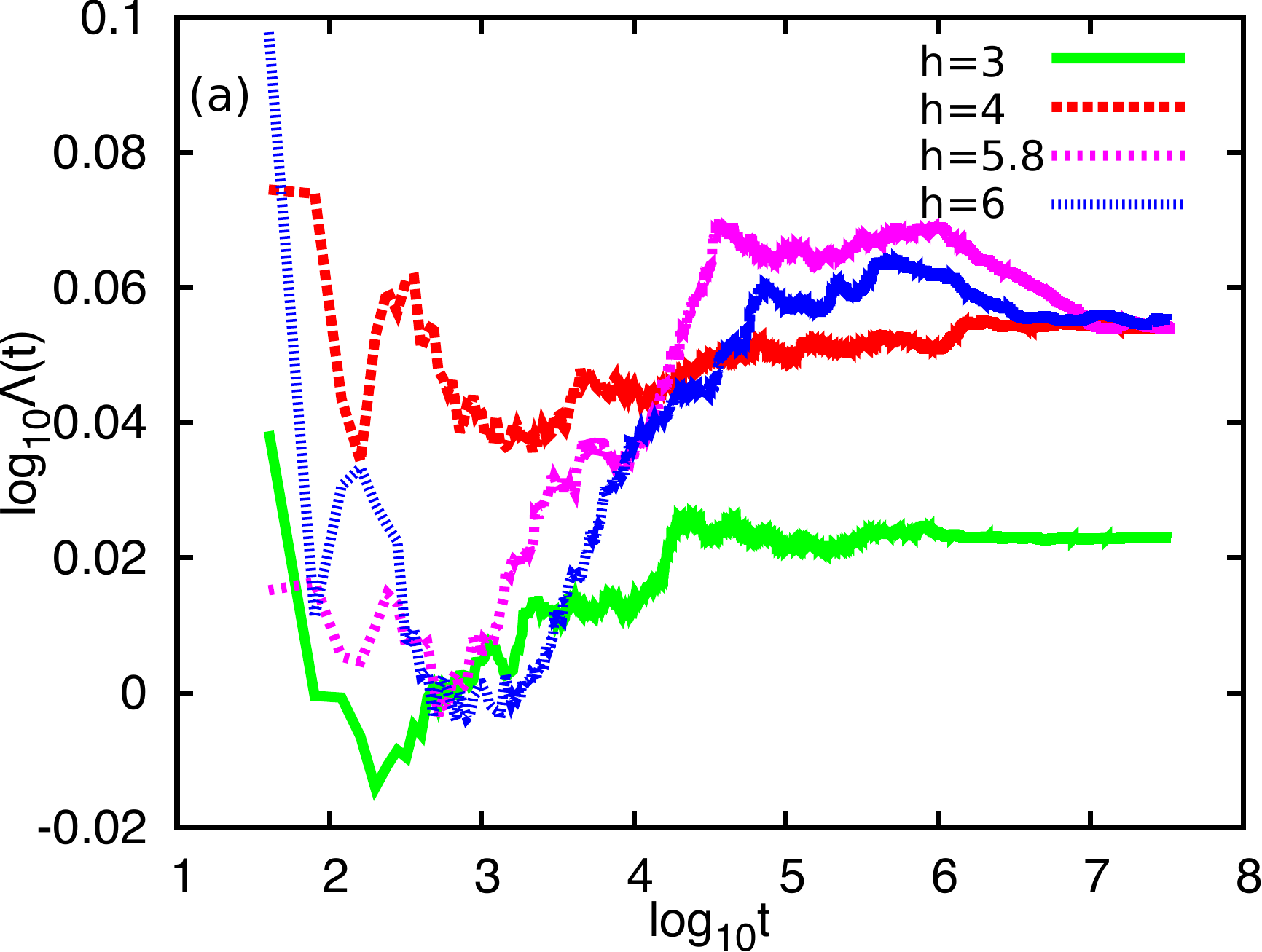}%[width=0.49\columnwidth]
\includegraphics[width=0.2cm,height=0.1cm]{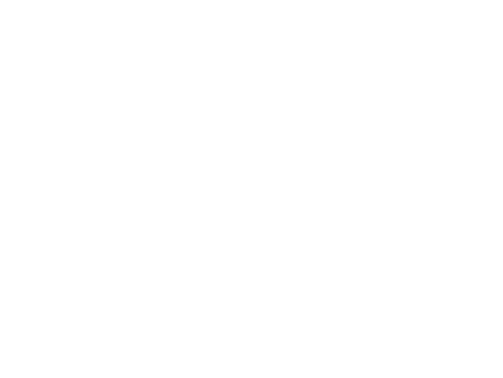}
\includegraphics[width=5.8cm,height=5.5cm]{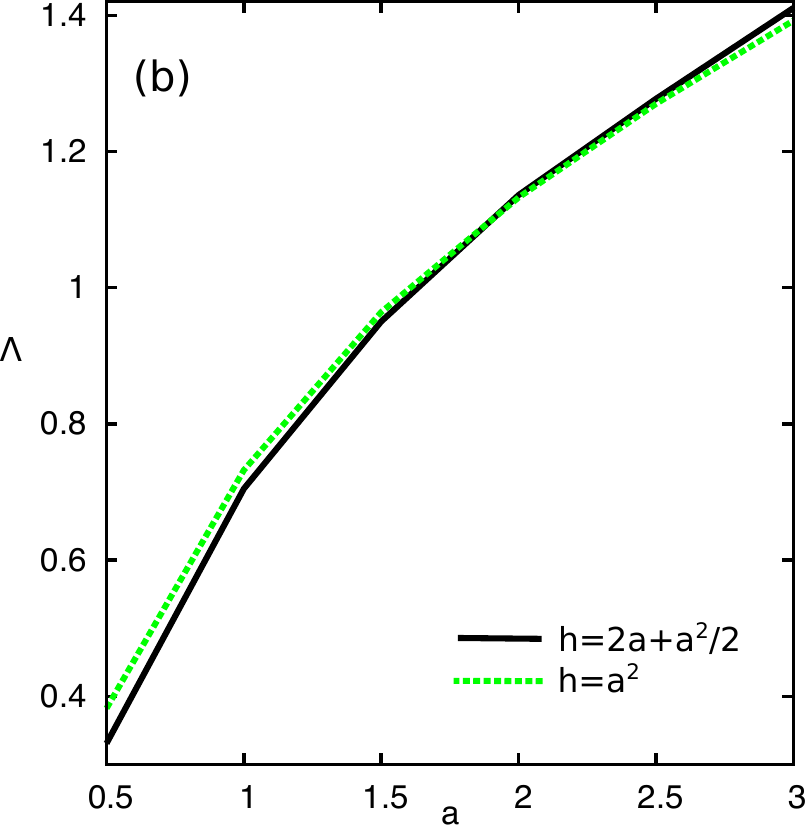}
\caption[The effect of initial state on the maximal Lyapunov characteristic exponent]{\label{fig:lce}{ (a) Time evolution of the  Lyapunov exponent for the fixed norm density $a=2$. The different $h$ values represent the Gibbs regime ($h=3$), the phase transition line ($h=a^2$), the ergodic to non-ergodic transition line ($h = 5.8$) and the non-Gibbs regime. (b) The maximal Lyapunov exponent for the $h=a^2$ and $h=2a+a^2/2$ lines.}}
\end{figure}

In Fig. \ref{fig:lce}(a), the evolution of mLCE is displayed for fixed norm density $a=2$, and energy density $h$ is picked from 3 to 6 in order to observe how LCE varies from Gibbs to the non-Gibbs regime. The mLCE saturates after $t=10^7$, and we calculate it by taking the average of mLCE for the range $t=10^7-10^8$. The positive value of mLCE signifies that the system is chaotic in both the Gibbs regime and the non-Gibbs regime. The comparison of mLCE at Gibbs regime, at $h=3$ and  $h=4$, reveals that the amount of chaos in the system is related to the energy density, and such dependency is also observed for the problem of quartic Klein--Gordon chain of coupled anharmonic oscillators \cite{Skokos:2013}.

Fig. \ref{fig:lce}(b) shows the mLCE for the various values of norm density which exhibits that mLCE increases for stronger interactions (larger norm density). 
As average norm density $a$ decreases to zero, the separation of $h=2a+a^2/2$ and $h=a^2$ lines in Fig. \ref{fig:lce}(b) becomes more clear. % due to the entrance to the nonergodic regime. 
%At a fixed norm $a<2$, the mLCE for the initial states with energy density $h=2a+a^2/2$ are less than as that of $h=a^2$. As average norm density $a$ decreases to zero, the separation of these two lines in Fig.\ref{fig:lce}(b) becomes more clear. 
The reason is the system slowly enters from ergodic to the non-ergodic regime from $a=3$ to $a=2$, and then as $a\rightarrow 0$, it gets closer to the integrable limit, which eventually reduces the amount of chaos in the system.  
%goes deeper in the nonergodic regime as $a\rightarrow 0$. 
%Due to reaching closer to the integrable limit, the chaosity of the system reduces. Although this transition is so slow, it is clear that two curves of mLCE meet at $a\approx 2$, where the system hit the weak nonergodic dynamics with $h=a^2/2+2a$. 
However, overall we did not observe a significant difference in mLCE to detect nonergodicity, as the method tests mainly if mLCE is positive.  Therefore, we commented in our 2018 article \cite{sergej} that the transition from ergodic to the non-ergodic regime is unnoticed by mLCE. %Since the dynamics are dimensionally independent for high energies ($\beta\leq 0$) \cite{Johansson:2004}, we expect to have similar results for higher dimensional discrete GPE.

\section{Ergodicity of Gibbs regime}\label{sec:ergodicGibbs}

For a mechanical system, the ergodic theorem predicts the average behavior of the system over long time periods. As stated in Sec. \ref{sec:ergodicity} previously, in an ergodic system, the infinite time average of a dynamical observable is equal to its ensemble average over the phase space.

In order to verify it, a dynamical observable has to be chosen which can be measured and has a meaning, sensitive to nonlinearity and interactions in the system, and randomly visits the available states in its phase space. To study the ergodicity of the GP lattice, we initially tested the inverse participation ratio as our non-local observable in the  phase space.  The participation ratio $P$, also described in Sec. \ref{sec:participationnumber}, represents the number of non-trivially excited sites $Z_{\nu}=\frac{|\psi_{\nu}|^2}{\sum_{\mu}|\psi_{\mu}|^2}$, and it is defined as $P^{-1}=\sum_{\nu} Z_{\nu}^2$. The time average of the inverse participation ratio, $P_{t}^{-1}$ is calculated from the direct simulation of Eq. (\ref{eq:eqsofmotion}).  Each $(a,h)$ point in the Gibbs phase can be defined with a $(\beta,\mu)$ pair, as described in Sec. \ref{sec:betamu_xy}. We use the Gibbs statistics with \ac{TIO} method to find the ensemble average of the participation ratio $\langle P_G \rangle$. The method is defined in detail in Sec. \ref{sec:tim} along with the calculation of $\langle P_G \rangle$. Recalling Eq. (\ref{eq:pgibbs}), the phase space (Gibbs) average of the inverse participation ratio  $\langle P_{G}\rangle^{-1}=-\frac{2}{N a^2 \beta \lambda_0}\frac{ \partial\lambda_0}{\partial g} $.

%In the thermodynamics limit, it gives $Z \approx (2\pi \lambda_0)^N$  where $\lambda_0$ is the largest eigenvalue of the transfer operator \cite{Rasmussen:2000,Aubry:1975,Scalapino:1972}. 

%Finally, we find the phase space (Gibbs) average of the participation number as $P_{TIO}^{-1}=-\frac{2}{N a^2 \beta \lambda_0}\frac{ \partial\lambda_0}{\partial g} $, as the details with the method are described in Appendix \ref{ap1}). 

 \begin{center}
    \begin{table}[h]%\
    \caption[Ergodicity test with inverse participation ratio]{The inverse participation ratio is averaged over phase $\langle P_{G}^{-1}\rangle$, and time $\langle P_{t}^{-1}\rangle$. The comparison is performed in Gibbs regime of GPL for $g=1, N=1024$.}
    \small
    \centering
    \begin{tabular}{p{0.6cm} p{0.7cm} p{1.1cm} p{1.3cm} p{1.3cm} c} %    \hline
    $\mu$ &  $\beta$ & a & h & $\langle P_{G}^{-1}\rangle$   & $\langle P_{t}^{-1}\rangle$ \\  
 \hline 
         5 & 0.1 & 1.304  & 1.275 & 0.00178 & 0.00185 \\
        2 & 0.1 & 1.943  & 2.514 & 0.00164 & 0.00168 \\
     0 & 0.2 & 2.016  & 1.717 & 0.00157 & 0.00152 \\     
       -2 & 1   & 3.885  & 0.809 & 0.000977 & 0.00103 \\    
        -4 & 1   & 5.923  & 6.714 & 0.000976 & 0.00100 \\
       -6 & 1   & 7.942  & 16.66 & 0.000977 & 0.000990 \\ 
          \hline   
    \end{tabular}
    \label{tab:betamu}
    \end{table}
    \end{center} 
\vspace*{-\baselineskip}
As depicted in Table \ref{tab:betamu}, we found that $\langle P_{G}\rangle \approx \langle P_{t}\rangle$ numerically for a large set of different initial conditions in the phase space. The fluctuations between two data are within the relative error $<3\%$ can be explained with the transfer integral method approximation, integration of the continuously varying Kernel function with discrete steps (see Sec. \ref{sec:tim}), and the finite time evaluation of $\langle P^{-1}(t)\rangle$. These results indicate the existence of an ergodic behavior in the Gibbs regime, i.e., the time spent by the variable $P^{-1}$ in a state is equivalent to the probability of finding this observable in the same state due to the homogeneous distribution of energy. This preliminary test, more importantly, shows GP lattice as a valid testbed for an ergodicity analysis. After this confirmation, we moved our interest to search the ergodicity breaking phenomena with a local observable in GP lattice that is explained in detail in the following section.

\section{Method to test ergodicity: Statistics of fluctuations}
\label{sec:pdf_method}

We use a novel method to investigate the statistical characteristics of equilibrium and non-equilibrium dynamics of 1D ordered GP lattice, as we published in 2018 \cite{sergej}. The success of this method is also shown in several other publications \cite{Danieli:2017,mithun2019,danieli2019}. 
At first, we choose proper functions $f$ of the phase-space variables for a given many-body system. The observable has to  have  equilibrium value(s) independent of the form of its initial trajectories, conserved in time. % Since the nonergodic behavior is expected to reveal itself as concentrated hot spots of localized norm or energy excitations, we select a local observable 
  %in its phase space,     
%\begin{enumerate}[label=(\roman*)]
%  \setlength\itemsep{0em}
%    \item linearly stable coherent states,
%    \item a strong connection with the integrability of the system,
%    \item high sensitivity to the excitation of these nonlinear states (stable local modes),
%    \item equilibrium value(s) in its phase space, independent of the form of its initial trajectory, conserved in time.
%\end{enumerate}

Concerning a thermalizing GP lattice system, the trajectory of the phase space will evolve under the constraint of fixed total energy, and the total norm such that the time average $\langle f \rangle _t$ is independent of the actual chosen trajectory up to a set of measure zero. The actual value of an observable $f(t)$ will depend on time $t$ along a typical trajectory. In this study, we mainly used a GP lattice with 1024 sites which correspond to 2048 independent dimensions considering that our phase space variable $\psi_\ell$ has both real and imaginary parts. Since energy and norm are conserved, we are dealing with a 2046 dimensional subset.

The global observables such as participation ratio can be insensitive to the local breather-like excitations (events) as the number of degrees of freedom increases. To follow all the trajectories in the phase space with a sensitive measurement, we chose a proper microscopic and local observable $f_\ell=g|\psi_\ell|^2$, $\ell=1,...,N$ which turn into integrals of motion in the infinite density limit.   We denote $g=1$, and $|\psi_\ell|^2\equiv a_\ell$, hence we call our local observable simply as $f_\ell=a_\ell$. We follow each trajectory in the phase space using an ergodic Poincar$\acute{e}$ section $\mathcal{F}_f$, and keep track of the instants  a trajectory pierces this hypersurface. The local norm densities  define $1024$ ergodic Poincar\'e manifolds $\mathcal{F}_\ell : f_\ell\equiv a$, where norm density $a=\langle \mathcal{A} \rangle/N$ obtained by the integral of motion of the system: total norm $\mathcal{A}$. %, while the parameters $g\equiv J=1$ kept constant as in Eq.\ref{eq:Hamiltonian2}. 
We integrate each observable $a_\ell(t)$ in time, and trace \& note each time $t_i^{(\ell)}$ the trajectory pierces any of the equilibrium hypersurfaces $\mathcal{F}_\ell$. The time intervals between each pair of sequential piercings by a trajectory are called recurrence times or excursion times and symbolized with $\tau$. These recurrence times are ideally used to highlight the certain conditions where the observables spend divergently long time periods away from $\mathcal{F}_f$. This implies the confinement of a trajectory in non-equilibrium states of the available phase space for excessive times and the consequent breaking of ergodicity. We define the excursion times as $\tau^{(\ell,\pm)}(i)=t^{\ell}_{i+1}-t_i^{\ell}$ where $t_i$, and $t_{i+1}$ for each trajectory stand for two consecutive piercing times with $i=1,\dots,M$. The total number of piercings $M \times N$ must be infinite in the thermodynamic limit ($N\rightarrow \infty$), or in the infinite time limit ($M\rightarrow \infty$) for an ergodic system. In our numerics, the number of piercings is controlled not only by the system size and the maximum integration time but also by the selected initial state $(a,h)$. The number of piercings may reach more than $10^{18}$ as one might see the hint in Fig. \ref{fig:y_x_1n5}. %This requires one to find a clever way to handle that many recurrence times.

 \begin{figure}[pht]  %\label{fig:IS_x2y6_4}%
\includegraphics[width=1\columnwidth]{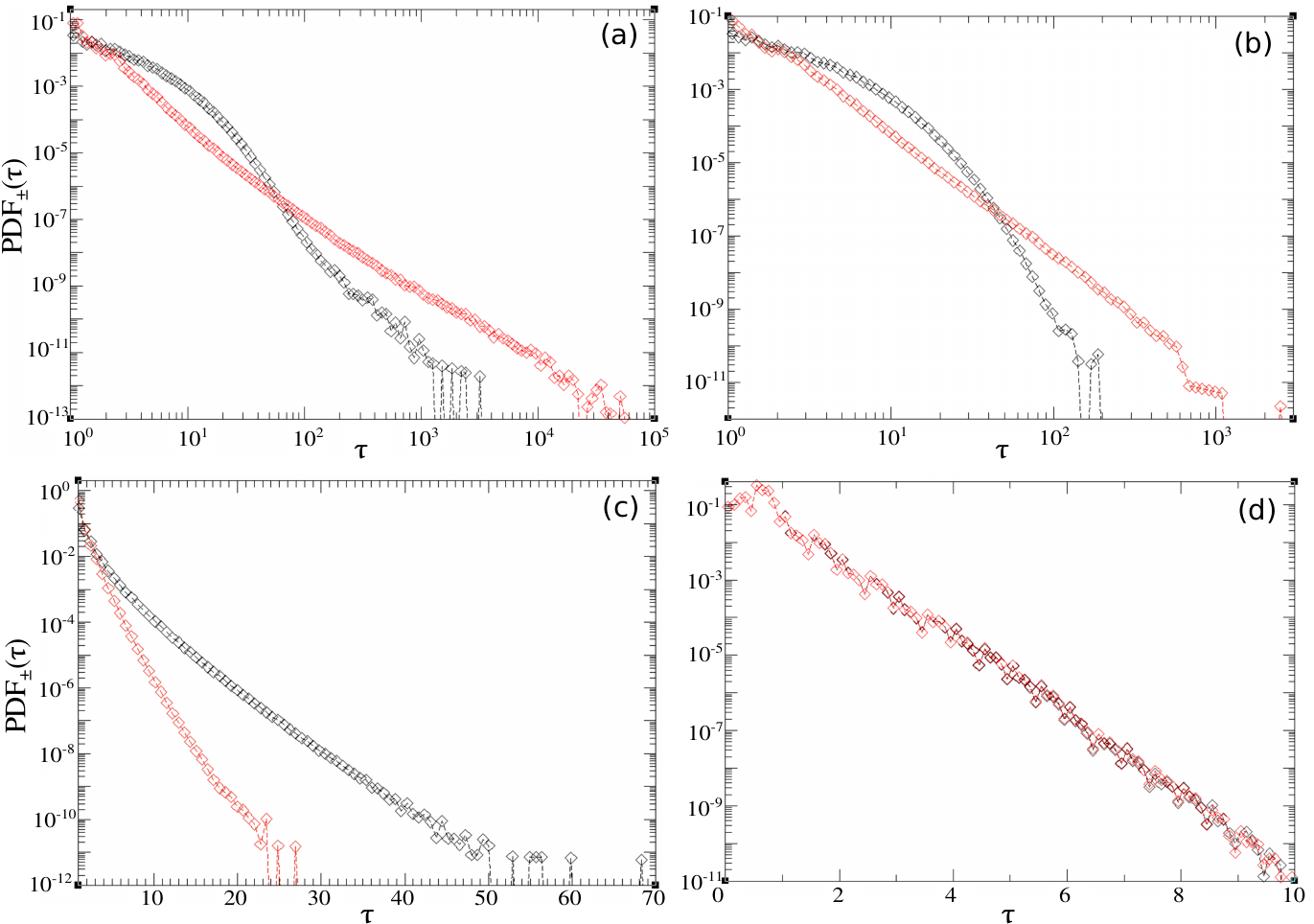}
\caption[Probability distribution function of recurrence times on constant norm]{{  Probability distribution function $\text{PDF}_{\pm}(\tau)$ with $a=2$. $\text{PDF}_{+}(\tau)$, and $\text{PDF}_{-}(\tau)$ are shown as red, and black squares, respectively with $N=2^7$. (a) $h=6$, (b) $h=4$, (c) $h=1$, (d) $h=-2$.}}
\label{fig:positivetau}
\end{figure}

We monitored the excursion times as $\tau_{+}$ and $\tau_{-}$ in order to determine whether they indicate a local depletion ($+$) or augmentation ($-$) of the norm density. We detected it for each trajectory by analyzing if $a_\ell$ is greater ($+$) or less ($-$) than $a$ throughout the excursion. We then obtain the probability distribution functions of the excursion times $\text{PDF}_{\pm}(\tau)$. They are presented in Fig. \ref{fig:positivetau} for different initial states $(a=2,h)$ as subfigures corresponding to zero temperature (d), finite temperature (c), infinite temperature (b), and negative temperature (a). 

%They are presented in Fig.\ref{fig:positivetau} for different initial states $(a=2,h)$ corresponding to the ground state $\beta=\infty$: Fig.\ref{fig:positivetau}(d), finite temperature: Fig.\ref{fig:positivetau}(c), infinite temperature $\beta=0$: Fig.\ref{fig:positivetau}(b), and negative temperature $\beta<0$: Fig.\ref{fig:positivetau}(a). 

Close to zero temperature in Fig. \ref{fig:positivetau}(d), we obtain almost matching results for $\text{PDF}_{+}(\tau)$, and $\text{PDF}_{-}(\tau)$ that signifies symmetry in the fluctuations of the observable value around its equilibrium manifold. At the ground state, all trajectories have the same phase $\phi_\ell=\phi$ and the same norm distribution $a_\ell=a$. The decay of $\log_{10}({\text{PDF}_{\pm}(\tau)})$ is linear, and the largest recurrence time corresponds to $\sim 10$ milliseconds in a real-life experiment, which means the system is absolutely clean of any breather formation.

As the temperature increases, $\text{PDF}_{+}(\tau)$, and $\text{PDF}_{-}(\tau)$ get separated (see Fig. \ref{fig:positivetau}(c)), while an algebraic decay of $\text{PDF}_{+}(\tau)$ in the tails with $\tau^{-\alpha}$ slowly takes place due to the formation of high-amplitude structures. These localized hot spots, which may also be called strong excitations or chaotic breathers (see Sec. \ref{discretebreathers}) in the phase space of GP lattice, can be realized as high local norm densities $a_\ell$. When an observable $a_\ell$ visits the hot regions of the phase space, it has to hold a value larger than the average norm density: $a_\ell>a$. After its trip, the trajectory comes back to the equilibrium value $a$ and this travel is recorded as $\tau^+$. We may expect the neighboring sites to possess $a_\ell < a$ due to the conservation of the total norm. Yet this effect is distributed over several sites via nearest-neighbor interactions. Hence, we see an enhancement on  $\tau_{-}$ 
%Hence, we see an enhancement on $\tau^{-\alpha}$, 
but it is an indirect extended impact of probable chaotic discrete breathers via the de-excitation of modes. 
Moreover, we are not aware of any low-temperature spots which cause similar glassy dynamics in ordered GP lattice. Therefore, we could not observe an algebraic decrease of $\text{PDF}_{-}(\tau)$.  Nevertheless, in Fig. \ref{fig:positivetau}(c), far away from the integrable limit and hot spot formations, the travels while $a_\ell<a$ are a bit longer, which may be due to the effect of closeness to the ground state. %These negative recurrence times might have been larger if the disorder was introduced, especially if the initial state was located in the Lifshits regime. 
An algebraic decay $\tau^{-\alpha}$ of $\text{PDF}(\tau)$ (see Fig. \ref{fig:positivetau}(a),(b)) hints at the formation of long breathers, henceforth we will only focus on the  $\text{PDF}_{+}(\tau)$ results, which are more efficient to calculate the exponent $\alpha$. 
%An algebraic decay $\tau^{-\alpha}$ of $\text{PDF}(\tau)$ hints the formation of long breathers, therefore from now on I will only present and discuss the  $\text{PDF}_{+}(\tau)$ results, which are more efficient to calculate the exponent $\alpha$. 

We attempt to fit the PDF tails with a power-law $\text{PDF}(\tau) \propto \tau^{-\alpha}$ to find the dependence of the exponent $\alpha$ on the densities $(a,h)$ in order to make use of the following scheme:

\begin{equation}
\langle \tau^m \rangle =\int_1^{\infty}\tau^m \frac{1}{\tau^{\alpha}}d\tau =\int_1^{\infty} \frac{1}{\tau^{\alpha-m}} d\tau
\end{equation}
\begin{itemize}
  \setlength\itemsep{0em}
\item $\alpha-m\leq 1 \qquad \quad \Rightarrow\quad \langle \tau \rangle, \langle \tau^2 \rangle$ diverges.
\item $1\leq \alpha -m \leq 2 \quad \Rightarrow \quad \langle \tau \rangle$ converges, but $\langle \tau^2 \rangle$ diverges.
\item $\alpha-m > 2 \quad \qquad \Rightarrow \quad \langle \tau \rangle, \langle \tau^2 \rangle$ converges.
%For $(x=1,y=0.56)$, $\alpha>4$
\end{itemize}
 
For $\alpha \leq 2$ non-ergodic dynamics takes place since the average of the excursion times $\langle \tau \rangle$ diverges. Thus, the time average of the observables, which is one of the essential properties of an ergodic system, does not exist. % make sense. 
The divergence of average indicates an infinite time taken by a trajectory $|\psi_\ell|^2$  to pierce back the equilibrium manifold. While it spends the infinite time in a non-equilibrium state, we may construe this behavior as the trajectory is sticky in some part of phase space close to breather formations. Consequently, in the other available states, the nonvisiting of trajectory for an infinitely long time causes a non-ergodic behavior. %% which consequently causes a non-ergodic behavior.  %Consequently, in the other available states, the nonvisiting of trajectory causes a non-ergodic behavior. 
The non-ergodic regime exists in the non-Gibbs phase of the densities diagram, exhibited above the red dashed line on Fig. \ref{fig:phasediagram_alpha}. 

%This nonergodic regime whose border is clearly shown as a red dashed line on Fig.\ref{fig:phasediagram_alpha}, lives in the nonGibbs regime.

When $\alpha\leq 3$ the average of recurrence times stays finite, yet their variance $\langle \tau^2 \rangle$ diverges, which may be interpreted as a disturbance to the ergodicity of the system. It indicates the trajectories might spend much longer times in out-of-equilibrium spaces, yet they all pierce back the equilibrium manifold at the end of their visits, so this regime is still ergodic. %, and the order of the equilibrium fluctuations of the observable value is broken. 
%Yet, all the trajectories come back to their equilibrium at the end of their visits, so this regime is still ergodic. 

This method has been successfully used to test ergodicity breaking phenomena in different systems such as classical Josephson junction chains \cite{mithun2019},  Fermi--Pasta--Ulam--Tsingou (FPUT) \cite{Danieli:2017} and Klein--Gordon (KG) lattices \cite{danieli2019}. %The general character of non-ergodic behavior is the power-law probability distribution, where the average of the recurrence times in the sticky region of the phase space diverges. The success of this analysis is more related to the selection of the most sensitive observable, and we measure the exponent from the power-law tail obtained from the tracking of a selected sensitive observable.

\section{Results} \label{sec:results3n6}

%The initial states with $\alpha=3$ can be always defined by Gibbs temperatures, and they are located near to the infinite temperature line for non-strong nonlinearity in Fig.\ref{fig:phasediagram_alpha}, shown as green dashed line.

%We did not exclude the presence of exponential cutoffs in the unresolvable part of $PDF$ at large values of $\tau$ (e.g., Fig.\ref{fig:y_x_1n5}), while computing $\alpha$ shown in Fig.\ref{fig:phasediagram_alpha}. For all of our tests, the location of an initial state in the densities diagram (a,h) is the solely effective matter on the results. Yet, its selected form -described in Sec.\ref{sec:intro}- is inconsequential on our $PDF(\tau)$ results in both ergodic and nonergodic regimes, according to our analysis. 

When the nonlinear part of the Hamiltonian dominates the linear part, the nearest neighbor interactions become negligible. The loss of the energy transfer through hoppings in the system triggers the localization of energy on lattice sites. With the stronger nonlinearity, the system reaches the integrable limit by manifesting the manner of independent oscillators which leads to nonergodicity. %eh

\begin{figure}[hbt!] 
    \centering
    \includegraphics[width=7.2cm,height=5.2cm]{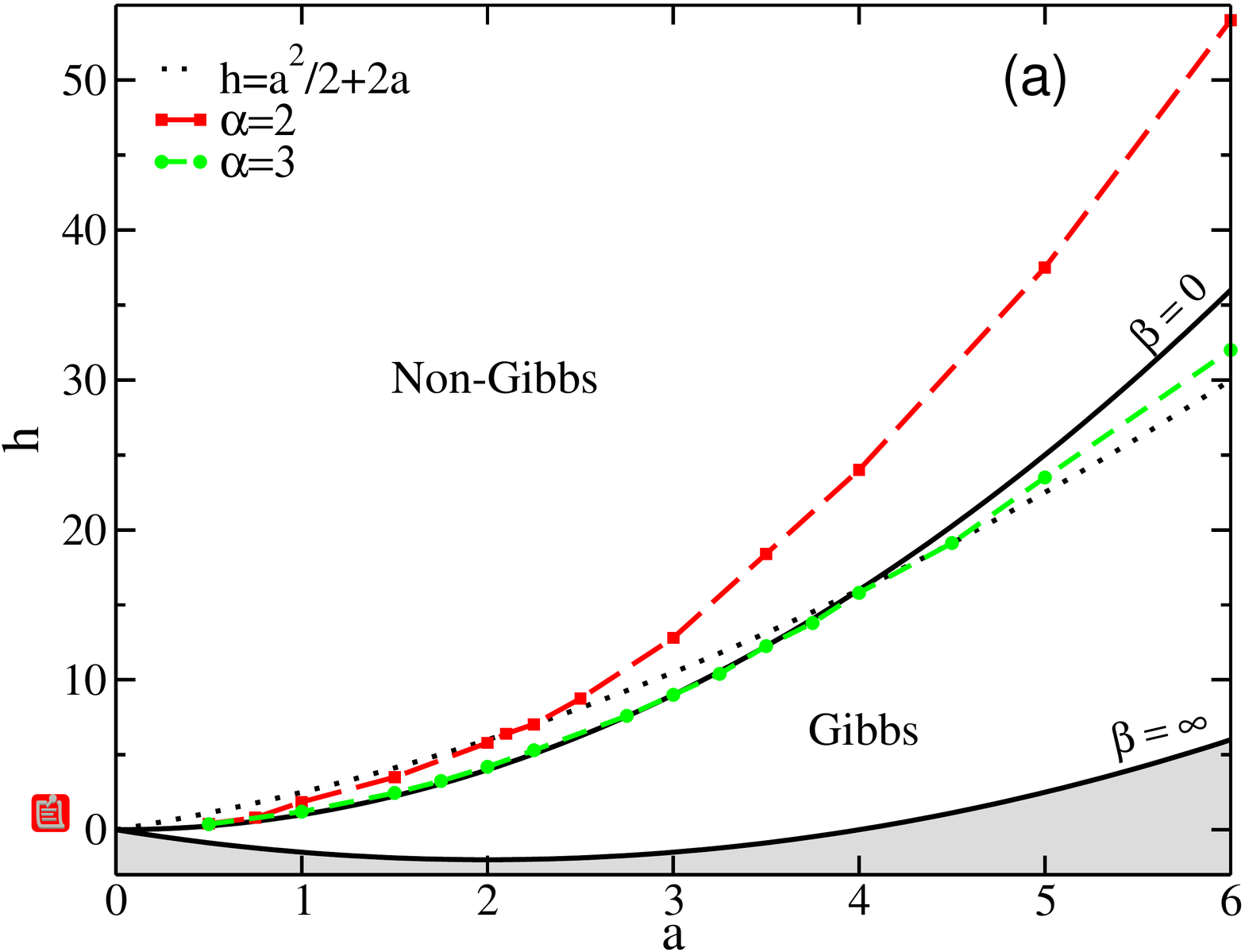}
        \includegraphics[width=6.2cm,height=5.2cm]{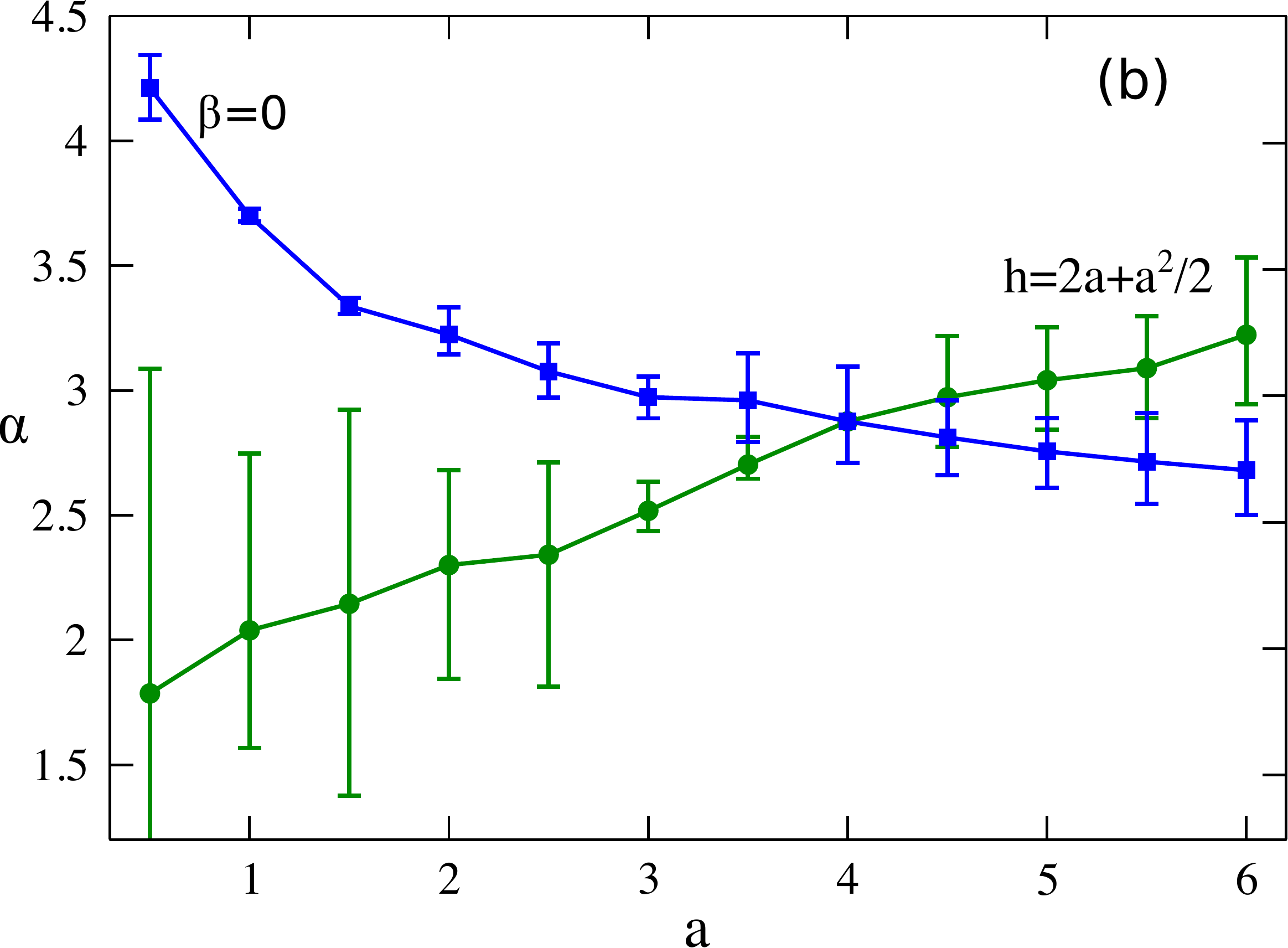}
    \caption[The ergodic to non-ergodic transition on the parameter space diagram]{{ The ergodic to non-ergodic transition on the parameter space diagram determined by the exponent $\alpha\leq 2$. (a) Phase diagram for the parameters ($a,h$). The Gibbs regime is defined in between the solid black lines:  $h=-2a+\frac{a^2}{2}$ ($\beta=\infty$) and  $h=a^2$ ($\beta=0$). Inhomogeneity in the norm distribution is required to reach above the black dotted line: $h=2a+\frac{a^2}{2}$. The statistics of fluctuations method (see Sec. \ref{sec:pdf_method}) is used to obtain $\alpha=2$ and $\alpha=3$ dashed lines, shown as red and green, respectively. (b) The exponent $\alpha$ of the power-law tail for the norm density $a$ calculated for $h=a^2$ and $h=2a+\frac{a^2}{2}$.} }
    \label{fig:phasediagram_alpha}
\end{figure}

The $h=2a+\frac{a^2}{2}$ line, has an initial state with homogeneous norm and fixed phase difference $\pi$ between neighboring sites $\psi_\ell=\sqrt{a}e^{i\ell \pi}$. %, which is subjected to a perturbation defined as in Sec.\ref{sec:initialstate}. 
While norm density $a$ increases on this black dotted line shown in Fig. \ref{fig:phasediagram_alpha};
\begin{itemize}
  \setlength\itemsep{0em}
    \item the system reaches to the Gibbs regime at $a=4$ with the infinite temperature, and then the temperature starts to decrease
    \item the system moves away from the integrable limit $a\rightarrow 0$,
    \item the number of breather solutions reduces,
    \item  the probability of short-time excursions enhances,
    \item the exponent $\alpha$ decreases.
\end{itemize}

On the other hand, if we shift our initial states $(a,h)$ in the direction of increasing norm density $a$  on $\beta=0$ line in Fig. \ref{fig:phasediagram_alpha};

\begin{itemize}
  \setlength\itemsep{0em}
%\item the temperature is infinite,
\item the nonlinear part of the Hamiltonian gradually dominates,
\item the lifetime and density of breather-like excitations increases,
\item the probability of larger excursion times grows until the hopping part becomes completely negligible,
\item the exponent $\alpha$  decays and converges to $2$ in the $a\gg J=1$ limit.
\end{itemize}

\begin{figure}[pht] 
\includegraphics[width=7.6cm,height=5.6cm]{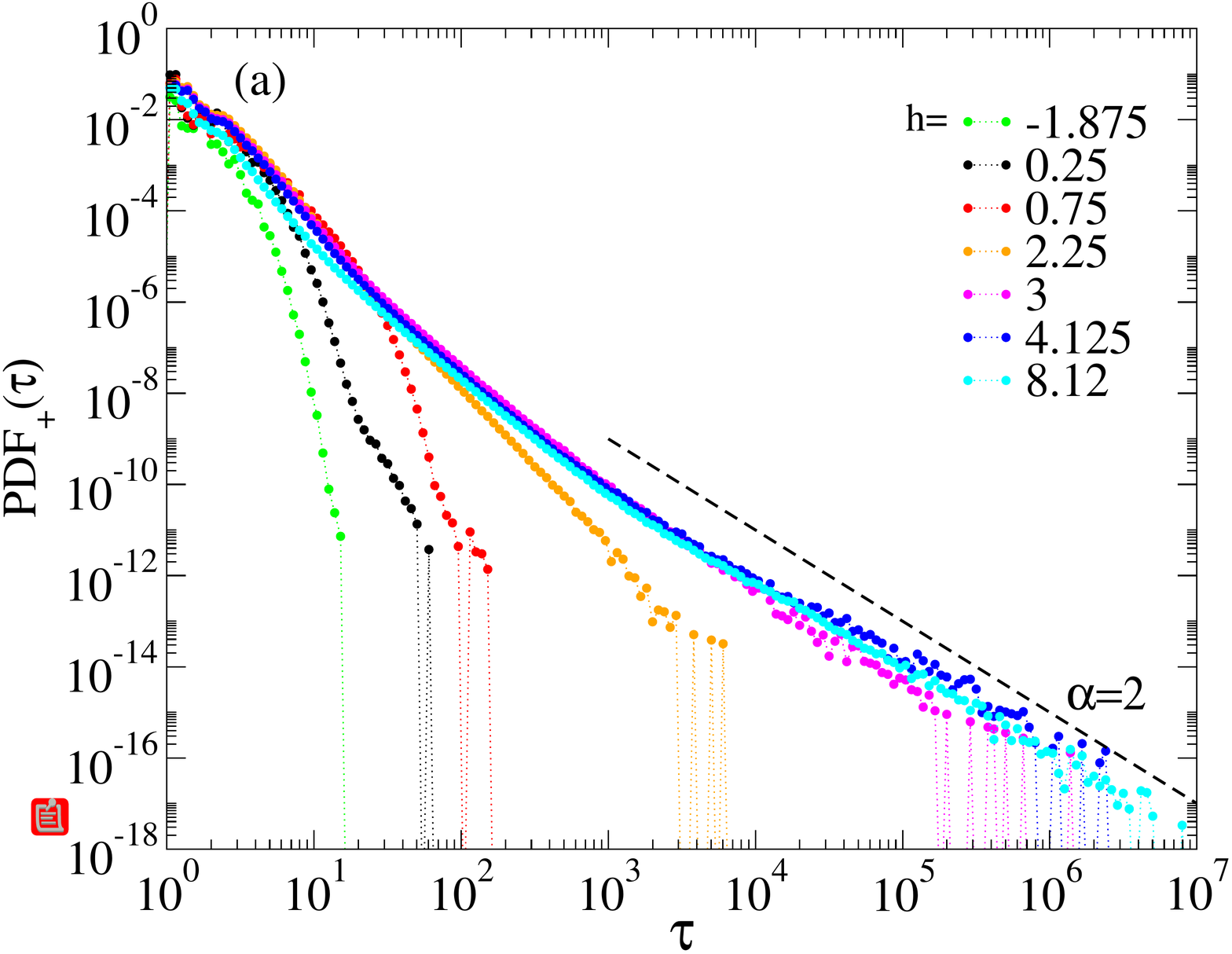}%[width=0.55\columnwidth]{x1n5const.pdf}%y_x_1n5.png}
\includegraphics[width=6.6cm,height=5.6cm]{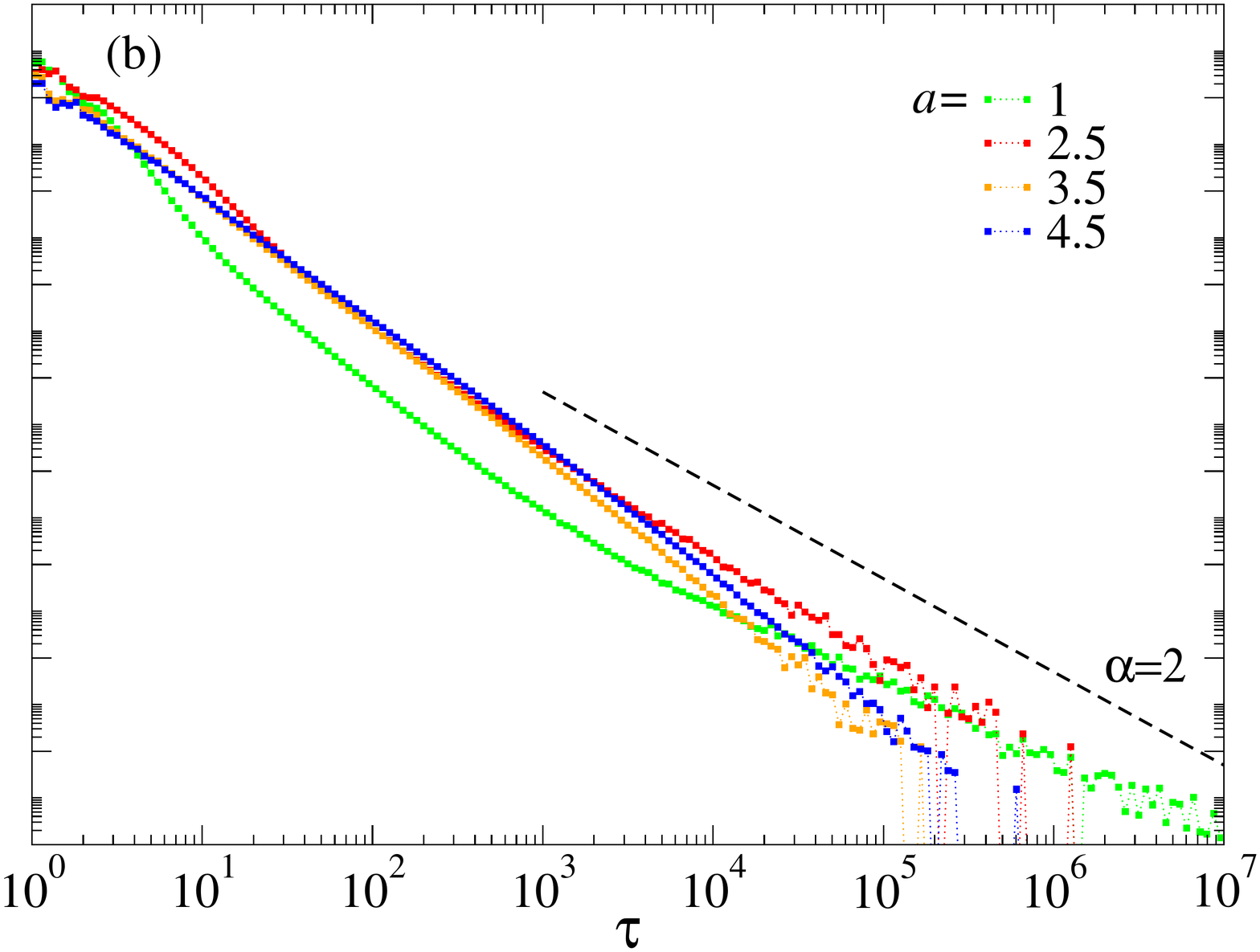}%PDF_yequals2xplurxsquareover2.pdf}
 \caption[Probability distribution function of recurrence times on norm inhomogeneity line, and on constant norm line]{\label{fig:y_x_1n5}{ (a) Probability distribution function $\text{PDF}_{+}(t_r)$) of different energy densities $h$ alternating from $-1.875$ to $8.12$, with constant norm density $a=1.5$ for all. (b) Probability distribution function $\text{PDF}_{+}(\tau)$) of norm densities $a$ differ from $1$ to $4.5$ on $h=a^2/2+2a$ line.}}
\end{figure}

The exponents, calculated from the power-law,  for the different $a$ and $h$ choices along the two lines $h=a^2$ and $h=2a+\frac{a^2}{2}$ are marked in Fig. \ref{fig:phasediagram_alpha}(b). The bar in the figure represents the standard deviation. It gets larger fluctuations when $\alpha \lesssim 2$, which is the case for $a<3$ on $h=2a+\frac{a^2}{2}$ line, since the very end of the PDF tails begin to get fatter due to the sticky dynamics of nonergodic regime (e.g., Fig. \ref{fig:y_x_1n5}(a)). The initial states with $\alpha\geq 3$ are always defined by Gibbs temperatures, where $\alpha=3$ curve is located near to the infinite temperature line for non-strong nonlinearity in Fig. \ref{fig:phasediagram_alpha}, shown as green dashed line. The Gibbs to non-Gibbs transition line $h=a^2$ does not show the divergence of the recurrence times $\alpha>2$; while, $h=2a+\frac{a^2}{2}$ line shows that the exponent crosses the value $\alpha =2 $ for small values of norm density. 
Thus, we interpret that the Gibbs-nonGibbs transition line does not separate the ergodic to non-ergodic dynamics in the non-strong interaction case.

\begin{figure}[pht] \label{PDFyxsquare}
 \includegraphics[width=0.5\columnwidth]{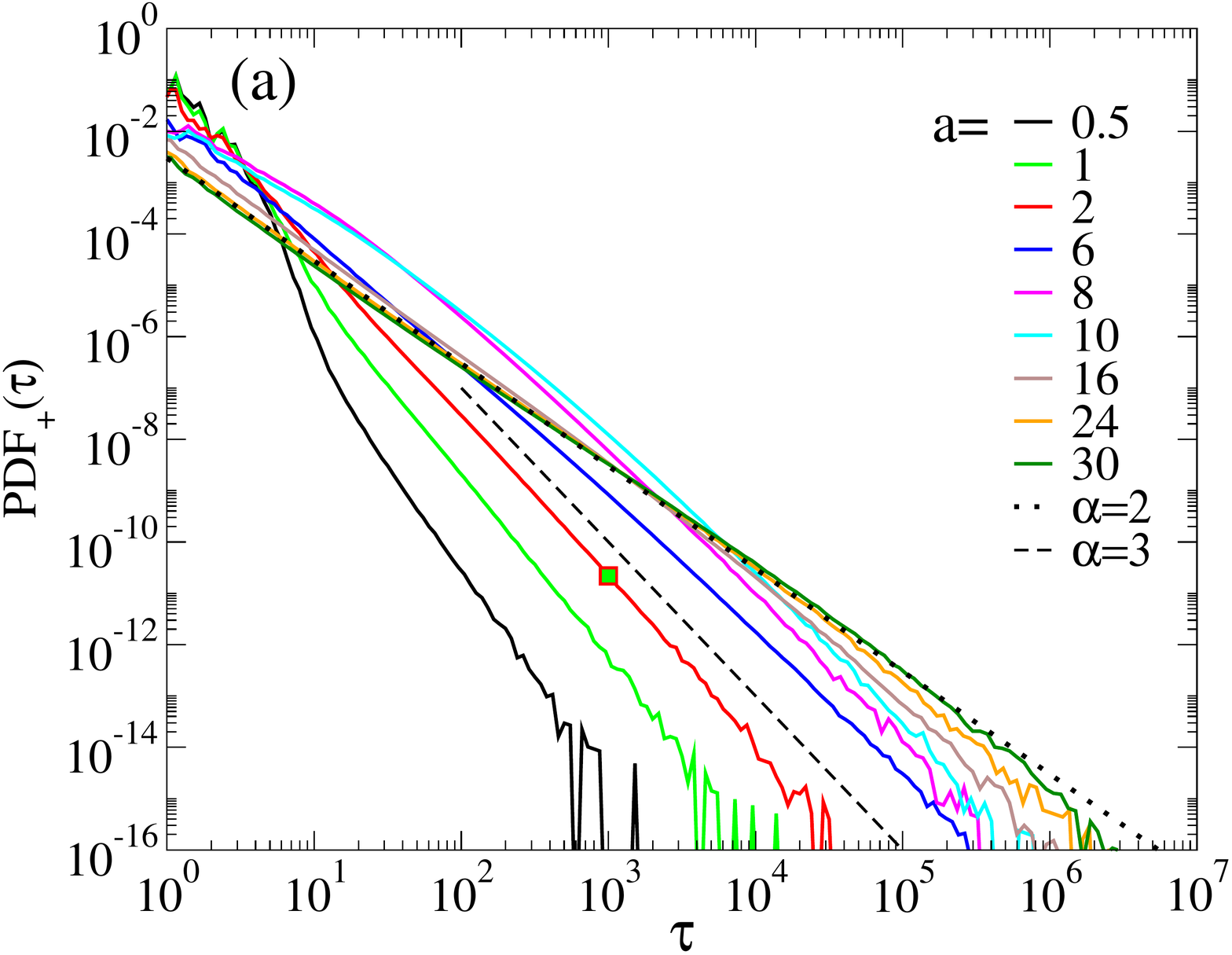}
 \includegraphics[width=7cm,height=5.6cm]{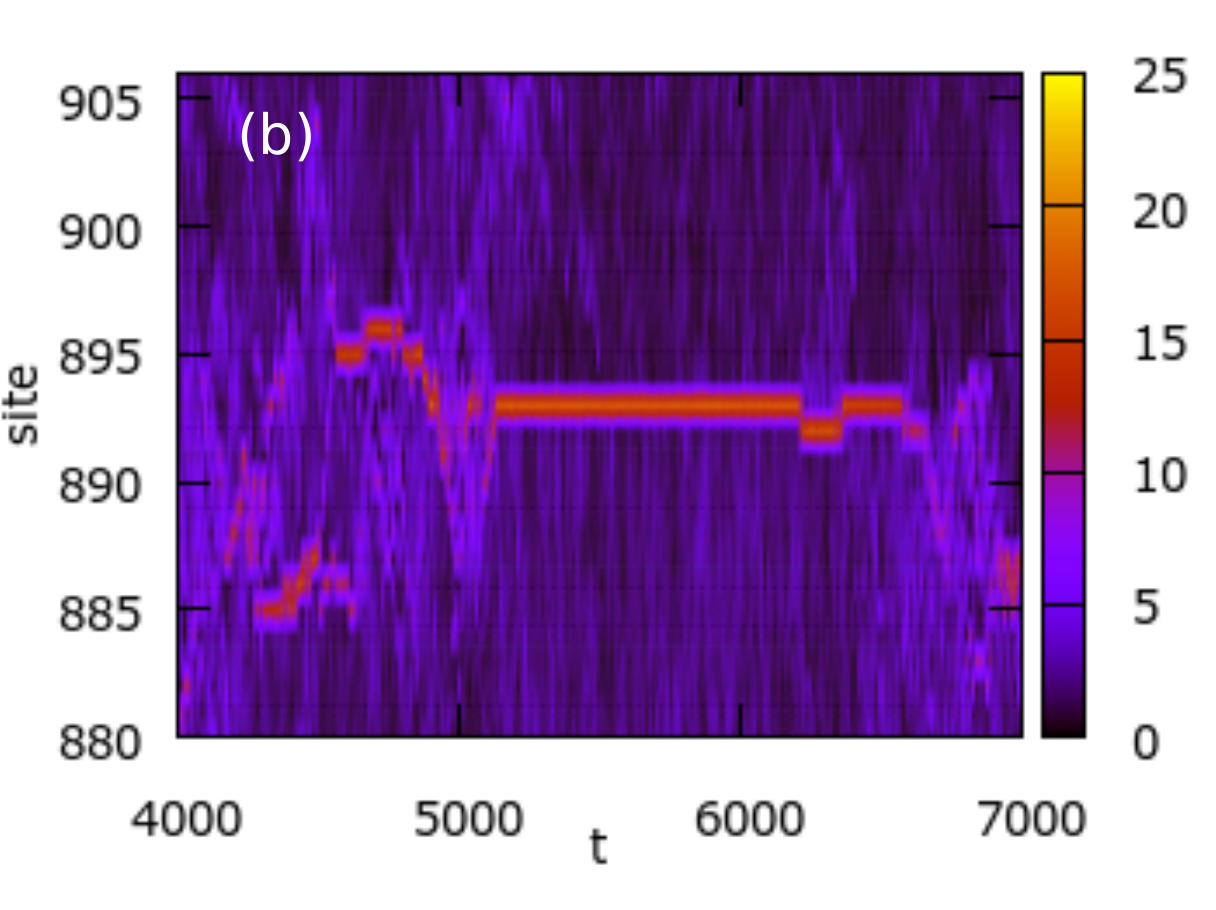}%.eps}
 \caption[Probability distribution function of recurrence times at infinite temperature]{\label{fig:PDFyxsquare}{ (a) Probability distribution function $\text{PDF}_{+}(\tau)$) on the $\beta=0$ line. The average norm densities $a=0.5,1,2,6,8,10,16,24,30$ correspond to black, red, green, blue, magenta, cyan, brown, orange, dark green lines. $\alpha=2$, and $\alpha=3$ are shown as black dotted, and dashed lines. $ g=1, h=a^2, N=2^{10}, dt=0.01$. The relative energy density is kept $<10^{-3}$ for all. (b) Evolution of density, $|\psi_\ell|^2$  in correspondence of one of the excursion times  marked with an orange square in (a).}}
\end{figure} 
%for the same parameter $a=2$ used in (a). The red color represents the maximum density of evolving breather and the corresponding recurrence time is marked with the orange square in (a).
While we increase the norm density on the $\beta=0$ curve linearly from $a=0.5$ to larger values, depicted as a blue line on Fig. \ref{fig:phasediagram_alpha}(b), the exponent of $\text{PDF}_+$ starts with $\alpha>3$, showing ergodic dynamics, and it exponentially decays to 2. For a large norm density in Fig. \ref{fig:PDFyxsquare}(a), the exponent $\alpha=2$ is observed for five-decades long. It means there are discrete breather formations that stay for longer times in highly interacting GP lattice, which will eventually break the ergodicity. Close to the strong nonlinearity limit in Fig. \ref{fig:PDFyxsquare}(a), all the PDF results gradually converge to the same PDF pattern, which must have $\alpha=2$ in the $a\gg J=1$ limit. This is the other integrable limit where the system consists of disconnected anharmonic oscillators, and the hopping terms get negligible. Close to that limit, the nearest neighbor interactions can create a short-range network of nonintegrable perturbations in the system. 
%This is the other integrable limit, in which the hopping terms get negligible. 
Nevertheless, we do not reach this integrable limit on $h=a^2/2+2a$ line with growing $a$. It is because the decrease in the temperature has a competing effect on increasing the number of hoppings.

The appearance of high amplitude breathers with the increasing norm density in GP lattice is reported in several articles \cite{Rumpf:2008,Rasmussen:2000,Johansson:2004}. Our method, in particular, explores the existence time of these high amplitude breathers, and relates the long excursion times with the formation of concentrated hot spots. In the equilibrium dynamics of the lattice system, an event can start with localization of norm on one of the sites due to the relative dominance of the nonlinear part of the Hamiltonian to the nearest neighbor interactions. These chaotic discrete breathers survive over the whole period of the excursion and then slowly diffuse their norm back into the other degrees of freedom until the end of the event. 
%%At the beginning of the event we observe the focusing of norm in one of the sites. These breatherlike excitations then survive over the entire duration of the excursion, only to dissolve their norm back into the other degrees of freedom at the end of the event.
In Fig. \ref{fig:PDFyxsquare}, we show the excursion of an event corresponding to the point marked with the green square. It is obvious that as long as the densities are concentrated in a site, represented by red color, interaction with the neighboring sites is frozen. This captured event at infinite temperature discloses how the discrete breather-like excitations can occur in time. However, here, the entire duration of the excursion is relatively short ($\sim 10^3$) compared to the chaotic discrete breathers (e.g., $\sim 10^6$) located in the nonGibbs regime,  responsible for the fat PDF tails with $\alpha\leq 2$. %We relate the long excursion times with the  formation of concentrated hot spots.%  generation of chaotic discrete breathers.
%concentrated hot spots or

%In the equilibrium dynamics of the lattice system,  the relative dominance of the nonlinear part can create temporary events by focusing of norm in one of the sites. These hot spots can survive over the entire duration of the excursion, and then dissolve their norm back into the other degrees of freedom at the end of the event.

%add DB to this paragraph!!

Fig. \ref{fig:y_x_1n5} shows that the exponent $\alpha$ decreases with increasing energy density $h$. In this case, larger $h$ values produce a smaller exponent until the observation of a hyperbolic PDF curve that does not fit a power-law anymore. The reason is that the trajectory does not converge to the equilibrium due to the freezing of high-amplitude breathers. Similarly, decreasing energy density $h$ for fixed $a$ generates a larger $\alpha$ until the exponential decay of PDF becomes dominant. In that limit, the algebraic decay of PDF will not be long enough to fit a power law due to the disappearance of long-time localized structures.
%such that we will not have long enough power law tail to fit due to the disappearance of long-time localized structures.

We did not exclude the presence of exponential cutoffs in the unresolvable part of PDF at large values of $\tau$ (e.g., Fig. \ref{fig:y_x_1n5}), while computing the $\alpha$ in Fig. \ref{fig:phasediagram_alpha}. For all of our tests, the location of an initial state in the densities diagram ($a,h$) is the only effective matter on the results. Yet, its selected form -described in Sec. \ref{sec:intro}- is inconsequential on our $\text{PDF}(\tau)$ results in both ergodic and non-ergodic regimes, according to our analysis.

\section{Conclusion}

In this chapter, we aim to clarify the equilibrium and non-equilibrium dynamics of the ordered semiclassical Gross--Pitaevskii model. 
We present an innovative strategy to study its dynamical properties in proximity to integrable limits. At equilibrium, we derive well-defined sectioning and interpret the algebraic tails of the probability density function of excursions in terms of temporal excitation of coherent states. By selecting local norm density as the observable of the phase space, we find out the statistics of its recurrence times in Gibbs and non-Gibbs regimes and analyze the non-equipartition phenomenon. This method allowed us to predict the existence of infinitely long-lasting fluctuations in a limited time when the system enters into a weakly non-ergodic phase.

We probed the \ac{PDF} of the recurrence times $\tau$ for a broad range of well-defined thermodynamic states, which may hold an algebraic decay $\text{PDF}\sim\tau^{-\alpha}$, where $\alpha\leq 2$ indicates breaking of ergodicity.
We show that the transition from ergodic to non-ergodic dynamics ($\alpha=2$) appears in a part of the phase diagram where the Gibbs distribution does not hold. In other words, a weakly non-ergodic regime lives "inside" the non-Gibbs part of the phase space where $\beta<0$. By saying this, we have to stress that the transition from ergodic to nonergodic does not happen on the $\beta=0$ phase transition line. However, if the system is in the strong interaction limit ($a\gg J=1$), the nearest neighbor couplings get negligible and the system enters an integrable limit. Hence, although the $\alpha=2$ and $\beta=0$ lines are separate in the weak interaction range we observed, they will meet in the strong interaction limit. This also confirms that any initial state defined with a finite Gibbs temperature will remain ergodic, independent of its average norm density $a$. Moreover, the thermalization of the Gibbs regime for all norm densities vindicates our preliminary test with the participation ratio in Sec. \ref{sec:ergodicGibbs}. In a similar manner, we see from our analysis that with the increase in the norm density on the constant infinite temperature line, the number density of breathers increases as well as their lifetimes due to the increase in the amplitudes.

We observed a gradual transition  from the ergodic to the non-ergodic regime, and notably found an ergodic subregion located within the non-Gibbs regime. An ergodic regime demands a defined temperature and a known distribution function that can describe its equilibrium dynamics. Thus, we claim that a yet unknown grand canonical distribution function should exist and is necessary to define this ergodic part of the non-Gibbs phase. Furthermore, we show that the transition from ergodic to the non-ergodic regime is unnoticed by the maximal Lyapunov exponent. mLCE is calculated as nonzero for both ergodic and weakly non-ergodic dynamics since the system still exhibits chaos except the discrete breather-like regions. In other words, the part of the lattice between the extreme local fluctuations is well thermalized and evolves in a chaotic manner \cite{Skokos:2013}.

This was the first attempt to discover the relation between statistical mechanics and dynamical properties of many-body systems in the framework of the Gross--Pitaevskii lattice. Our theory estimates the out-of-equilibrium times irrespective of the perturbation considered. We observed a high probability of short recurrence times due to chaos for all of our samples, which indicates the dominance of chaos throughout the lattice, which is also verified by positive Lyapunov exponents for all Gibbs-nonGibbs regions. The divergently long recurrence times have less probability which discloses that nonergodicity exists, yet only in a relatively smaller region of the lattice. Therefore, we call it weak nonergodicity.

We detected the duration of long-lasting excursions out of equilibrium caused by the formation of chaotic breathers in weakly non-integrable GP lattice. The extreme events start by the localization of norm at a site, generate concentrated hot spots, persist for a period, and then dissolve into the thermalized surrounding of the lattice. These localized excitations can have a long life depending on the amplitude of the excitation which causes anomalous fluctuations in time of observables. The amplitude of large fluctuations relative to their surroundings and their number of occurrences on the lattice both can have an impact on their survival times. These outcomes are tested for a number of PDF fat tail excursions. 

%The surrounding part of the lattice is thermalized and chaotic, therefore the excitation dissolves 

%These anomalous fluctuations in time of observables are also called as concentrated hot spots due to the localized norm excitations. 

%The extreme events start by the localization of norm at a site, can persist for a large time and end when the excitation decays into the thermalized surrounding.

The outcome of our computer simulations with the statistics of fluctuation method close to the integrable limits is in analogy with the results obtained for FPUT, KG, and Josephson junction chains \cite{Danieli:2017,danieli2019,mithun2019}. Similar observations may indicate that our method is generic for the many-body Hamiltonian dynamics near the integrable limits, where both the average  and standard deviation of recurrence times diverge.   Moreover, our work and findings might guide new proper ways to assess weak ergodicity-breaking phenomena in a large number of classical and quantum many-body systems. In the future, this study may be useful for the understanding of anomalously slow diffusion processes in out-of-equilibrium states.

%\item Our PDF observation near to the integrable limits is in line with similar results obtained for classical chains of Josephson junctions [40], indicating the emergence of a weak nonergodicity. We found that both the average  and standard deviation of recurrence times diverge in the same integrable limits which indicates the emergence of fat tails. These similar observations for classical chains raises the question of how general our findings are. Other challenges concern collecting evidence that the above observed scenarios of many-body Hamiltonian dynamics approaching integrable limits is generic, as well as the impact of quantization on the slow ergodization dynamics. %conclusion of danieli

%\item  Our results are in analogy with the weak nonergodicity phenomena studied in glass systems [14], continuous-time random walks [15–17], as well as in other many-body systems [19]. We expect them to be applicable also to larger spatial dimensions, and to other lattice models with similar integrable limits. We also speculate that spatial disorder, which induces Anderson localization, at small densities will again lead to weakly nonergodic dynamics at (then small but) finite densities. %from mithun2018.pdf:

%\item Future expectations are two-fold: on the one hand, we expect the dynamical glass to explain anomalously slow diffusion processes in out-of-equilibrium states in classical systems; on the other hand, we anticipate the quantum analog of the dynamical glass to be a prelude to many-body localization in quantum many-body systems.% Campbell_popularsummary.pdf

\cedp
\chapter{Disorder: Ground state statistics and Bogoliubov excitations}
\label{chapter4}
\ifpdf
\graphicspath{{Chapter4/Figs/}{Chapter4/Figs/PDF/}{Chapter4/Figs/}}
\else
\graphicspath{{Chapter4/Figs/}{Chapter4/Figs/}}
\fi

One of the main concerns of this thesis is the interplay of disorder and nonlinearity, as explained in Chapter \ref{Chapter1}. 
%The motivation of this thesis mainly concerns the interplay of disorder and nonlinearity, as explained in Chapter \ref{Chapter1}. 
While most of the theoretical predictions of statistical mechanics consider clean systems, real-world experiments and events are hugely affected by the disorder owing to its ubiquitous nature. After the discovery of Anderson localization, the studies of disordered systems gained more attention. %We use a good candidate Gross-Pitaevskii lattice (GPL) to study disorder in a fundamental level. 
Beginning from this chapter, we find out how the statistical and thermodynamic properties of the GP lattice change in the presence of disorder. We are particularly interested in the statistical characteristics of the new states which become available since disorder alters the ground state of GP lattice.

%We presume that disorder alters the ground state of GPL, and new states  become available. We are particularly interested in the statistical mechanics of these new states.

%introduce a random potential to the Gross-Pitaevskii lattice in order to find out how disorder affects its statistical and thermodynamical properties.  and find out its relation with interactions

\section{Motivation}\label{sec:intro}%Introduction

The experiments on trapped ultra-cold bosonic atoms in an optical lattice potential \cite{Greiner2002,Inouye1998,PhysRevLett.80.2027,Andrews1997,Orzel2001,Denschlag_2002,PhysRevLett.95.070401,PhysRevLett.94.090405,PhysRevA.72.021604,Jaksch:1998}  have motivated the studies describing a Bose-condensed gas with two main tools: the Bogoliubov method and Gross--Pitaevskii (GP) dynamics \cite{PhysRevA.53.4245,PhysRevA.53.909,PhysRevA.55.R1581,PhysRevLett.77.3489,PhysRevA.54.R3722,PhysRevA.51.4704,PhysRevA.51.1382,PhysRevA.55.4338,PhysRevA.53.R1950,PhysRevLett.76.6,PhysRevLett.76.1405,PhysRevA.53.2477,PhysRevLett.77.2360,PhysRevA.53.R1,PhysRevA.53.R1954,PhysRev.106.1135,PhysRevLett.91.030401}. The numerical results from the excitation spectra of GP equation are found to agree well with the direct experimental measurement of the Bogoliubov spectrum of an atomic \ac{BEC} at $T=0$ \cite{PhysRevLett.77.1671}, and the conditions to describe the 1D interacting Bose gas model by a discrete GP equation are found in \cite{Polkovnikov2002}. Moreover, the elementary excitations of an imperfect BEC have been extensively studied to explain superfluidity \cite{Chalker,Pavloff,Fontanesi_PRA,Fontanesi_PRL,Huber,Kramer2003}. Seventy-five years after Bogoliubov's microscopic theory of weakly interacting Bose gases \cite{bogoliubov1947theory}, we investigated this problem with disordered GP lattice and observed new properties of the Bogoliubov modes. 

After the theory of superfluidity was released by Bogoliubov in 1947,  Pierre-Gilles de Gennes published a book on superconductivity with an explanation of Bogoliubov equations \cite{de2018superconductivity} in 1966, and the method later called \ac{BdG}. The BdG method is extensively used in homogeneous systems, and its application to systems with random potential is rather new, which brings novelty to our study. %makes our study more interesting. 

%On the theoretical side, most efforts have been devoted to the study of the ground state, and its elementary excitations with several analytic approximations, considering both weak \& strong interaction regimes. 

\subsection{Phases of interacting Bose gas in a disordered environment} %Phases of 

Considering a Bose condensate, the phase transition from Mott insulator (MI) to superfluid (SF) is direct in the absence of any disorder, yet it is not if there is an uncorrelated random potential. For disordered ultra-cold bosons, another phase exists in between Mott insulator and superfluid, called Bose-glass (BG) \cite{BG_Fisher,LuganPRL98,Deng,BG_Falco,BG_Nelson,BG_Semmler,BG_Yong}.  
The existence of these three phases -MI, SF, BG- gets affected in case of strong enough disorder. As the disorder becomes stronger, the Mott insulator will disappear, and we will be left with two phases: Bose glass and superfluid \cite{Deng}.

Bose-glass, known as fragmented BEC, has no gap, zero superfluid density, localized and yet it has finite compressibility \cite{BG_Fisher}. It does not have long-range phase coherence because of exponentially decaying correlations. On the other hand, the Mott insulator has a gap for particle-hole excitations and has zero compressibility \cite{cetoli}.

In 1998, Prokof'ev et al. \cite{PhysRevLett.80.4355} showed a clear phase diagram obtained by the IS-DMRG method, differentiating superfluid, Bose-glass, and Mott insulator phases. They had also mentioned the Lifshits regions in the Bose-glass regime when disorder dominates over interactions.  Although disorder strongly affects the occurrence of the superfluid phase
transition, Bose glass to superfluid phase transition has been observed in cavity polaritons, and even in highly disordered systems such as CdTe, and GaN \cite{Malpuech2002}.

In 2007, Lugan  et al.\cite{LuganPRL98} showed an approximate phase diagram for Bose gas in 1D disorder, exhibiting three phases: Lifshits-glass (Anderson glass), Bose-glass, and BEC. The diagram separated the \ac{LG} and Bose-glass, as in the following year, Deng et al. \cite{Deng} suggested a numerical study is necessary to separate them clearly. %Lifshits glass (LG) is also referred as Anderson glass (AG).
However, the transition from Lifshits to Bose-glass is found continuous and smooth according to the participation ratio calculations in 1D disordered Bose gas \cite{cetoli}. Similarly, we also found a smoothly varying participation ratio of particle densities for the GP lattice in the whole interaction range (see Fig. \ref{fig:p_a_mu}). Considering the techniques to separate different phases of Bose gas such as superfluid fraction, compressibility, localization, and disorder; there is no sharp difference between Lifshits and Bose-glass regimes \cite{Deng,cetoli,BG_Fisher}. Lifshits regime occurs in the limit of vanishing interaction, and one can distinguish LG and BG only by the inhomogeneity of their density profile. Thus, we can conclude that LG is a region that lives "inside" the Bose-glass, and is not a phase in its own right \cite{cetoli}.  %Thus, we can conclude that Lifshits glass is a region that lives "inside" the Bose-glass, and is not a phase in its own right \cite{cetoli}. 

\subsection{Overview}

The statistical properties for the clean GP lattice is found by Rasmussen et al. \cite{Rasmussen:2000}. Yet, the statistics of \textit{disordered} GP lattice remained unresolved. In this chapter, we examine its thermodynamic properties for the ground state in the weak and strong interaction regimes, and find out the characteristics of BdG excitations above the ground state. 
This chapter is organized as follows. 
In the following section, we present the disordered GP lattice model. Next, we discuss its dynamics and statistical properties at the \ac{GS} displaying both analytical and numerical calculations. In Sec. \ref{sec:localizationproperties}, we introduce a small perturbation to the GS, and analyze the localization properties of the elementary BdG modes. Finally, in Sec. \ref{sec:conclusionch4} we comment on our results and its possible future outlook. % the relation between our current results and our future projects.

%In the first part of this chapter, I will present the zero-temperature dynamics and statistical properties of the discrete GP model. For the weakly interacting particles, we observed and confirmed the existence of the Lifshits-glass regime in GP lattice, while it has been found experimentally and theoretically for ultra-cold Bose gas \cite{LuganPRL98,Fontanesi_PRA,Fontanesi_PRL,Scalettar,LuganPRL2007,Kati2020,Deng}. In 2000, Rasmussen et al. \cite{Rasmussen2000} found the statistical properties for the clean GP lattice. Yet, the statistics of \textit{disordered} GP lattice remained unresolved. In this chapter, we examine its statistics for the ground state, both analytically, and numerically. On the theoretical side, most efforts have been devoted to the study of the ground state, and its elementary excitations with several analytic approximations, considering both weak \& strong interaction regimes. 

\section{Disordered Gross--Pitaevskii lattice model}\label{sec:disorderedGPHamiltonian}
%Ground state statistics}
%From this chapter, 
We now introduce disorder to the Hamiltonian of GP lattice with $N$ sites, defined in Sec. \ref{sec:gpl}:

\begin{equation}\label{eq:Hamiltoniandisorder}
\mathcal{H}=\sum_{\ell=1}^N \frac{g}{2}|\psi_{\ell}|^4 + \epsilon_{\ell} |\psi_{\ell}|^2 -J (\psi_{\ell} \psi_{\ell+1}^* +\psi_{\ell}^*\psi_{\ell+1}),
\end{equation}
where $\epsilon_\ell$  represents the random on-site energies which are chosen uniformly from the interval $[-\frac{W}{2},\frac{W}{2}]$, while $W$ is the disorder strength. Their variance $\sigma^2(\epsilon)=W^2/12$. 

The nonlinear coupling parameter $g$ is fixed to 1 in all numerical measurements unless stated otherwise. The time $t$ is measured in the unit of the inverse of the hopping strength $J$ which is set to be one without losing generality. This model is used to describe nonlinear optical waveguide arrays~\cite{Eisenberg}, nonlinear photonic lattices~\cite{Lahini} and dilute-cold  bosonic gases~\cite{Trombettoni2001,Inguscio11}. The equations of motion, generated by $\dot{\psi_\ell}={\partial H}/{\partial i \psi_\ell^*}$, is
\begin{equation}\label{eq:eomdisorder}
i\dot{\psi_\ell}=\epsilon_\ell \psi_\ell + g|\psi_\ell|^2\psi_\ell-(\psi_{\ell+1}+\psi_{\ell-1})
\end{equation}
which conserves the total norm ${\mathcal A}\equiv\sum_{\ell}|\psi_\ell|^2$ and the total energy $\mathcal{H}$. The partition function then is
\begin{equation}
{\cal Z}=\int^{\infty}_0 \int^{2\pi}_0 \prod_\ell d \phi_\ell d {\cal A_\ell } \exp[-\beta({\cal H}+\mu {\cal A })],
\end{equation}
where $\beta$ and $\mu$ are Lagrange multipliers associated with the total energy and the total norm respectively.

\begin{figure}[hbt!]  
\centering
\includegraphics[width=0.51\textwidth]{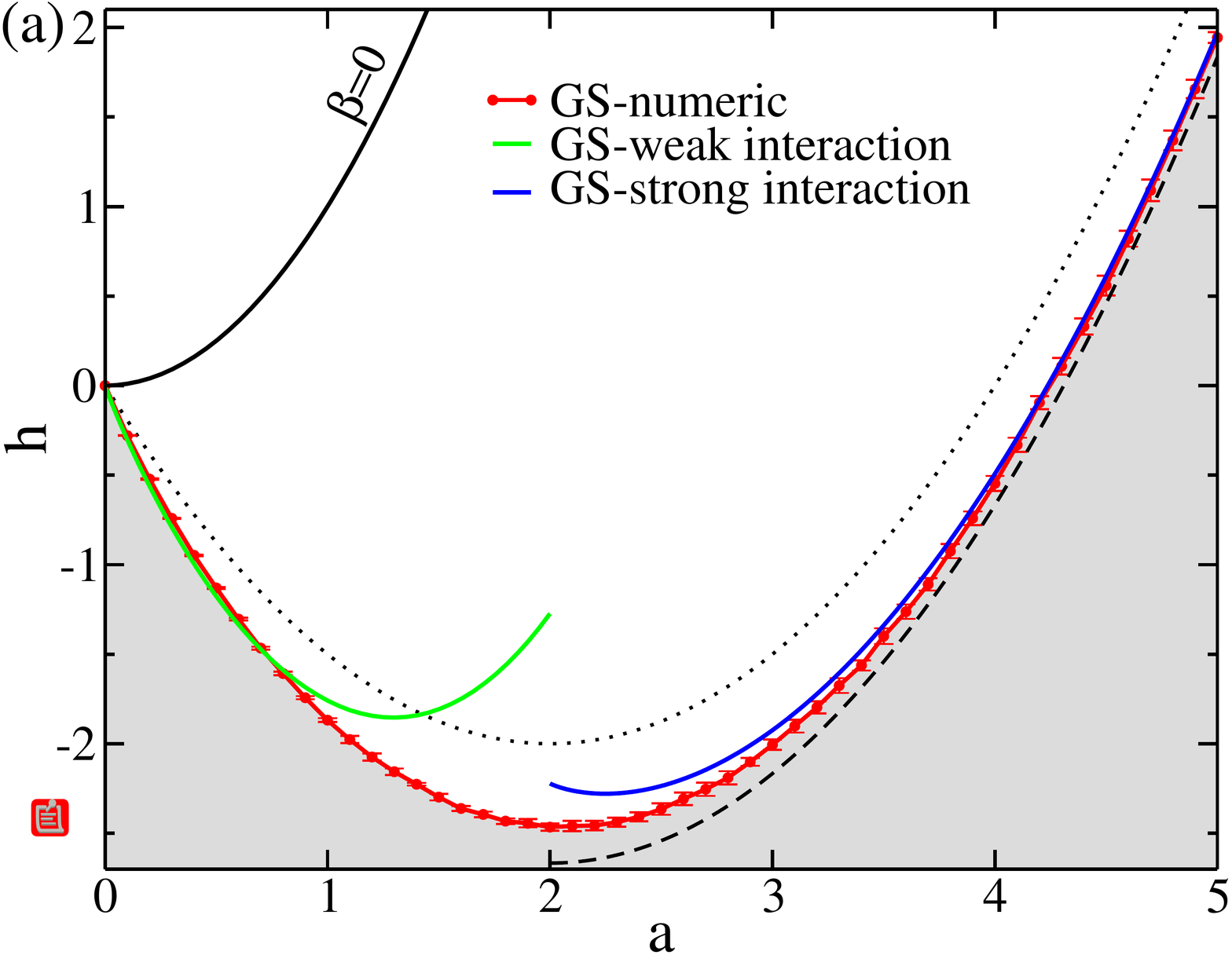}%phasech4.pdf}%phaseanalytic.pdf}
\includegraphics[width=0.51\textwidth]{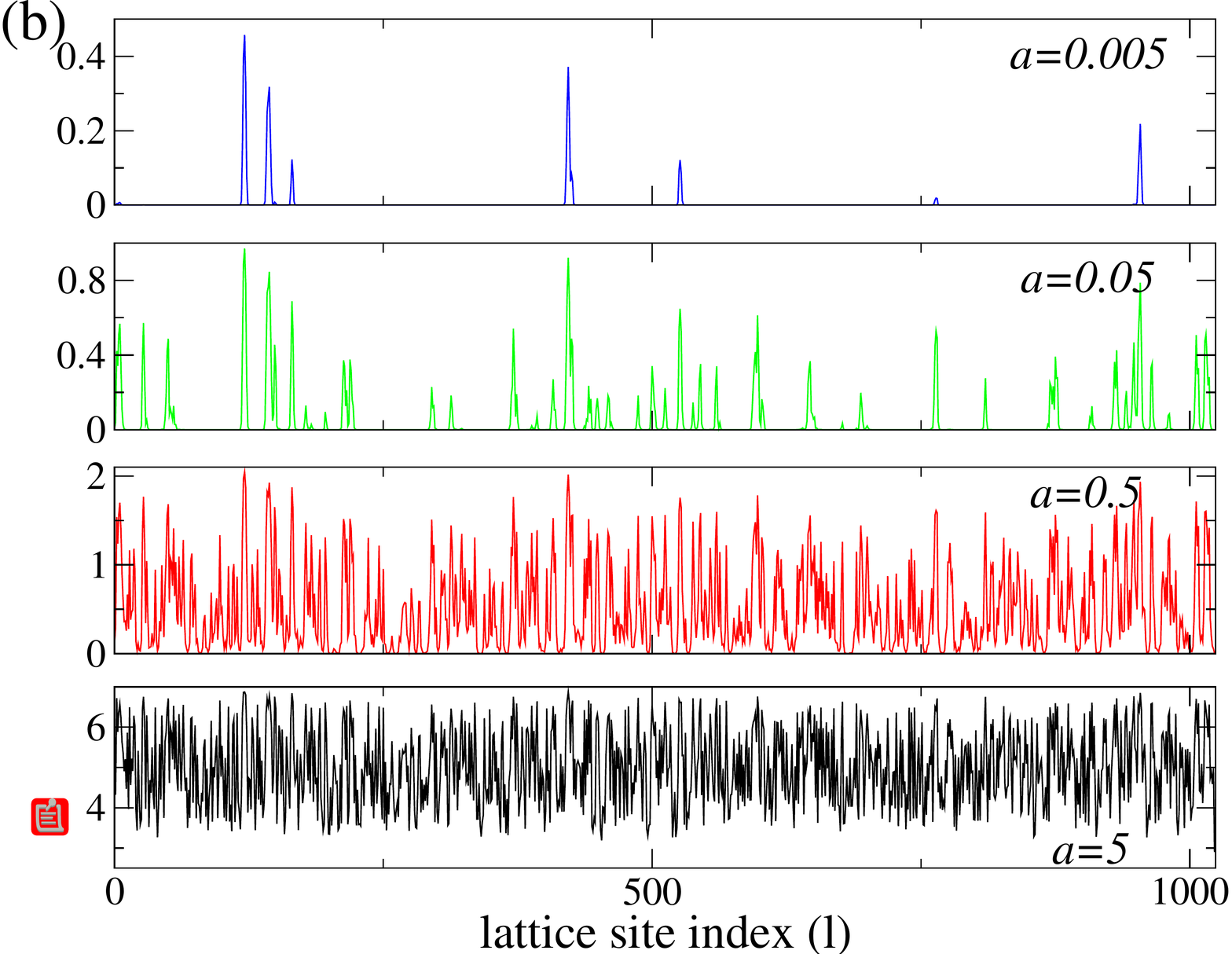}%an0005to5_2.pdf}%an0005to5_.pdf}
\caption[Phase diagram of disordered GP lattice]{(a) Phase diagram of energy and norm densities of disordered GP lattice with $W=4$, $g\equiv J=1$. GS is defined analytically for both the weakly interacting regime (green line) with Eq. (\ref{eq:hfistul}), and the strongly interacting regime (blue line) with Eq. (\ref{eq:h_large_a}). The connected red circles represent the numerically found GS for $N=10^4$, averaged over $N_r=50$ disorder realizations, where the error bars are the standard deviation of $N_r$ realizations. The $\beta=0$ line, black solid curve, is not affected by disorder (see Appendix \ref{ap:betazero}). The dashed black line is Eq. (\ref{eq:GSanalytic2}) with $\zeta_\ell=2$ for strong interaction limit. The dotted black line, $h=-2a+a^2/2$, is the GS for zero disorder as a reference (see Sec. \ref{sec:zerotemperatureorder}). (b) GS local norm density $G_\ell^2$ vs lattice site index $\ell$ for one fixed realization is shown for $a=0.005, 0.05, 0.5, 5$ from top to bottom.}% different average norm densities. Top to bottom: $a=0.005, 0.05, 0.5, 5$.}
\label{fig:phase}
\end{figure}

\section{Ground state statistics}

The dynamics of the system is depending on the two energy and norm densities $h=\mathcal{H}/N$ and $a = \mathcal{A}/N$. For a fixed value of $a$, the GP model has a ground state of minimum energy which is characterized by the lowest possible value of $h$. 
The ground state can be obtained subject to the imposed constraint  $\mathcal{A}$ using the method of Lagrange multipliers:

%We can find the ground state,  subject to the imposed constraint  $\mathcal{A}$ by the method of Lagrange multipliers:
\begin{equation}\label{eq:lagrange}
\mathcal{L}=\mathcal{H}-\mu A, \qquad \Delta \mathcal{L}(\psi_{\ell},\psi_{\ell}^*,\mu)=0. 
\end{equation}
Using Eq. (\ref{eq:Hamiltoniandisorder}) in Eq. (\ref{eq:lagrange}) gives
\begin{equation}\label{eq:eq4}
    \mu \psi_{\ell} =\frac{ \partial \mathcal{H}}{ \partial \psi_{\ell}^*}=\epsilon_{\ell} \psi_{\ell} +  g|\psi_{\ell}|^2\psi_{\ell}-J(\psi_{\ell+1}+\psi_{\ell-1}),
\end{equation}
where $\mu$ is the chemical potential (Lagrangian multiplier). Here, we can define the GS participation number $P$ via multiplying both sides of Eq. (\ref{eq:eq4}) by $\psi_{\ell}^*$, and averaging over all sites using the definitions of $\mathcal{H}$ (\ref{eq:Hamiltoniandisorder}), $\mathcal{A}$, and the generic participation ratio $P=\mathcal{A}^2 / \sum_\ell |\psi_\ell|^4$. Thus $P$ in terms of $a$ and $h$ at the ground state can be found as
%We multiply Eq.\ref{eq:eq4} by ${\psi_{\ell}}^*$, sum over $\ell$ with the definitions of $\mathcal{H}$ (Eq.\ref{eq:Hamiltoniandisorder}), $\mathcal{A}$, and the participation ratio $P=\mathcal{A}^2 / \sum_\ell |\psi_\ell|^4$. We then average over all sites, and obtain 
%and averaging over all sites; we get

\begin{equation}\label{eq:Pstep3}
 P^{-1} N = \frac{2}{g a^2} (\mu a -h). 
\end{equation}
To find the ground state dynamics, we compare Eq. (\ref{eq:eq4}) and (\ref{eq:eomdisorder}) which yields
\begin{equation}\label{eq:psilt}
    \psi_{\ell}(t)=G_\ell e^{-i\mu t}, 
\end{equation}
% can be chosen to be purely real and non-negative
where  $G_\ell=\psi_{\ell}(0)\equiv \sqrt{a_\ell}>0$. %Now, we found the dynamics at the ground state such that all sites have the same phase, $\mu$. 
Inserting Eq. (\ref{eq:psilt}) in Eq. (\ref{eq:Hamiltoniandisorder}) gives
\begin{equation}\label{eq:Hreal}
\mathcal{H}=\sum_{\ell=1}^N \frac{g}{2}G_\ell^4 +\epsilon_\ell G_\ell^2 -2J G_\ell  G_{\ell+1} 
\end{equation}
where we can choose the initial $G_\ell$ as real, owing to the fact that the phase difference between all sites kept as zero during time evolution (see Appendix \ref{ap:zerotemperaturedisorder}). We obtain energy density $h=\mathcal{H}/N$ via averaging Eq. (\ref{eq:Hreal}) over all sites:

\begin{equation}\label{eq:h_avg1}
h= \left\langle \epsilon_\ell G_\ell^2 \right \rangle +\frac{g}{2} \left \langle G_\ell^4 \right\rangle -2J  \langle G_\ell G_{\ell+1} \rangle. 
\end{equation}
Now, to define $G_\ell$, we insert Eq. (\ref{eq:psilt}) into Eq. (\ref{eq:eq4}), divide it by $G_\ell$, and obtain
\begin{equation}\label{eq:a_generalrule}
G_\ell^2=(\mu -\epsilon_\ell+J\zeta_\ell)/g \geq 0    
\end{equation}
with the GS field $\zeta_\ell=(G_{\ell+1}+G_{\ell-1})/G_\ell$. We note that the ground state solution for the ordered case $W=0$ is simply $G_{\ell}^2=a$. Using Eq. (\ref{eq:Hamiltoniandisorder}) we arrive at the analytical dependence $h=ga^2/2-2a$, which is shown as a dotted black line in Fig. \ref{fig:phase}(a). In the presence of disorder, there are three competing energy density scales: the kinetic energy density $J=1$, the disorder energy density $W$, and the interaction energy density $ga$ (we remind that we set $g=1$). %For all studied cases $W \sim J$. 
We distinguish the regime of weak interaction $ga \ll W$ and strong interaction $ga \gg W$.
By fixing the norm density, we numerically minimize the energy by varying the real and nonnegative variables $G_\ell$ for a given disorder realization as in Ref. \cite{Kati2020} (see also Sec. \ref{sec:GroundStateRenormalization}). The resulting ground state density distribution $G_{\ell}^2$ is plotted in Fig. \ref{fig:phase}(b) for four different norm densities $a=0.005, 0.05, 0.5, 5$, with one and the same disorder realization. For each outcome, we compute the total energy and the corresponding energy density. We then toss a new disorder realization and repeat the process $N_r$ times. We finally compute the average energy density $h$ and its standard deviation. The resulting dependence $h(a)$ is shown with red circles in Fig. \ref{fig:phase}(a) (the error bars are the standard deviation).

\begin{figure}[hbt!]  
\centering 
\includegraphics[width=0.49\textwidth]{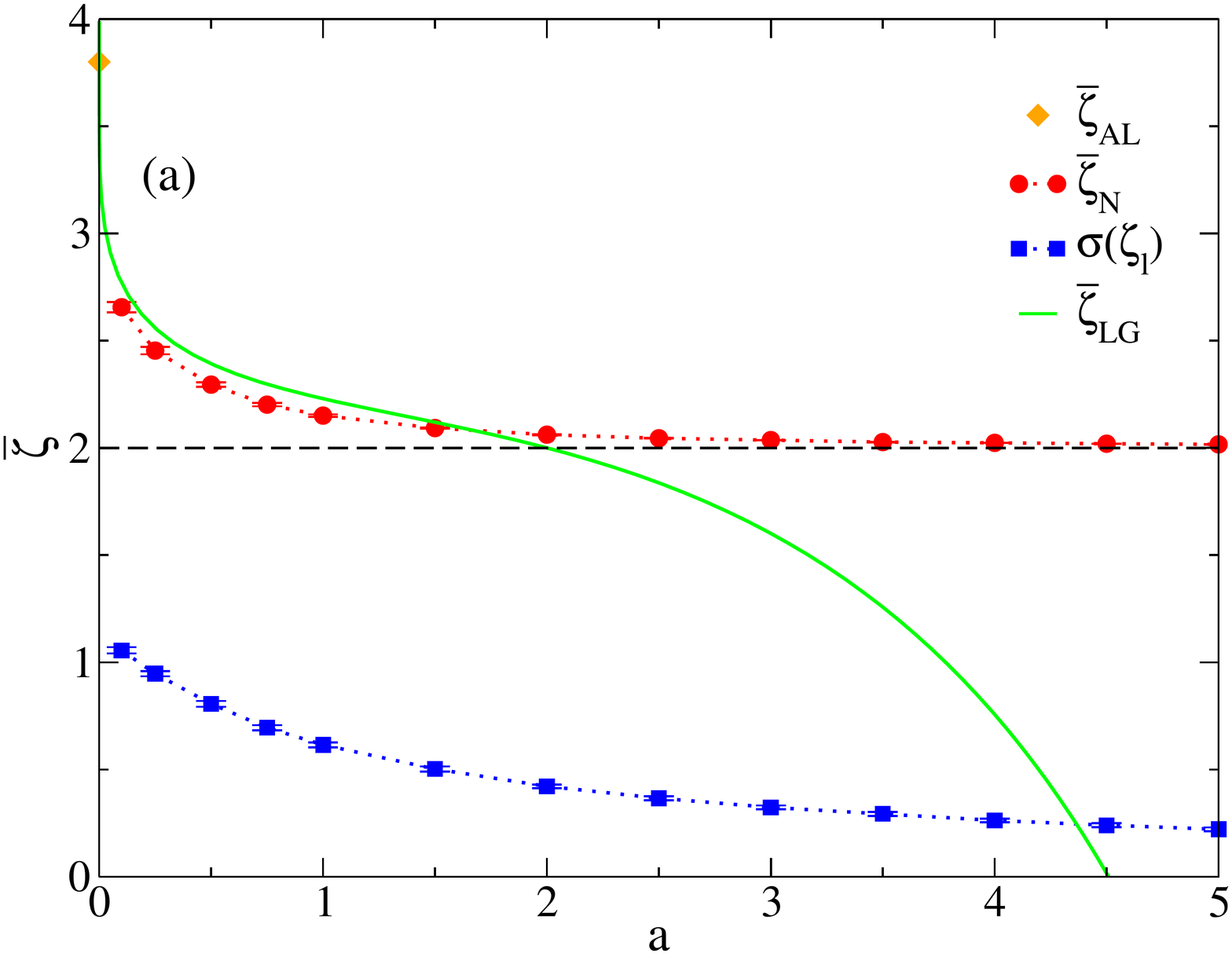}
\includegraphics[width=0.49\textwidth]{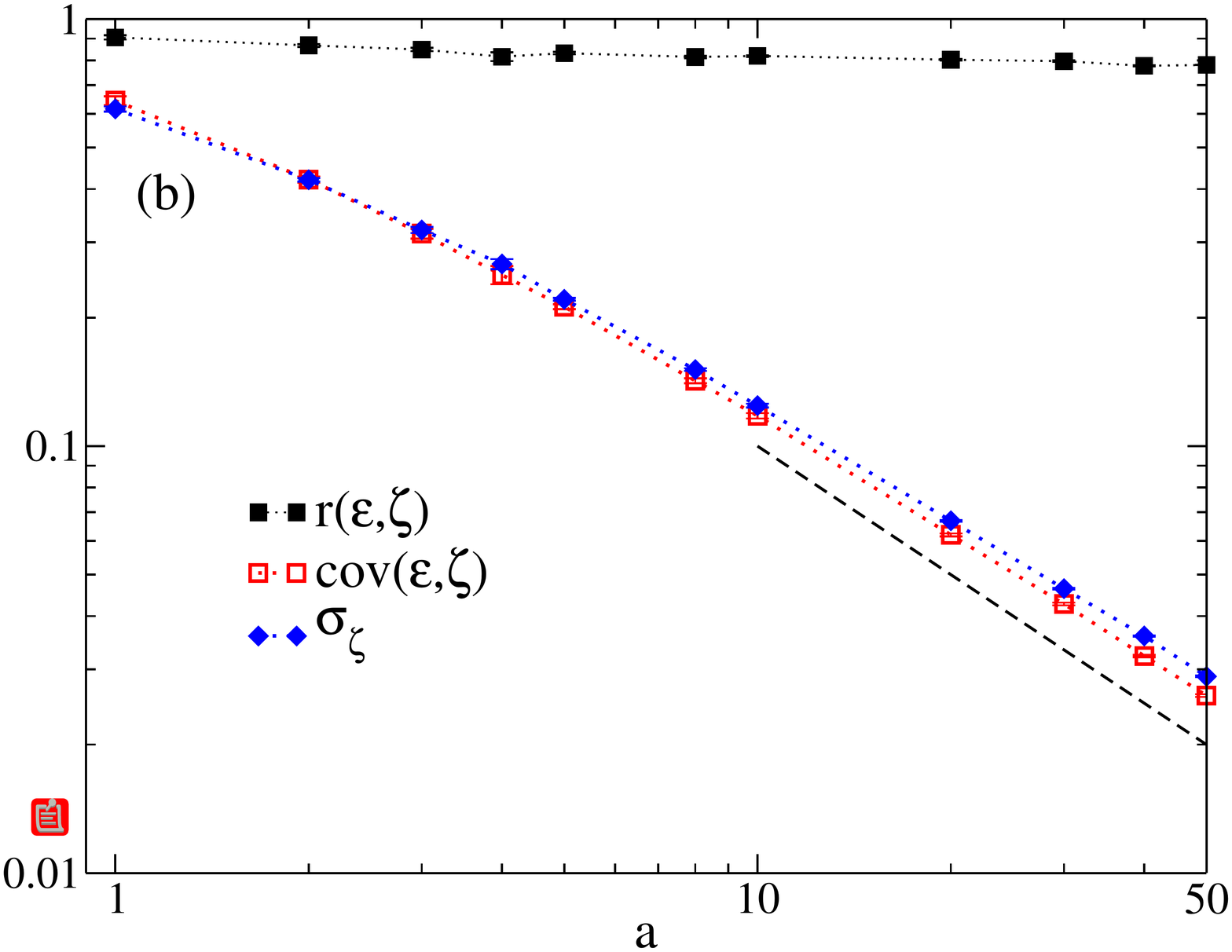}
\caption[The ground-state field versus norm density]{(a) The average GS field $\bar{\zeta}$ vs norm density $a$ for $W=4$. $\bar{\zeta}_{AL}$ is the Anderson approximation by Eq. (\ref{eq:cosh}). $\bar{\zeta}_N$ is the numerical result, and $\sigma_\zeta$ is the standard deviation of $\zeta_\ell$. $\bar{\zeta}_{LG}$ is found by using Eq. (\ref{eq:a_mikhail2}) in Eq. (\ref{eq:muavg}). The black dashed line is the strong interaction approximation $\bar{\zeta}=2$. (b) $\sigma_{\zeta}$ (standard deviation), $\mathrm{cov}(\epsilon,\zeta)$ (covariance) and $r(\epsilon,\zeta)$ (correlation) versus
$a$ on logarithmic scales.  For both plots: black dashed line is $\sim 1/a$, $N=1000$, $N_r=100$, $W=4$, $g=1$, $J=1$.
Dotted lines connect the data and guide the eye.} 
 \label{fig:zeta_a_corr}%fig:corr
\end{figure}

Averaging both sides of Eq. (\ref{eq:a_generalrule}) over all sites in an infinite system yields the chemical potential

\begin{equation}\label{eq:muavg}
\mu= g {a} -J \bar{\zeta},   
\end{equation}
where $\bar{\zeta}=\langle \zeta_\ell \rangle$ is the average ground state field, which is calculated as a function of norm density $a$ in Fig. \ref{fig:zeta_a_corr}(a). Thus, the GS field is %$a=\mathcal{A}/N$, $\langle \epsilon_\ell \rangle=0$, and 
\begin{equation}\label{eq:zeta}
\zeta_\ell=\frac{G_{\ell+1}+G_{\ell-1}}{G_\ell}. 
\end{equation}
  Inserting Eq.~(\ref{eq:muavg}) into Eq.~(\ref{eq:a_generalrule}) leads to
\begin{equation}\label{eq:gglsquarex}
    g G_\ell^2=ga-\epsilon_\ell +J \delta\zeta_\ell \geq 0
\end{equation}
where $\delta \zeta_\ell=\zeta_\ell-\overline{\zeta}$ describes the fluctuations of the GS field.

%%, shown as blue diamond, black square, and red open square, respectively. 

%$\sigma_{\zeta}$, correlation $r(\epsilon,\zeta)$ \& covariance  of $\zeta_\ell$ with $\epsilon_\ell$ $cov(\epsilon,\zeta)$ are plotted vs $a$ in log-log scale. Black dashed line is $\sim 1/a$.

Fig. \ref{fig:zeta_a_corr}(b) shows an almost constant correlation $r$ between the onsite potential $\epsilon_\ell$ and $\zeta_\ell$ where their covariance $cov(\epsilon_\ell,\zeta_\ell)$ and the standard deviation of $\zeta_\ell$ both decrease with a factor $1/a$, as norm density $a$ increases. The same decrease rate of both quantities keeps the value of correlation stable $\approx 0.8$ due to the relation %$r(\epsilon_\ell,\zeta_\ell)=cov(\epsilon_\ell,\zeta_\ell)/\sigma_\zeta$. Here covariance is $\langle \epsilon_\ell \sigma_{\zeta} \rangle$. 

\begin{equation}\label{eq:corr}
r(\epsilon_{\ell},\zeta_{\ell})=\frac{cov(\epsilon_\ell \sigma_{\zeta})}{\sigma_{\epsilon}\sigma_{\zeta}}
=\frac{\overline{(\epsilon_{\ell} - \bar{\epsilon_{\ell}} )(\zeta_{\ell}-\bar{\zeta_{\ell}} )}}{\sigma_{\epsilon}\sigma_{\zeta}}.
\end{equation}

\subsection{Ground state renormalization}\label{sec:GroundStateRenormalization}

The zero-temperature ground state line of the ordered case $h=-2a+a^2/2$ is renormalized in the presence of disorder. This happens in the regime of small norm
density (i.e., weak nonlinearity) $a < 1$ due to the presence of Lifshits states which are sparsely distributed AL eigenstates with eigenvalues close to the bottom of the AL spectrum, i.e., their distance from the bottom $\Delta_{\mu} = \mu+2+W/2 \ll 1$. Such Lifshits states exist due to rare disorder fluctuations with 
$\epsilon_\ell+W/2 < \Delta_{\mu}$ over a simply connected chain segment of length $L=\pi/\sqrt{\Delta_{\mu}}$. The average distance between such regions
$d_L \approx (W/\Delta_{\mu})^L$.
As a result, one can expect a set of disjoint puddles of norm distribution in real space for small norm density $a$.
Note also that for any finite system the ground state is bounded by $h=-(2+W/2)a+a^2/2$ which is generated by the disorder realization $\epsilon_\ell=-W/2$. Contrary, in the large norm density limit (i.e., for strong nonlinearity) the ground state correction becomes weak since the nonlinear terms
$a^2/2$ are of leading order and disorder has a minor impact.

In order to numerically compute the ground state, we note that $\psi_{\ell}$ can be gauged into real variables as all the phases $\phi_{\ell} = \phi_{{\ell}'}$ to minimize
the Hamiltonian (\ref{eq:Hamiltoniandisorder}). The remaining task is to minimize a real function $\mathcal{H}$ defined in Eq. (\ref{eq:Hreal}) for real variables $\psi_\ell(t_0)\equiv G_{\ell}$ for a given disorder realization. We choose an initial set of $G_{\ell}$ under the constraint $Na=\sum_{\ell} {G_{\ell}}^2$. We define a window of $\ell_w=3-5$ adjacent sites and minimize the energy varying the amplitudes on these consecutive sites using the Nelder-Mead simplex algorithm \cite{lagarias}. As the algorithm changes the total norm in general, we perform a homogeneous
renormalization of all amplitudes $G_{\ell}$ to restore the required norm density $a$. We then shift the window by one lattice site and repeat the procedure, until
the whole lattice with $N$ sites has been covered by minimization windows. The procedure is repeated  around 10-40 times until the full convergence is obtained.
The chemical potential 
\begin{equation}\label{eq:mu_computation}
    \mu=\epsilon_{\ell} +g G_\ell^2 -  (G_{\ell+1}+G_{\ell-1})/G_{\ell}
\end{equation}
is defined through local relations (see Eq. \ref{eq:a_generalrule}) and yields a ratio of the standard deviation to mean which is less than $10^{-3}$, indicating the quality of our ground state computation.
Finally, we repeat the procedure for 50 different disorder realizations and compute the average ground state energy density $h$ and its standard deviation.
The result is shown as red solid circles in Fig. \ref{fig:phase} with their standard deviation for $N=10^4$, and $W=4$. Optimizing small parts of the lattice at a time immensely reduced the computational time, enabling us to reach the ground states for larger system sizes, e.g., $N=10^5$.

\subsection{Weakly interacting regime}%Lifshits-glass regime}

%Lifshits regime is defined as possessing finite but weak interactions s.t. the chemical potential lies in the Lifshits tail of the energy spectrum \cite{LuganPRL98} in 1D disordered Bose gas, while we also observed a similar picture in the inset of Fig.\ref{fig:p_a_mu} for disordered GP lattice. 
%. $a< W/2g$ and $a<J$,

In the small norm density limit $ga \ll \max \{ W,J\}$, the particles have to group in the rare regions of a size $L\geq 1$ to satisfy Eq. (\ref{eq:a_generalrule}). Let $\rho_1$ be the probability of having excitation with $\epsilon_\ell<\mu+J\zeta_\ell$ on one site. Then, we can write the probability $\rho(\mu)$ for the exponential behavior of band tails as \cite{Kramer93}

\begin{equation}\label{eq:a_mikhail}
\rho(\mu)= {\rho_1}^L= \left[ \int_{-W/2}^{\mu+2J} \rho d\epsilon \right]^L= {\left[ \frac{\mu+2J+W/2}{W} \right]}^L,  
\end{equation}
where $\rho_1$ is the probability per site in a region with length $L$. The dimensionless wave number
\begin{equation}
k= \sqrt{(\mu+2J+W/2)/J},   
\end{equation}
where $\sqrt{2m}/\hbar \rightarrow 1/\sqrt{J}$, $E\rightarrow \mu+2J$, and $V_0=-W/2$ in the generic definition of wave number: $k=\sqrt{2m(E-V_0)}/\hbar$. %The ground state $G_\ell\simeq \sin(k\ell)$. 
While $G_\ell\simeq 0$ outside of the connected cluster of particles, the size of the region which allows trapping of a single particle is
\begin{equation}\label{eq:L_mikhail}
L= \frac{\pi}{k}=\pi \sqrt{\frac{J}{\mu+2J+W/2}},% \pi/\sqrt{(\mu+2J+W/2)/J},  
\end{equation}
where we assume $L$ as half of the wavelength. In the small particle density regime; the norm can be approximated as the total number of excitations $N_0$, multiplied by their amplitudes $A_0$, and their width $L$:
 
 \begin{equation}\label{eq:a_small}
\mathcal{A}= \sum_\ell {\psi_\ell}^2=N_0 A_0 L = N \rho(\mu) A_0 L,    
\end{equation}
where $N_0$ is found by probability $\rho(\mu)$ of these rare fluctuations multiplied by $N$.
Similarly, one can find $P$ as

\begin{equation}\label{eq:pa}
P= \frac{(\sum_\ell {{\psi_\ell}^2})^2}{\sum_\ell {\psi_\ell}^4}= \frac{({A_0} L   {\rho(\mu)} N)^2}{ {A_0}^2 L  \rho(\mu)N}=N \rho(\mu)L.  
\end{equation}
From Eq. (\ref{eq:a_small}), we may write the norm density in the weak interaction regime as

\begin{equation}\label{eq:a_mikhail2}
a=A_0 L \rho(\mu)\approx \frac{2}{g} {\left[ \frac{\mu+2J+W/2}{W} \right]}^L,
\end{equation}
%V={A_0} L
where $g A_0 L \approx 2$ is considered as a fitting parameter of the transcendental equation, found via inserting  Eq. (\ref{eq:muavg}) at $a=2/g$, with $\bar{\zeta}\approx 2$ into Eq. (\ref{eq:a_mikhail2}). %The factor 2 was also numerically discovered for the size of the Anderson states in \cite{Kati2020} (see Sec.\ref{sec:initialstatewave}). 

We insert Eq. (\ref{eq:pa}) into Eq. (\ref{eq:Pstep3}), and obtain $h$ for weak interaction regime shown in Fig. \ref{fig:phase}:
\begin{equation}\label{eq:hfistul}   
h=\mu a-\frac{g a^2 }{2\rho(\mu)L}. 
\end{equation}

We can consider $G_\ell$ has an Anderson localized structure in the $a\rightarrow 0$ limit of weakly interacting regime. Hence, we can use the definition $G_{\ell} \propto e^{-\ell/\xi}$ in Eq. (\ref{eq:zeta}), and find $\bar{\zeta}$ as
\begin{equation}\label{eq:cosh}
\bar{\zeta}=e^{1/\xi}  + e^{-1/\xi} =2 \cosh{1/\xi},   
\end{equation} 
where $\xi$ is the localization length, calculated by the standard transfer-matrix approach \cite{Kramer93,Krimer} as $\xi(\lambda=-2-W/2) \approx 0.8$ at the edge of the band, which gives  $\bar{\zeta}\approx 3.8$ by Eq. (\ref{eq:cosh}). Similarly, it is $\bar{\zeta}(a\rightarrow0)\approx 4$ by the weak interaction approach (see Fig. \ref{fig:zeta_a_corr}(a)).

\subsection{Strongly interacting regime}

In the strong coupling regime, we have $\max \{ W,J\}\ll ga$. 
First, we  use Eq.~(\ref{eq:gglsquarex}) when $J/(ga)\ll 1$

%For large norm densities; i.e.,  $a\gg J$, we expect that $\bar{\zeta}$ in Eq. \ref{eq:zeta} will be approximately 2, since $G_{\ell+1}\approx G_{\ell-1}\approx G_\ell$ in that range. Similarly, we observed in Fig. \ref{fig:zeta_a_corr}(a) that as $a$ increases, $\bar{\zeta}\rightarrow 2$, while the standard deviations diminish $\sigma_\zeta \rightarrow 0$. Thus, we can take $\zeta_\ell \approx \bar{\zeta}\approx 2$ for large $a$ range in Eq. \ref{eq:muavg}, and obtain the following relation

\begin{align}
{G_\ell}\simeq\sqrt{a-\frac{\epsilon_l }{g}}.\label{eq:gglsquare}
\end{align}
that requires $a>W/2g$. 
%\subsubsection{Energy}
We can find the average energy density analytically as follows. %With the assumption $\delta\zeta_\ell \approx 0$, 
We use Eq. (\ref{eq:gglsquare}) in Eq. (\ref{eq:h_avg1}) that yields %, and get

\begin{equation}\label{eq:h_largea_2}
 h \approx \frac{g}{2}\left(a^2-\frac{W^2}{12 g^2}\right)  -2J\left\langle \sqrt{\left(a-\frac{\epsilon_{\ell}}{g}\right)\left(a-\frac{\epsilon_{\ell+1}}{g}\right)} \right\rangle. 
\end{equation}
We assume that different sites are uncorrelated, i.e., for arbitrary $f(x)$:
%\begin{equation}\label{eq:corr}
$    \langle f(\epsilon_i)f(\epsilon_j)\rangle=\langle f(\epsilon_i)\rangle\langle f(\epsilon_j)\rangle$, 
%\end{equation}
for $i\neq j$. Then
\begin{equation}\label{eq:fsquare}
    \frac{1}{N}\sum_i \langle f(\epsilon_i)f(\epsilon_j)\rangle = \left(\int d\epsilon \rho(\epsilon)f(\epsilon)\right)^2. 
\end{equation}
By employing $f(\epsilon)=\sqrt{a-{\epsilon}/{g}}$ in Eq. (\ref{eq:fsquare}), we get
\begin{align}
&\left\langle \sqrt{\left(a-\frac{\epsilon_{\ell}}{g}\right)\left(a-\frac{\epsilon_{\ell+1}}{g}\right)} \right\rangle = \left[\frac{1}{W}\int_{-\frac{W}{2}}^{\frac{W}{2}}d\epsilon \sqrt{a-\frac{\epsilon}{g}}  \right]^2 \nonumber \\
 &= \left( \frac{2}{3W}\left[\left(a+\frac{W}{2}\right)^{3/2}-\left(a-\frac{W}{2}\right)^{3/2}\right] \right)^2. \label{eq:T}
\end{align}
So, now we can rewrite $h$, using Eq. (\ref{eq:T}) in Eq. (\ref{eq:h_largea_2}) as
\begin{equation}\label{eq:h_large_a}
h\approx  \frac{g}{2}\left(a^2-\frac{W^2}{12 g^2}\right)      -\frac{8J}{9 W^2}\left[\left(a+\frac{W}{2g}\right)^{3/2}-\left(a-\frac{W}{2g}\right)^{3/2}\right]^2,
\end{equation}
for $a\geq W/2g$.

%\begin{align}\label{eq:pa_large}
%{p(a)}^{-1}&= 1-\frac{2J\bar{\zeta}}{ga}+\frac{\sigma^2}{g^2 a^2}\\
% &+\frac{16J}{9gW^2 a^2}\left[\left(a+\frac{W}{2g}\right)^{3/2}-\left(a-\frac{W}{2g}\right)^{3/2}\right]^2   \nonumber 
%\end{align}
%for $a\geq W/2g$.

\begin{figure}[htp]%[htb] 
\centering 
\includegraphics[width=0.65\textwidth]{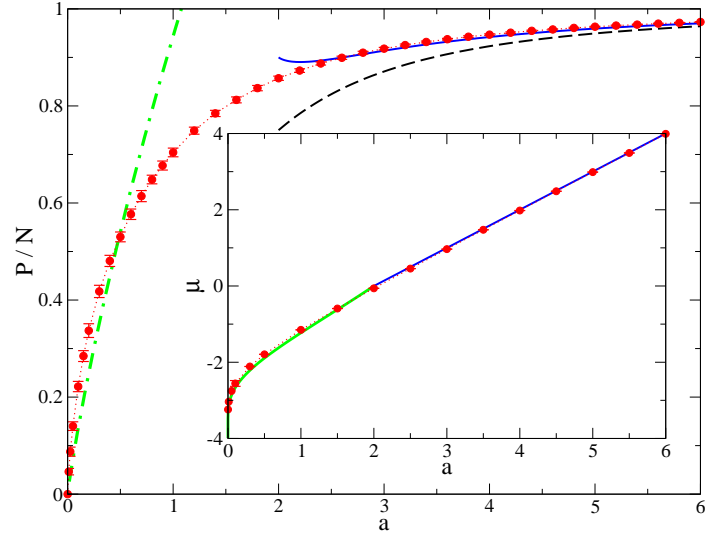}
\caption[Participation ratio of norm densities at the ground state]{ Average participation ratio density $P/N$ vs. norm density $a$. Inset: Chemical potential $\mu$ vs. norm density $a$. The numeric results for $P$ and $\mu$ are shown as red circles, whereas weak interaction approximations (Eq. \ref{eq:pa}, Eq. \ref{eq:a_mikhail2}) are shown as green line, and strong interaction approximations (Eq. \ref{eq:pa_large}, Eq. \ref{eq:muavg}) are shown as blue line; respectively in main and inset figures. The strong interaction approximation of $P$ in Eq. (\ref{eq:P_sergej}) is shown as dashed black curve. All data are averaged over $N_r=100$, and shown with their standard deviations.}
 \label{fig:p_a_mu}
\end{figure}
%\caption[participation ratio of norm densities at the ground state]{ $P({G_\ell}^2)/N$ vs. $a$. The numeric $P/N$ is found by Eq.(\ref{eq:P_eigen}), shown as red circles. The Lifshits approximation in Eq.\ref{eq:pa} is represented as dash-dotted green line. The strong interaction approximations Eq.(\ref{eq:P_sergej}) and Eq.(\ref{eq:pa_large}), are shown as dashed black, and blue curves, respectively.Inset: The numerical $\mu$ vs $a$ is represented as red circles. The Lifshits approximation of $\mu$ is obtained by Eq.(\ref{eq:a_mikhail2}), shown as green line. The strong interaction approximation in Eq.(\ref{eq:muavg}) is shown as blue line.

%\subsubsection{Participation ratio}
We can find the participation ratio of norm density at ground state for the strongly interacting regime by two different analytic approximations:

\begin{enumerate}
    \item Using Eq. (\ref{eq:gglsquare}) in the participation ratio definition for ground state gives%, and obtain
\begin{equation}\label{eq:P_sergej}
P =\frac{\mathcal{A}^2}{\sum_{\ell}^N {|G_{\ell}|}^4}\approx \frac{N a^2}{\left\langle \left (a-\frac{\epsilon_\ell}{g}\right)^2 \right\rangle}=
\frac{N}{{1+\frac{W^2}{12 g^2 a^2}}}, \qquad a\geq W/2g 
\end{equation}
where we inserted  the variance $\langle {\epsilon_\ell}^2 \rangle=W^2/12$. 
\item Inserting $h$ from Eq. (\ref{eq:h_large_a}), and $\mu$ from Eq. (\ref{eq:muavg}) into Eq. (\ref{eq:Pstep3}) produces%, thus we obtain

\begin{align}\label{eq:pa_large}
P^{-1} N&= 1-\frac{2J\bar{\zeta}}{ga}+\frac{W^2}{12 g^2 a^2}\\
 &+\frac{16J}{9gW^2 a^2}\left[\left(a+\frac{W}{2g}\right)^{3/2}-\left(a-\frac{W}{2g}\right)^{3/2}\right]^2,  \qquad a\geq W/2g.  \nonumber 
\end{align}
\end{enumerate}

These two approximations of $P$ at large norm densities are exhibited in Fig. \ref{fig:p_a_mu}. The difference between them comes from the analytic hopping term calculation and $\bar{\zeta}=2$ approximation in Eq. (\ref{eq:T}) included in approximation 2 (\ref{eq:pa_large}), which is absent in approximation 1 (\ref{eq:P_sergej}). Using Eq. (\ref{eq:pa_large}) in Eq. (\ref{eq:Pstep3}), we can write the energy density again as follows 
\begin{equation}\label{eq:GSanalytic2}
    h=\frac{g a^2}{2}-J\bar{\zeta}a-\frac{W^2}{24} 
\end{equation}
for strongly interacting regime. 
In both Eq. (\ref{eq:h_large_a}), and Eq. (\ref{eq:GSanalytic2}) we used $\delta\zeta_\ell=0$ approximation, yet Eq. (\ref{eq:GSanalytic2}) requires an additional approximation for $\bar{\zeta}$. We assumed $\bar{\zeta}\approx2$ in the strong interaction limit, and shown it in the phase diagram as black dashed line.

Note that in the strong-coupling regime we can further expand Eq.~(\ref{eq:gglsquare}) in small parameter $W/(ga)$
\begin{align}
{G_\ell} \simeq \sqrt{a} \left[ 1 - \frac{\epsilon_l }{2ga} 
%+O\left( \frac{JW}{(ga)^2} \right)
  \right].\label{gglsquare}
\end{align}
Then the resulting ground-state energy is very close to that of Eq.~(\ref{eq:h_large_a}).

%Using Eq.\ref{eq:gglsquare} in the participation ratio definition for ground state gives%, and obtain
%This equation \ref{eq:GSanalytic2} in which we used $\delta\zeta_\ell=0$  requires an additional approximation for $\bar{\zeta}$.

%\begin{equation}\label{eq:P_sergej}
%P =\frac{A^2}{\sum_{\ell}^N {|G_{\ell}|}^4}\approx \frac{N a^2}{\left\langle \left %(a-\frac{\epsilon_\ell}{g}\right)^2 \right\rangle}=
%\frac{N}{{1+\frac{W^2}{12 g^2 a^2}}},  
%\end{equation}
%for $a\geq W/2g$, where we inserted  the variance $\langle {\epsilon_\ell}^2 \rangle=W^2/12$. 
%Eq.\ref{eq:gglsquare} and

%Another approximation is as follows. %We multiply Eq.\ref{eq:eq4} by ${\psi_{\ell}}^*$, sum over $\ell$ with the definitions of $\mathcal{H}$ (Eq.\ref{eq:Hamiltoniandisorder}), $\mathcal{A}$, and the participation ratio $P=\mathcal{A}^2 / \sum_\ell |\psi_\ell|^4$. We then average over all sites, and obtain 
%and averaging over all sites; we get

%\begin{equation}\label{eq:Pstep3}
% P^{-1} N = \frac{2}{g a^2} (\mu a -h).
%\end{equation}

%Inserting $h$ from Eq.\ref{eq:h_large_a}, and $\mu$ from Eq.\ref{eq:muavg} into Eq.\ref{eq:Pstep3} produces%, thus we obtain

%\begin{align}\label{eq:pa_large}
%P^{-1} N&= 1-\frac{2J\bar{\zeta}}{ga}+\frac{W^2}{12 g^2 a^2}\\
% &+\frac{16J}{9gW^2 a^2}\left[\left(a+\frac{W}{2g}\right)^{3/2}-\left(a-\frac{W}{2g}\right)^{3/2}\right]^2   \nonumber 
%\end{align}
%for $a\geq W/2g$.

\section{Localization properties of elementary excitations}
\label{sec:localizationproperties}
As mentioned above in Sec. \ref{sec:intro}, the elementary excitations are initially introduced by Bogoliubov \cite{bogoliubov1947theory} as a way to explain superfluidity in BEC. Bogoliubov-de Gennes equations are usually found by introducing a small perturbation to the equilibrium wave function, which has to be the ground state for the GP lattice with the disorder.

%Introducing a small perturbation to the ground state, we solve and find the dynamics of Bogoliubov-de Gennes \cite{bogoliubov1947theory,Fetter1971,de2018superconductivity} excitations for disordered GP lattice. 

\begin{equation}\label{eq:delta}
\psi_\ell (t) \approx (G_\ell+\delta_\ell(t)) e^{-i\mu t}. 
\end{equation}
By inserting the definition of $\psi_\ell (t)$ into the equations of motion \ref{eq:eomdisorder}, and linearizing the solutions around the ground state, we obtain
\begin{equation}
i\dot{\delta_{\ell}}=(\epsilon_{\ell}-\mu) \delta_{\ell} + g {G_{\ell}}^2( \delta_{\ell}^* + 2 \delta_{\ell}) - J(\delta_{\ell+1} + \delta_{\ell-1}) \label{eq:idotdelta_l}
\end{equation}
after we eliminate the second order perturbation terms. We can define $\delta_{\ell}$ as
\begin{equation}\label{eq:delta_l1}
\delta_{\ell}(t)= \chi_{\ell} e^{-i\lambda t} - \Pi_{\ell}^* e^{i\lambda t}. 
\end{equation}
We place Eq. (\ref{eq:delta_l1}) into Eq. (\ref{eq:idotdelta_l}), and let $ \epsilon_{\ell} -\mu = \Tilde{\epsilon_{\ell}}$. Then, we obtain the exactly solvable, linear BdG equations:
\begin{align}\label{eq:chipi}
\lambda \chi_{\ell} =&\Tilde{\epsilon_{\ell}}\chi_{\ell} -J(\chi_{\ell+1}+\chi_{\ell-1}) -g G_{\ell}^2(\Pi_{\ell} -2\chi_{\ell}) \\
 \lambda \Pi_{\ell} =&-\Tilde{\epsilon_{\ell}}\Pi_{\ell} +J(\Pi_{\ell+1}+\Pi_{\ell-1})+ g G_{\ell}^2( \chi_{\ell} -2\Pi_{\ell}). \nonumber
\end{align}
Hence, the eigenvalue problem is $M\vec{R}=\lambda \vec{R}$ with
\begin{equation}
M=
\begin{bmatrix} 
c_1 & -b_1 & -J & 0 & 0 & 0 & \dots\\
b_1  & -c_1 & 0 & J & 0& 0& \ddots\\
-J  & 0 & c_2 & -b_2 & -J & 0 & \ddots \\
0 & J & b_2 & -c_2 & 0 & J & \ddots\\
0 & 0 & -J & 0 & c_3 & -b_3 & \ddots\\
0 & 0 & 0 & J & b_3 & -c_3 & \ddots\\
\vdots & \ddots  & \ddots& \ddots& \ddots& \ddots& \ddots
\end{bmatrix} 
\end{equation}
where 
\begin{equation*}
 b_{\ell}=g G_{\ell}^2, \qquad  c_{\ell}=\Tilde{\epsilon_{\ell}}+2g {G_{\ell}}^2,
\end{equation*}
and

\begin{equation}\label{eq:R}
\vec{R}=\{\dots,{\chi}_{\ell-1}, {\Pi}_{\ell-1},{\chi}_{\ell}, {\Pi}_{\ell}, {\chi}_{\ell+1}, {\Pi}_{\ell+1}\dots \}. 
\end{equation}
We find the eigenvalues ($\lambda$), and eigenvectors ($\chi_\ell,\Pi_\ell$) by diagonalizing $2N\times 2N$ matrix $M$ of Eq. (\ref{eq:chipi}). To do that, we first implant $G_\ell$ which is exactly found by the numerical method explained in Sec. \ref{sec:GroundStateRenormalization}.

\begin{figure}[hbt!] 
     \centering
     \includegraphics[width=0.7\textwidth]{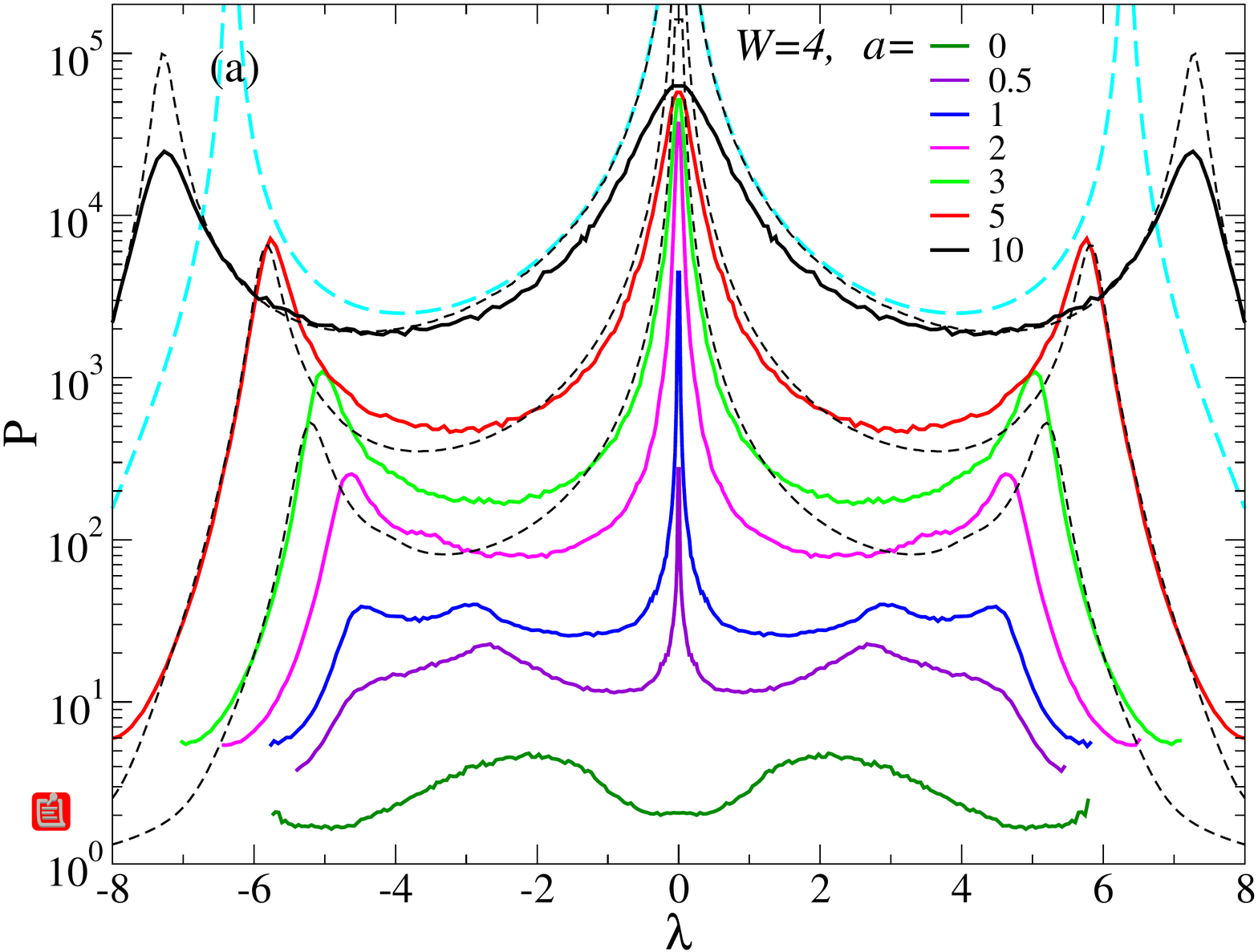}\\%PlaW4N1d5tez.pdf}\\%PlaW4N100000.pdf}%PlaW4N100000simple.pdf}%
    \includegraphics[width=0.7\textwidth]{Plax1WvaryN50000.pdf}%Plax1Wvaryx.pdf}%Plax1Wvary.pdf}
     \caption[The participation ratio of BdG modes vs. energy]{The average participation ratio $\bar{P}$ of BdG modes versus the energy $\lambda$. (a) Solid lines - numerical computation with $N=10^5$, $W=4$ and $a=0$ to 10 from bottom to top. Thin dashed lines - transfer matrix calculation results of the localization length $1.5 \xi(\lambda)$ for $a=3,5,10$ with $10^8$ number of iterations. Cyan thick dashed line - analytic result (\ref{invloc_fin}) in the strong interaction and small energy approximation for $a=10$ (we plot $1.5\xi$).  (b) Solid lines - numerical computation with $a=1, N=10^5$ and $W=0.5$ to 4 from top to bottom.}
     \label{fig:PlaN20000log}
     \end{figure}

The BdG equations in Eq. (\ref{eq:chipi}) have particle-hole symmetry $\left[\lambda_\nu, \left\{{\chi_\ell}^\nu,{\Pi_\ell}^\nu\right\}\right] \longleftrightarrow \left[-\lambda_\nu,\left\{{\Pi_\ell}^\nu,{\chi_\ell}^\nu\right\}\right]
$ that gives the solution $\chi_\ell =\Pi_\ell$ for $\lambda_\nu = 0$. Here $\nu$ is the mode number. 
Since the perturbation $\delta_\ell$ -defined in Eq. (\ref{eq:delta_l1})- is time independent at $\lambda_\nu=0$, the time evolution of the perturbed part in Eq. (\ref{eq:delta}) is proportional to the ground state $G_\ell$ such that $\psi_\ell=(G_\ell+\delta_\ell)e^{i\mu t}$. 
The lowest-energy solution of the equations (\ref{eq:chipi}), with eigenvalue $\lambda=0$, has the shape of the ground state, i.e., $\chi_{\ell}=\Pi_{\ell} \propto G_\ell$ which is delocalized in space for any $G_\ell>0$. % first mentioned for bosons in \cite{Fetter1971}. 
For real eigenvalues, we numerically obtain real and normalized eigenvectors, i.e., $\sum_{\ell=1}^N |\chi_{\ell,\nu}|^2 +|\Pi_{\ell,\nu}|^2=1$. BdG modes to nonzero values of $\lambda$ are expected to be Anderson localized due to the presence of disorder and the one-dimensionality of the system.
We numerically calculate the participation ratio of each mode as

\begin{equation}\label{eq:P_eigen}
P_\nu= \frac{\left(\sum_{\ell}^N  n_{\ell,\nu}  \right)^2}{\sum_{\ell}^{N}n_{\ell,\nu}^2} \; , \; n_{\ell,\nu} = |\chi_{\ell,\nu}|^2 +|\Pi_{\ell,\nu}|^2
\end{equation}
where $\nu=1,\dots,N$ is the mode number. The results for different norm densities at $W=4$ are plotted in Fig. \ref{fig:PlaN20000log}, showing a distinct side peak behavior, and an anomalous growth at $\lambda=0$. The participation ratio enhances for all $\lambda_\nu$ as interaction strength increases or disorder decreases. 

%Inset: The transfer matrix result for $a=10$ is zoomed to show the emergence of a small side peak.

We divide the $\lambda$-axis into small bins of size 0.05 and average the
participation ratios in each bin to obtain the dependence $\bar{P}(\lambda)$. For a system size as large as $N=10^5$, the standard deviation of  $\bar{P}(\lambda)$ of different disorder realizations  is $< 0.1$ of its value. Hence, we used three random disorder realizations of $N=10^5$ to smooth the curves without any filter. The resulting curves are plotted  for different norm densities with $W=4$ in Fig. \ref{fig:PlaN20000log}(a), and for different disorder strengths with $a=1$ in Fig. \ref{fig:PlaN20000log}(b). We observe symmetric
curves $\bar{P}(\lambda)=\bar{P}(-\lambda)$ due to the particle-hole symmetry of the BdG
eigenvalue problem. Increasing norm density $a$ has a competing effect with increasing disorder strength $W$. It is because the system approaches the strong interaction limit when $a\gg W$. In Fig. \ref{fig:PlaN20000log}(b), the energy of the side peak $\lambda_{sp}$ shifts with the increase in $a$, which is later explained analytically with Eq.(\ref{invloc_fin}). On the other hand, $\lambda_{sp}$ stands still at approximately the same energy while we decrease $W$ in Fig. \ref{fig:PlaN20000log}(b), since $\lambda_{sp}$ is independent of $W$ as in the analytical relation (\ref{invloc_fin}). 

The dispersion relation -see (\ref{k0_1})-
\begin{equation}
\lambda^2-[ga +4J\sin^2{({q}/2)}]^2+(ga)^2=0
\end{equation}
hints the interplay between energy ($\lambda$) and momentum ($q$). According to the approximated dispersion relation $E=2(1-\cos(q))$ in (\ref{Edef}) with $E=\lambda^2/(2gaJ)$, we expect to observe the side peak approximately at $q\rightarrow \pi/2$ in the strong interaction limit.

%According to the dispersion relation -see (\ref{k0_1})-
%\begin{equation}
%\lambda^2-[ga +4J\sin^2{({q}/2)}]^2+(ga)^2=0,
%\end{equation}
% we expect to observe a minor peak  at $q\approx \pi/2$,  which is more obvious in the weak interaction regime in Fig. \ref{fig:PlaN20000log}. Later in the strong interaction regime, it is dominated by the side peak formations.

Let us rewrite Eq. (\ref{eq:chipi}) using Eq. (\ref{eq:a_generalrule}) simply as
\begin{align}\label{eq:chipinew}
\lambda \chi_{\ell} =& J \zeta_{\ell} \chi_{\ell} -J(\chi_{\ell+1}+\chi_{\ell-1}) -g G_{\ell}^2(\Pi_{\ell} -\chi_{\ell}), \\
 \lambda \Pi_{\ell} =&- J \zeta_{\ell} \Pi_{\ell} +J(\Pi_{\ell+1}+\Pi_{\ell-1})+ g G_{\ell}^2( \chi_{\ell} -\Pi_{\ell}). \nonumber
\end{align}
where $J\zeta_\ell=\Tilde{\epsilon_{\ell}}+g G_\ell^2$. 
Next we use the decomposition
%substitute
$\chi_\ell=(S_\ell+D_\ell)/2$ and $\Pi_\ell=(S_\ell-D_\ell)/2$ %into Eqs.~(\ref{eq:chipi}) and
we arrive at the (still) exact set of equations
\begin{align}
\lambda S_{\ell}&=(J\zeta_\ell+2g G_{\ell}^{2}) D_{\ell} -J(D_{\ell+1}+D_{\ell-1}), \label{Sl} \\
\lambda D_{\ell}&=J\zeta_\ell S_{\ell}- J(S_{\ell+1}+S_{\ell-1}) \label{Dl}.
\end{align}
Inserting (\ref{Dl}) into (\ref{Sl}) yields

\begin{equation}
\label{onlyS}
\frac{\lambda^2}{J} S_{\ell} = (J\zeta_\ell+2g G_{\ell}^{2})(\zeta_\ell S_{\ell}- S_{\ell+1}-S_{\ell-1}) - J(\zeta_{\ell+1} S_{\ell+1} +
\zeta_{\ell-1} S_{\ell-1} - S_{\ell-2} - 2S_\ell - S_{\ell-2} ).\nonumber
\end{equation}

BdG modes to nonzero values of $\lambda$ are expected to be Anderson localized due to
the presence of disorder and the one-dimensionality of the system.

\subsection{The interaction dependent exponent near zero energy}%of $\bar{P} (|\lambda| \rightarrow 0) \rightarrow \infty$}% on phase transition}

\begin{figure}[hbt!] 
     \centering
     \includegraphics[width=0.7\textwidth]{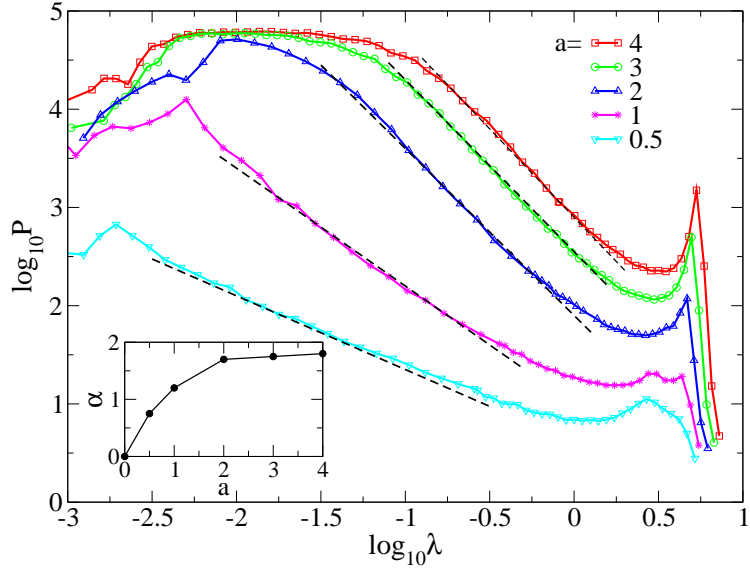}%logPloglaN20000alpha.pdf}
     \caption[The exponent of divergence of localization length near zero energy]{The log-log plot of $\bar{P}$ of BdG modes versus eigenvalues ($\lambda$) for different norm densities $a=0.5,1, 2, 3, 4$, $W=4, N=10^5, N_r=1$. The calculated slope with $\sim 1/ \lambda^{\alpha}$ are shown as black dashed lines, and plotted versus $a$ in the inset.}
     \label{fig:PlaW4alpha}
\end{figure}

All curves show a clear divergence $\bar{P} (|\lambda| \rightarrow 0) \rightarrow \infty$ which is
only limited due to finite-size effects. This divergence agrees with the above result
that the BdG mode at zero energy $\lambda=0$ must be delocalized and thus have an infinite participation ratio. The divergence has been addressed in a number of publications \cite{Pavloff,Fetter1971,Chalker,Deng,Fontanesi_PRA,Fontanesi_PRL} which results in $\bar{P} \sim 1/|\lambda|^{\alpha}$ with $\alpha=2$ in the strong interaction regime \cite{Pavloff,Ishii,Ziman}, whereas $\alpha=1$ signifies  the  transition  from  a  superfluid  to  an  insulator \cite{Chalker,Fontanesi_PRA,Fontanesi_PRL}. In Fig. \ref{fig:PlaW4alpha} we exhibited how the exponent $\alpha$ depends on the interactions. The strong interaction limit $a\gg W$ can be reached by fixing $W$, and gradually increasing norm density $a$. Correspondingly, we observed a slow convergence to the $\alpha=2$ value, as we increase $a$. %We did not calculate the exponent for $a=5$, since the $\bar{P}(\lambda)$ curve is not long enough to fit due to the dominance of the finite-size effect. 
A similar trend in the increase of $\alpha$ is observed in \cite{Fontanesi_PRL} for 1D disordered Bose gas. %, yet we could calculate further $\alpha$ results through using a much larger system size. 

%The universal properties of superfluids, near the quantum phase transition can be characterized by the exponent of the diverging localization lengths near zero energy. There has been extensive work on the localization of eigenstates in one-dimensional disordered boson systems \cite{Ziman,Ishii,Slevin_2014,Fetter1971,Chalker,Pavloff} to examine the transition from a superfluid to an insulator. The exponent ($\alpha$) of divergent localization length of Bogoliubov modes at low energies  has been studied in \cite{Fetter1971,Ziman,Chalker,Pavloff,Slevin_2014,Huber} for boson systems in a random potential. In 2008, Gurarie et al. \cite{Chalker} claimed that the exponent $\alpha=1$ signifies the transition from a superfluid to an insulator, while  $\alpha=2$ is a well-known result in the limit of strong interaction \cite{Pavloff,Ishii,Ziman}. %To examine the superfluid to Bose-Glass phase transition, Fontanesi et al. studied power-law divergence of the density of states at low energy \cite{Fontanesi_PRL,Fontanesi_PRA}. Nevertheless, they were not able to observe the exponents ($\alpha$) of divergent participation number -localization length- for a long enough range of energy spectrum. The finite size of their simulations limited their analysis for strongly interacting regime (see Fig.\ref{fig:fontanesi_figure}).

\subsection{Side peak behavior of BdG spectrum}

The participation ratios at nonzero energies show 
an expected finite height peak in the weak interaction regime which is a continuation
of the zero interaction limit (Fig. \ref{fig:PlaN20000log}). In that limit, the BdG eigenvalue equations decompose into two copies
of the tight-binding chain with the onsite disorder. The largest localization length and participation ratio are then obtained in the centers of their spectra $\lambda=\pm \mu$ which host the largest density of states \cite{Kramer93}. Upon crossing over from the weak interaction to the strong interaction regime, we observe a second peak developing at larger absolute $\lambda$ values, which has a finite height that yet appears to grow swiftly (Fig. \ref{fig:PlaN20000log}). This new side peak results in an unexpected enhancement of the participation ratio, localization length, and size of BdG modes at finite energies $\lambda$ in the strong interaction regime.

%\subsection{Finite momentum localization length singularity for strong interactions}

In order to analytically assess the observed side peak of the BdG modes in the strong interaction regime
$g a \gg W$, we use the exact equations (\ref{onlyS}) with the approximated GS field (\ref{eq:gglsquare}) and compute the localization length
$\xi(\lambda)$ with a transfer matrix method (see  Appendix \ref{LLTMM}).

The participation ratio ($P$) and localization length ($\xi$) are proportional to each other according to the following relations. From the generic description of localization length $\psi_\ell \sim  e^{-|\ell|/\xi}$, we find the total norm 
\begin{equation}
\mathcal{A}= \sum_\ell |\psi_\ell|^2 \equiv 2 \int_{0}^{\infty} |\psi_\ell|^2 d\ell \sim 2 \int_{0}^{\infty} e^{-2|\ell|/\xi}  d\ell =  \xi.
\end{equation}
Similarly
\begin{equation}
 \sum_\ell |\psi_\ell|^4 \sim  2 \int_{0}^{\infty} e^{-4|\ell|/\xi}  d\ell ={   \xi} /2.
\end{equation}
Thus we find the participation ratio %is generically defined as
\begin{equation}
    P=\frac{\mathcal{A}^2}{\sum_\ell{|\psi_\ell|^4}} \sim \frac{ \xi^2}{ \xi /2} = 2 \xi.
\end{equation}
Although here their proportionality factor is roughly 2, we consider the relation $P_\nu\approx1.5 \xi_\nu$, which is found numerically in \cite{Kati2020}. 
%The participation ratio is proportional to the localization length with a relation $P_\nu\approx1.5 \xi_\nu$ \cite{Kati2020}. 
%%Hence, we compare $1.5 \xi_\nu$ (black dashed lines) with the exact $\bar{P_\nu}$ solutions (solid lines)  in Fig. \ref{fig:PlaN20000log}.

\begin{figure}[hbt!]%[tp]
\centering
\includegraphics[width=0.7\textwidth]{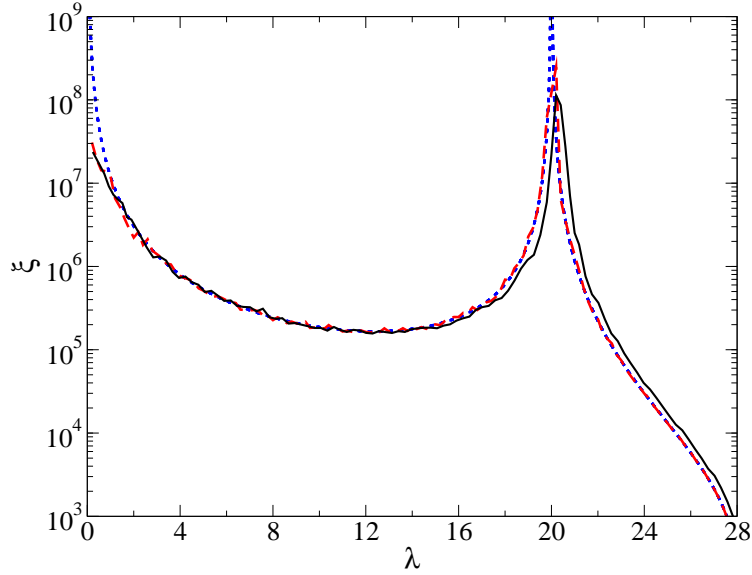}
\caption[The localization length vs. energy for strongly interacting regime]{
The localization length $\xi(\lambda)$ for $a=100$, $g=J=1$, $W=4$.
Black solid line: exact BdG equations (\ref{onlyS}), approximate GS field (\ref{eq:gglsquare}), transfer matrix (\ref{tmm1new}) calculation results with $10^8$ iterations. Red thick dashed line: approximate BdG equations
(\ref{Slap}), approximate GS field (\ref{eq:gglsquare}), transfer matrix (\ref{tmm2new}) calculation results with $10^8$ iterations. Blue thick dotted line - analytical result (\ref{invloc_fin}) with (\ref{Edef}).
}
\label{largea}
\end{figure}

The resulting curves of $1.5 \xi_\nu$ are plotted in Fig.\ref{fig:PlaN20000log}(a) for $a=3,5,10$
and show almost full quantitative agreement with the numerical results from the exact equations and the numerically exact GS for $a=10$, while the
agreement is less quantitative but still qualitative as the value of $a$ is reduced. Therefore we can use the
approximate GS field (\ref{eq:gglsquare}) with the exact equations (\ref{onlyS}) as a reliable reference for even larger values of $a > 10$, which are
not accessible by brute force numerical computations. The resulting dependence $\xi(\lambda)$ is shown in Fig.~\ref{largea} for $a=100$. The
side peak is not only remaining in place but is also increasing its height relative to the background.

In the strong interaction limit, $ga \gg W$, deeper insight on the localization properties of BdG modes can be obtained. We simplify the exact Eq. (\ref{onlyS}) keeping only the leading order term on the RHS of (\ref{onlyS}), and arrive at
\begin{align}\label{Slap}
  \frac{\lambda^2}{J}S_\ell=2g G_{\ell}^{2}(\zeta_{\ell} S_{\ell} -S_{\ell-1} -S_{\ell+1}).
\end{align}
Note that $\lambda=0$ and $S_l=G_l$ is still a valid solution.
We again compute the localization length $\xi(\lambda)$ using (\ref{Slap}) and the GS field approximation (\ref{eq:gglsquare}). The resulting curve
for $a=100$ agrees quantitatively with the exact equation result in Fig.~\ref{largea}, confirming the validity of our equation approximation.

Defining the dimensionless energy $E$ as $E=\lambda^2/(2gaJ)$
%\begin{align}\label{Edef}
%E=\lambda^2/(2gaJ)=2 (1-\cos q)
%\end{align}
we cast equation (\ref{Slap}) into the standard form
\begin{align}\label{stand}
 (\tilde{E}+\varkappa_\ell)S_\ell=S_{\ell-1} +S_{\ell+1}
\end{align}
with
\begin{align}
\tilde{E}=\langle \zeta_\ell\rangle-E\langle a/G_{\ell}^{2}\rangle\;,\;
\varkappa_\ell=\zeta_\ell-E\frac{a}{G_{\ell}^{2}}-\tilde{E}.\label{kapp}
\end{align}
In the strong interaction regime $ga \gg W$, $\tilde{E}=2-E$ and the perturbing random potential $\varkappa_\ell$ is small,
%in the strong-coupling regime $W/(ga) \ll1$
with its expectation value being zero: $\langle\varkappa_\ell\rangle =0$.
The disorder field $\epsilon_\ell$ is uncorrelated at different sites: $\langle\epsilon_n\epsilon_m\rangle =\delta_{nm} W^2/12$.
This holds as well for the ground state field (\ref{eq:gglsquare}) in the strong coupling regime. However, the ground state field $\zeta_\ell$ has a finite range of correlations due to the presence of nearest neighbor terms $G_{\ell\pm1}$ in its definition (\ref{eq:a_generalrule}). As a consequence, the random potential $\varkappa_\ell$ is also correlated.
Anderson localization with correlated disorder was studied in many publications (see Ref.~\cite{Lifshits88:book} for continuum models and Refs.~\cite{Griniasty88,Luck89,Izrailev99,Titov05} for lattice models).
The localization length of model (\ref{stand}) is calculated following Sec.~5.2.1 of the review \cite{Izrailev12} (see Appendix \ref{LLCD} for details):
\begin{align}\label{invloc_fin}
\xi=\frac{96g^2a^2}{W^2}\frac{4-E}{E(2-E)^2}\;.
\end{align}
In the vicinity of $E=\lambda^2/(2gaJ)=0$, we obtain the localization length divergence 
\begin{equation}\label{eq:analyticlocalizationlengthk0}
{\xi}\simeq \frac{192g^3a^3 J}{W^2\lambda^2}    
\end{equation}
Notably, we discover an additional divergence of the localization length at finite
energy $E=2$, i.e.
%$q=\pi/2$ and
$\lambda = \pm 2\sqrt{gaJ}$,
%momentum $q=\pi/2$ (i.e. $\lambda= 2\sqrt{gaJ}$)
as
\begin{align} \label{sidepeak}
\xi \simeq \frac{24g^3a^3J}{W^2 (2\sqrt{gaJ} \pm \lambda)^2}\;, \; |\lambda \pm 2\sqrt{gaJ}| \ll 2\sqrt{gaJ} \;.
\end{align}
The above singularity is the explanation for the observed side peak. We plot (\ref{invloc_fin}) in
Fig.~\ref{largea} for $a=100$ and find quantitative agreement with the localization length data from transfer matrix evaluations of the exact and approximate equations while using the approximate GS field dependence as induced by the disorder.

Eq. (\ref{eq:analyticlocalizationlengthk0}) is also found in Appendix \ref{ap:analyticlocalizationlength} in which we observed that the side peaks of the localization length of BdG modes vanish if we neglect the correlation between GS field ($\zeta_\ell$) and disorder ($\epsilon_\ell$).

\begin{figure}[hbt!] 
\centering
\includegraphics[width=0.7\textwidth]{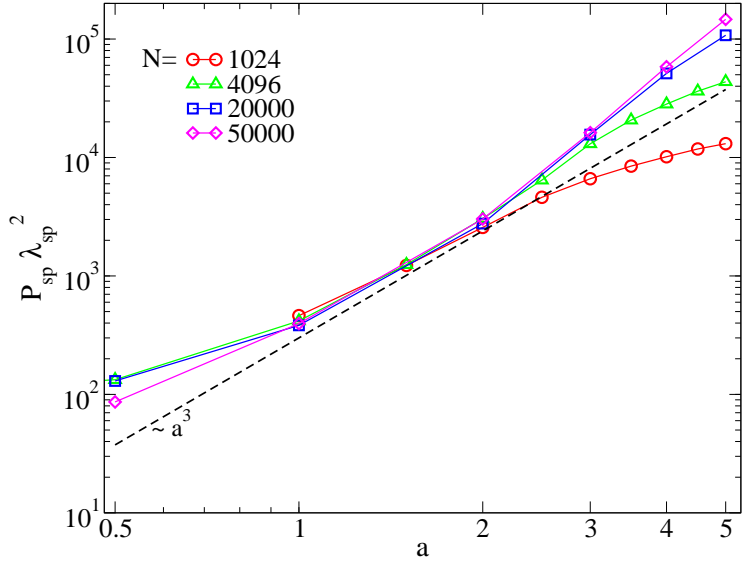}%pdpN4096log.pdf}%logPdp.pdf}
\caption[The participation ratio of BdG modes at the side peak vs. norm density]{The log-log scaled plot of $P\lambda^2$ at the location of the side peaks in Fig. \ref{fig:PlaN20000log} versus norm density $a$, shown for $N=1024, 4096, 20000, 50000$ with  $W=4$. The slope with $\sim a^{3}$ is shown as black dashed lines.} %N_r=1
\label{fig:plasquare_sidepeak}
\end{figure}

In Fig. \ref{fig:plasquare_sidepeak}, the value $P_{sp}$ represents the participation ratio of modes averaged in the bin ($\lambda_{sp}$) where the side peak occurs. We plotted $P_{sp}{\lambda_{sp}}^2$ vs $a$ for different sizes $N$, as there is a power-law increase with $\sim a^3$, which is confirmed by the analytical calculation of localization length in Eq. (\ref{eq:analyticlocalizationlengthk0}). 

\subsection{Generalizations}

Let us generalize to any lattice dimension with some hopping network or generalized discrete Laplacian:
\begin{equation}\label{eq:eomGen}
i\dot{\psi_{\ell}}=\epsilon_{\ell} \psi_{\ell} +  g|\psi_{\ell}|^2\psi_{\ell}-\mathcal{D}(\psi_\ell)
\end{equation}
The discrete Laplacian
\begin{equation}\label{DL}
\mathcal{D}(\psi_\ell) = \sum_m J(\ell-m)\psi_m\;.
\end{equation}
We assume $J(m) \geq 0$ to ensure the nonnegativity of the ground state field $G_l$. Note that the Hamiltonian (\ref{eq:Hamiltoniandisorder}) is obtained with the choice  $J(m)=J(\delta_{m,1}+\delta_{m,-1})$.
It follows from the definition (\ref{DL}) that
\begin{equation}\label{eq:gglsquare_gen}
    g G_\ell^2=ga-\epsilon_\ell + \delta\hat{\zeta}_\ell \geq 0
\end{equation}
and
\begin{equation}\label{zetal}
\hat{\zeta}_l = \frac{1}{G_l} \mathcal{D}(G_l)
\end{equation}
Note that with this definition the field $\hat{\zeta}_l$ includes the strength of the hopping network, as opposed to previous notations.

The \emph{exact} equations for $S_\ell$ and $D_\ell$ take the form
\begin{align}
\lambda S_{\ell}&=(\hat{\zeta}_\ell+2g G_{\ell}^{2}) D_{\ell} -\mathcal{D}(D_l)  \label{Sl2}\\
\lambda D_{\ell}&=\hat{\zeta}_\ell S_{\ell}- \mathcal{D}(S_l)  \label{Dl2}.
\end{align}
The approximate expression for the field $G_\ell$ in the strong-coupling case is still given by Eq. (\ref{eq:gglsquare}), with all corrections due to the change in the hopping network and even the dimensionality appearing in higher-order corrections.

Since both $\hat{\zeta} \sim J$ and $\mathcal{D} \sim J$, we arrive at the generalized strong interaction BdG equations similar to the
above considered one-dimensional case with nearest neighbor hopping as
\begin{align}\label{Slap2}
  \lambda^2 S_\ell=2g G_{\ell}^{2} \left( \hat{\zeta}_{\ell} S_{\ell} -\mathcal{D}(S_l)\right)  .
\end{align}
Equations (\ref{eq:gglsquare}), (\ref{DL}), (\ref{zetal}), and (\ref{Slap2}) constitute the generalization of the BdG equations in the strong coupling limit to any lattice dimension and hopping
network. It remains to be studied whether these equations also result in a strong enhancement of transport properties of disordered BdG modes
at certain finite energies and momenta due to ground state correlations.

%\subsection{Transfer matrix method}%Participation number $\&$ localization length}

 %, and obtain a good match between them for low energies. 
%%However, the localization length of Eq. (\ref{sldl}) gives a side peak that has a relatively much smaller amplitude. Its side peak behavior can be seen more clearly in the inset of Fig. \ref{fig:PlaN20000log}(a). This shows that the assumption $\zeta_\ell=2$ causes the disappearance of a huge part of the side peak. We conjecture that the reason must be the nondecaying correlation between $\zeta_\ell$ and $\epsilon_\ell$ - $r(\zeta_\ell,\epsilon_\ell)\sim 0.8$ - shown in Fig. \ref{fig:zeta_a_corr}(b). When we assume $\zeta_\ell=2$, we also assume that the correlation $r(\zeta_\ell,\epsilon_\ell)$ is zero, which is incorrect. Moreover, we tested the exact participation ratio of the ground state of a Hamiltonian in which $\zeta_\ell=2$ and $W=4$, and found matching results with the TMM method, both of which have no clear side peak for $a<5$. %This is also the confirmation of our numerical results. %This demonstrates how significant the fluctuations of $\zeta_\ell$ is.

\section{Conclusion}\label{sec:conclusionch4}

By this chapter of the thesis, the disorder is introduced to the one-dimensional classical GP equation. At first, we found the new disorder-induced ground state and calculate its exact statistical properties numerically. Then, we obtain analytical expressions for the thermodynamic properties of the ground state, i.e., the chemical potential and the participation ratio density, and compare them with direct numerical calculations. For small ground state density, we identify a Lifshits regime where disorder dominates over the interactions. In this regime, the ground state consists of separated islands of Lifshits states, analogous to the fragmented Bose--Einstein condensates in the Lifshits-glass (Anderson glass) phase \cite{LuganPRL98,Fontanesi_PRA,Fontanesi_PRL,Scalettar,LuganPRL2007,Kati2020,Deng}. On the other hand, for large ground-state density, the interaction dominates and screens the disorder. In that scenario, we observe a relatively homogeneous distribution of norm density at the ground state, similar to the unfragmented Bose--Einstein condensate.

%In the first part of this chapter, I will present the zero-temperature dynamics and statistical properties of the discrete GP model. For the weakly interacting particles, we observed and confirmed the existence of the Lifshits-glass regime in GP lattice, while it has been found experimentally and theoretically for ultra-cold Bose gas \cite{LuganPRL98,Fontanesi_PRA,Fontanesi_PRL,Scalettar,LuganPRL2007,Kati2020,Deng}. In 2000, Rasmussen et al. \cite{Rasmussen2000} found the statistical properties for the clean GP lattice. Yet, the statistics of \textit{disordered} GP lattice remained unresolved. In this chapter, we examine its statistics for the ground state, both analytically, and numerically. On the theoretical side, most efforts have been devoted to the study of the ground state, and its elementary excitations with several analytic approximations, considering both weak \& strong interaction regimes. 

The main purpose of this study is to find the statistical behavior of BdG mode excitations above the ground state of GP lattice. We introduce a small fluctuation to the ground state wave function and solve its equations of motion which are known as the BdG equations. We compute their localization properties by measuring participation ratios and localization length. We observed the localization length $\xi$ of the elementary excitations at different norm densities and disorder strengths. 

We confirmed the divergence of the localization length at zero energy ($\lambda=0$), due to the delocalization of modes $\chi_\ell = \Pi_\ell \propto G_\ell$, as it has been theoretically expected for BEC \cite{Fetter1971}. We found the well-known $\lambda^{-\alpha}$ power-law divergence of the localization length near to zero energy. We found $\alpha=2$ analytically and observed numerically with the transfer matrix method in the weak disorder or large norm density limits, as it is commonly observed in the models of harmonic chains \cite{Ishii}, acoustic and electromagnetic waves \cite{PhysRevB.34.4757,PhysRevA.31.3358}, tight-binding Hamiltonian \cite{Ziman}, and BEC \cite{Pavloff,Chalker}, in the weakly disordered environment. Moreover, we show how interaction and disorder affect the value of $\alpha$ which may shed a light on the phase transition from Bose-glass (fragmented BEC) to BEC, which is expected to occur at $\alpha=1$ \cite{Chalker}.

In the strong interaction regime, a novel \ac{BM} anomaly develops with a strong increase of localization length at finite momentum $q\sim \pi/2$. We call this enhancement of BM as side peaks and postulate that its existence originates from the correlation between GS field $\zeta_\ell$ and onsite potential $\epsilon_\ell$. To elucidate their occurrence, we perform a systematic perturbation approach that gives approximate eigenvalue equations which are valid in the strong interaction regime. That eigenvalue problem corresponds to a one-dimensional tight-binding chain with nearest-neighbor hopping and correlated on-site disorder. We derive analytical expressions for the localization length as a function of energy. We then finally obtain a singularity and length divergence at finite energy, which precisely corresponds to the numerically
observed side peak. Therefore we conclude that a weakly excited disordered condensate in the regime of strong interaction will allow for almost ballistic transport of excitations for selective finite energies and momenta. We also generalize the strong interaction equations for Bogoliubov-de Gennes modes for more complicated and higher dimensional networks.
This surprising new feature of the localization lengths is conjectured to influence the studies on the disordered BEC and certainly requires a future assessment and open to a further explanation. %, and may affect the field on the disordered BEC.

\cedp
\chapter[Disorder: Density resolved wave spreading]{Density resolved wave spreading}
\label{chapter5}

\ifpdf
    \graphicspath{{Chapter5/Figs/Raster/}{Chapter5/Figs/PDF/}{Chapter5/Figs/}}
\else
    \graphicspath{{Chapter5/Figs/Vector/}{Chapter5/Figs/}}
\fi

In the previous chapter, we introduced the equilibrium statistical and dynamical properties of the new disorder-induced  ground state of Gross--Pitaevskii (GP) lattice, and the elementary excited modes close to it. In this chapter, we will present the non-equilibrium dynamics of disordered GP lattice considering the states close to the  ground state, and in Gibbs \& non-Gibbs regimes. %the states far from it. %in Gibbs \& nonGibbs regimes. %far from it.

%the dynamics of disordered GP lattice close to the new ground state, along with the statistical properties of the ground state.
%the statistical properties of the disorder-induced ground state, and its dynamics close to the 
%dynamics of disordered GP lattice close to the new ground state 

Above all, this study is the first density resolved wave packet spreading  performed in the disordered GP lattice.  
%This chapter is particularly devoted to our study on the first density resolved wave packet spreading. %,  performed in the disordered Gross-Pitaevskii (GP) lattice.  
GP wave spreading is controlled by the relation of two conserved quantities - energy and norm. This removes ambiguity from previous attempts and greatly improves the possibility to observe different spreading regimes. In this chapter, we share direct evidence for the observation of the GP regime of \ac{SC} sub-diffusive spreading and reconfirm the regime of \ac{WC} sub-diffusive spreading. We further quantify the ground state which is impacted by the finite strength of disorder.  Close to the ground state,  wave packets fragment and get trapped in a disorder-induced phase of disconnected insulating puddles of matter.  Based on our analysis, we identify a Lifshits phase which shows a significant slowing down of sub-diffusive spreading.

%old abstract
%We observe an evidence of weak and strong chaos in the spatiotemporal evolution of thewave packets on DNLS chain i.e. possessing properly excited initial states defined with a pairof centain norm and energy densities. A technique of local derivatives on logarithmic scales is used in order to quantitatively visualize the slow crossover processes

\section{Introduction}

%\begin{itemize}
%\item Consider Discrete Nonlinear Schr\"odinger Equation (DNLS) with disorder.
%\item Linear limit: Anderson localization (AL), wave packets first spread, then get stuck.
%\item Nonlinear terms destroy AL, wave packets spread subdiffusively with second moment $m_2 \sim t^{\alpha}$.
%\item Weak chaos: $\alpha=1/3$ was observed; strong chaos: $\alpha=1/2$ was NOT observed.
%\item DNLS conserves two integrals of motion: energy $H$ and norm $A$.
%\item DNLS has a Gibbs and a non-Gibbs phase in energy-norm density space ${h, a}$.
%\item Spreading wave packets have size $L$, norm density $a_w=A/L$ and energy density $h_w=H/L$.
%\item Spreading wave packets follow line in density space which hits origin.
%\item Previous studies were not density resolved, mixed different density pairs.
%\item Results: we evolve density resolved wave packets.
%\item Results: we observe strong chaos $\alpha=1/2.$
%\item Results: non-Gibbs phase is not important, but self-trapping line is.
%\end{itemize}

Disorder is inevitable naturally in all materials due to the impurities or defects caused by external fields, or a secondary incommensurate lattice \cite{Laptyeva2014}. It does not only impair transport properties of waves but also leads to a ubiquitous phenomenon of wave physics with exponential localization of eigenstates in linear waves, propagating in a medium with uncorrelated random impurities, labeled “Anderson localization” (AL). The complete suppression of the usual wave propagation has been manifested by a bevy of experimental observations; including localization of light waves \cite{opticAL}, photonic crystals \cite{photon_AL}, sound waves \cite{Sound}, microwaves \cite{Microwave}, and atomic matter waves \cite{matterwave}. However, in the presence of nonlinearity, delocalization can arise as a consequence of many-body interactions and ultimately lead to chaotic dynamics, which destroys Anderson localization through incoherent spreading \cite{PhysRevLett.107.240602,Shepelyansky94, Molina98, Pikovsky08,Flach102}.

The spreading of wave packets in disordered nonlinear lattices has been studied extensively using the interplay of disorder and one conserved quantity, in both numerical simulations~\cite{Shepelyansky94, Molina98, Pikovsky08,Flach102, Skokos:2009, Yu,VakulchykFistulFlach2019} and analytical treatments~\cite{Wang,Fishman08,Fishman09,Basko2011,Basko2014}. Although GP lattice has two integrals of motion, the relation between two conserved quantities has never been used for wave spreading.

A recent study based on the nonlinear diffusion equation shows that for weak disorder ($W\leq 4$) and moderate nonlinearity, the strong chaos regime exists in the disordered GP chain~\cite{Basko2014}. On the numerical simulation side, although strong chaos in the Klein-Gordon chain was observed \cite{Laptyeva2010},  the onset of strong chaos was less clear in the GP chain. Hence, one of the triggers of our study is if it is possible to spot strong chaos in GP lattice with a density resolved system.

%The existence of strong chaos is rigorously proven in \cite{flach2009universal}     

On the other hand, in the energy density-norm density phase diagram of GP lattice, there is a region where thermalization is absent due to the presence of long-lived breather-like excitations~\cite{Rasmussen:2000,Johansson:2004,Rumpf2004,Rumpf:2008,Rumpf2009,Iubini:2013}. States in this region cannot be described by a standard Gibbs ensemble at positive temperatures which are formally corresponding to “negative temperature” states \cite{Rasmussen:2000}. We also presumed that a novel area in the two-densities phase space will be accessible at low temperatures because the ground state of the system has to be shifted in the presence of disorder (see Appendix C, \cite{Basko2011}). The existence of two compelling regions has tempted some inquiries on how the course of spreading would be if a wave packet acquires such initial states. The first thing one may want to explore is whether the wave will be halted, or if it will show a distinct spreading. % Will the wave be halted, or will it show a distinct spreading behavior? 

While it is widely assumed that wave packets can spread to the whole lattice with strong and/or weak chaos in nonlinear disordered media; our study finds that two cases strongly restrict this behavior, and result in an AL-like localization: \ac{ST} that is mainly located in the non-Gibbs regime, and \ac{LP} discovered in the low-temperature area of phase space.

Although the effect of self-trapping has been observed in GP lattice, it was not yet clear since all previous studies ignored the presence of two relevant densities and were usually unwillingly averaging over different energy densities. Recently, it is found that self-trapping mainly depends on the relation of norm and energy density, which has been shown as an inhomogeneity line above which no microcanonical state with constant norm density is allowed \cite{sergej}.

Self-trapping and Lifshits phases are not present in the systems with one integral of motion  \cite{Basko2011,Basko2014}. Therefore, one has to transport the properties of these systems to a density resolved system \cite{Johansson:2004,magnus}.

In this chapter, we will present the norm density and the energy density resolved wave packet spreading in a disordered GP chain. We prepare multiple-site excitations of given norm density and energy density. We aim to investigate the spreading of wave packets in different regions of the equilibrium phase diagram of the GP chain. We find 1) evidence of strong chaos 2) a new localization regime (LP), 3) signatures of non-Gibbs phase from the spreading of wave packets (ST) 4) the onset of weak chaos spreading appear in the time window of the simulations 5) an improvement in the efficiency of spreading as the absolute value of initial energy density approaches zero. %3) a ground state depending on the disorder strength 4

\section{Initial densities \& subdiffusive regimes}%Initial State densities and the temperature }%Computational Details} Direction of Wave Spreading on Densities Diagram}%

The initial states are described as an excited region ($\psi_\ell>0$) in the middle of the lattice, which occupies a size $L_0\ll N$. We leave the rest of the lattice sites unexcited ($\psi_\ell=0$) so that the spreading process begins with a wave packet length $L_0$. The initial wave packet is defined with two densities: norm density $a=\mathcal{A}/L_0$ and energy density $h=\mathcal{H}/L_0$. %, where $L_0$ is the length of the wave packet at the beginning of spreading, and $g=1$. %In  $a(t)=\langle {\cal A} \rangle /L(t)$ is the norm density and $h(t)=\langle  H \rangle /L(t)$ is the energy density of the wave packet, where $L(t)$ is the width of the wave packet at time $t$. 
%For convenience we introduce the rescaled norm density $x\equiv g a$ and rescaled energy density $y\equiv \nu h$, where $a=\langle {\cal A} \rangle /N$ is the norm density and $h=\langle  H \rangle /N$ is the energy density. 
In the course of spreading, both densities $a(t)= {\mathcal A}  /L(t)$ and $h(t)= \mathcal{H}  /L(t)$ will approach zero keeping their ratio $h/a$ fixed, since the length of wave packet $L(t)$ increases in time, i.e.,  $L(t\rightarrow\infty)\rightarrow\infty$, while the total norm and energy stay constant. 
Thus we can define the energy density during the period of wave evolution as $h(t)=\frac{\mathcal{H}}{\mathcal{A}}a(t)=\frac{h}{a}a(t)$. %, where $h/a$ is kept constant during the evolution. 
Hereby, we assert that all the statistical wave dynamics results should depend on the \textit{direction} of spreading that connects from $(a,h)$ to the origin $(0,0)$. While both densities follow the chosen \textit{direction}, the wave packet experiences the following cases during its spreading, depending on the selected initial energy density: 

%Thus, during spreading on the chosen direction, the wave experiences the following results, for the selected initial energy density: 

\begin{enumerate}[label=(\roman*)]
  \setlength\itemsep{0em}
    \item $h>0 \Rightarrow $ entering the non-Gibbs regime,
    \item $h=0\Rightarrow $ heating up to $\beta\rightarrow 0$,
    \item $h<0\Rightarrow $ cooling down to $\beta\rightarrow\infty$.
%    \item $h<-2a+a^2/2\Rightarrow $ forming Lifshits states, %dominant spreading.
\end{enumerate}
Case (i) may cause the self-trapping of waves, examined in Sec. \ref{sec:selftrapping}, whereas case (iii) may result in the formation of Lifshits states, which is analyzed in Sec. \ref{sec:lifshits}. Nevertheless, case (ii) exhibits a fast and uniform wave packet spreading with weak or strong chaos, presented in Sec. \ref{sec:weakandstrongchaos}.

This scheme shows the necessity to perform a density resolved wave packet spreading study. The relation between temperature and  $h/a$ are discussed in detail in Appendix \ref{ap:betamu_a_h} for the weak disorder limit.

    \begin{figure}[hbt!] 
    \centering
    \includegraphics[width=0.75\columnwidth]{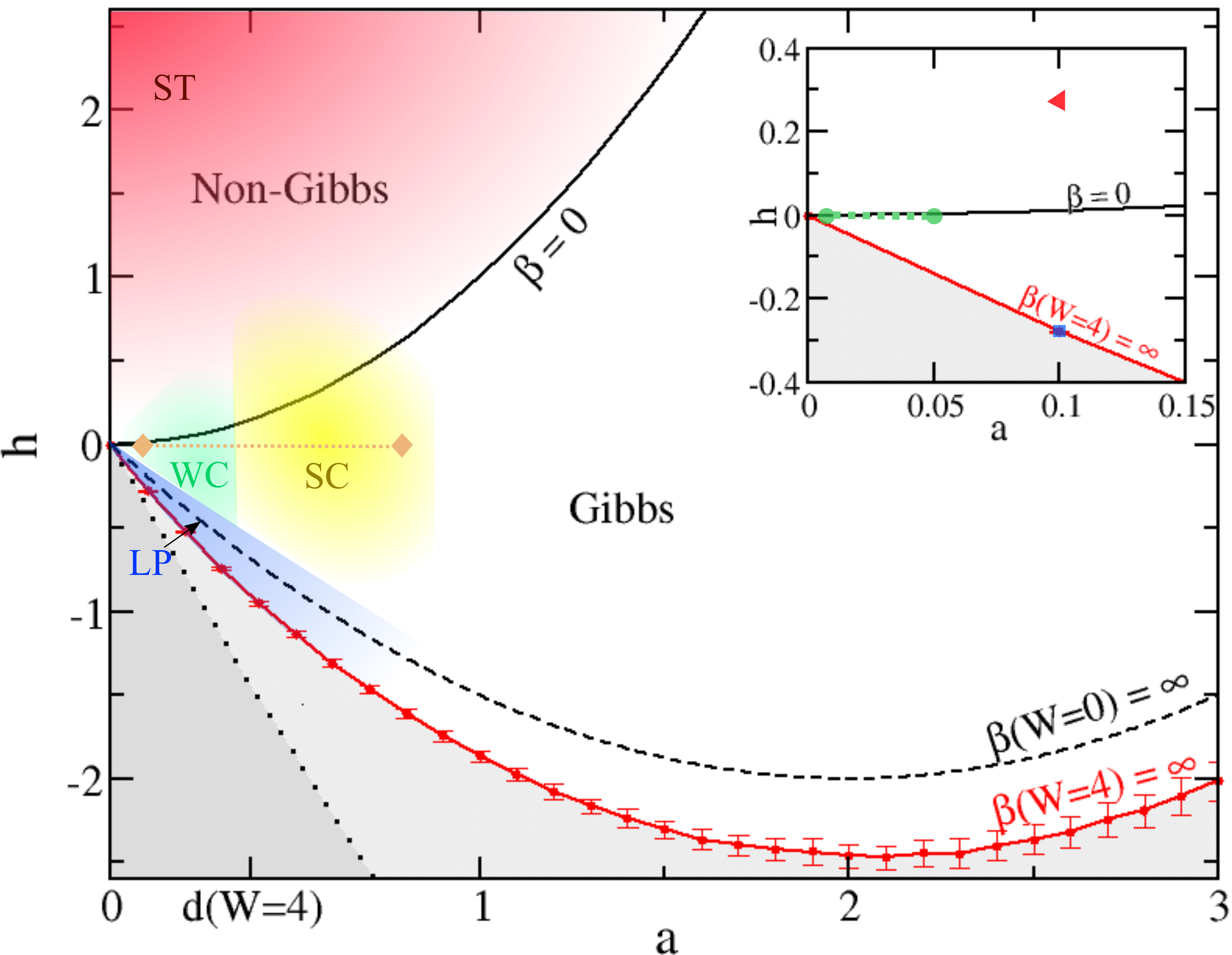}
    \caption[Phase diagram of the disordered microcanonical GP system]{Phase diagram of the disordered microcanonical GP system. Red connected circles - the renormalized ground state for $W=4, N=10^3$ averaged over $N_r=100$ disorder realizations. The black solid line $h=a^2$ corresponds to infinite temperature $\beta=0$ for any strength of disorder. Black dashed line - ground state $h=-2a+a^2/2$ for $W=0$.  
The four shaded areas correspond to SC, WC, ST, and LP.
The tick label $d(W=4)$ marks the position where the norm density equals the average level spacing $d$. The dotted black line represents the absolute minimum energy line
$h=-(2+W/2)a + a^2/2$ reachable for finite systems.} 
 	\label{f:phased}
    \end{figure}

%\subsection{Wave spreading in DNLS}
\section{Initial state and computational details}\label{sec:initialstatewave}

We introduce $\psi_{\ell}=\sqrt{a_{\ell}} \exp(i \phi_{\ell})$ to the Hamiltonian  Eq.~(\ref{eq:Hamiltoniandisorder}) of disordered GP lattice, explained in Sec. \ref{sec:disorderedGPHamiltonian}, we obtain % Eq.~\ref{eq:Hamiltoniandisorder} becomes
\begin{equation}
 H = \sum_{\ell}\left[-2\sqrt{a_{\ell}a_{\ell+1}}\cos(\phi_\ell-\phi_{\ell+1})+\epsilon_{\ell}a_\ell+ \frac{g}{2}a_\ell^2 \right].   
\end{equation}
%We choose the initial state with a homogeneous norm distribution $\psi_l=\sqrt{a}e^{i\phi_l}$, and fix the phase differences 
We consider the initial wave packet of the norm and energy densities $\{a,h\}$ as we set $g=1$ and $\mathcal{A}=L_0$ for convenience. The only exceptional case is the Anderson localization (AL) in Fig. \ref{f:BGSTALNM_R}(a) where we used $g=0$. The size of the initial wave packet $L_0$ is chosen to be the localization volume of the band center state of the corresponding linear system, $V \approx 3\xi(0)$, where $\xi$ is the localization length, found as $\xi(\lambda_\nu) \leq \xi(0) \approx 96/W^2$, for weak disorder $W \leq 4$ and $g=0$ \cite{spreading}. We numerically found the participation ratio of normalized AL eigenstates as $P_\nu=\sum_\ell |\psi_\ell|^4\approx1.5\xi_\nu$, whose random amplitude fluctuations result in another factor of 2 for the size of the state \cite{Kati2020}. The average spacing of eigenvalues of normal modes within the range of a localization volume is then $d\approx \Delta/V$, where $\Delta=W+4$ is the spectrum width \cite{Bodyfelt:2011}. We prepare a wave packet on $L_0\approx V$ consecutive sites in the center of a disordered lattice (see Table \ref{table:initiallength})

\begin{table}[h!]
\centering
\caption{Length of initial wave packets $L_0$ for chosen $W$}
\begin{tabular}{ |p{1cm}||p{1cm}|p{1cm}|p{1cm}|p{1cm}|p{1cm}|p{1cm}|  }
 \hline
 $W$ & 1 & 2 & 3 & 4 & 6 & 8 \\ 
 \hline
$L_0$ & 361 & 91 & 37 & 21 & 10 & 6   \\
  \hline
 \end{tabular}
  \label{table:initiallength}
 \end{table}

The energy density of the initial wave packet can be written as 
\begin{equation} 
h=- \frac{2 a}{L_0} \sum_\ell \cos(\Delta\phi_\ell) + \frac{g}{2} a^2 + \frac{g a}{L_0}\sum_\ell \epsilon_\ell,
\end{equation}
where $\Delta \phi_\ell=\phi_\ell-\phi_{\ell+1}$. 
For each disorder realization of on-site potentials {$\{\epsilon_\ell\}$}, we choose properly random phases $\{\Delta\phi_{\ell}\}$ such that $h$ takes a given value upon a controllable small fluctuation. We use  open boundary conditions for the initial wave packet. All simulations begin with a wave packet defined in the middle of the lattice and end before the wave packet spreads to its edges.

%One has to use open boundary conditions for the initial wave packet. 

The initial states defined in the range $|h|\geq a^2/2 +2a$ are accessible only for the wave packets with an inhomogeneous distribution of norm density. We find the local norm density $\psi_\ell$ for a given $h$ value by fixing $a$ and the disorder realization. We optimize the total energy of a wave packet with size $L$ to the desired value with the Nelder-Mead simplex algorithm, described in Section \ref{sec:GroundStateRenormalization} in detail. On the other hand, an initial state defined in the range $a^2/2-2a\leq h\leq a^2/2 +2a$ can be realized with a homogeneous norm distribution. Hence, we chose $\psi_\ell=\sqrt{a}e^{i\phi_\ell}$, and fix the phase differences $\Delta \phi= \arccos{\left(\frac{h}{2a}-\frac{a}{4}\right)}$ for $L_0-1$ sites. We then adjust the phase on the latest site to tune the total energy such that $\mathcal{H}=L_0 h$. We omit the disorder realizations for which the adjustment can not be realized. %If homogeneous norm density is used as suggested above, one has to determine precisely the phase difference of the latest site to control the energy fluctuation. 
Disregarding this boundary effect caused serious energy differences in different realizations, in previous studies.

We integrated   Eq. (\ref{eq:eomdisorder}) with a symplectic method $\text{SBAB}_2$ (Appendix \ref{sec:symp}), and performed a vast of numerical simulations, conserving total norm and total energy. We characterized the wave packet spreading both in real space and normal mode space, and detect no distinction in the feature of dynamics in both spaces. As we present our method of computation in real space, one can transform their results into a normal mode space, by referring to \cite{Flach102}. 
The wave spreading is characterized by two main ingredients: the second moment $m_2\equiv \sum_\ell (\ell-\bar{\ell})^2 |\psi_\ell (t)|^2/{\mathcal{A}^2}$
to measure the width of the wave packet which contains the contribution from the edges of the wave packet, and the participation ratio $P\equiv {\mathcal{A}^2}/{\sum_\ell |\psi_\ell|^4 }$ to measure the volume of the wave packet, which contains bulk information of the wave packet. 
Here
$\bar{\ell}={\sum_\ell \ell |\psi_\ell (t)|^2}/{\mathcal{A}^2}$ is the center of the wave packet. 
The second moment $m_2$ of a spreading wave is analogous to the speed of spreading. At large times, it increases exponentially as $t^\alpha$ with  $\alpha\leq 2$, while $\alpha=2$ is the expected ballistic evolution \cite{matterwave}.

If $P \sim \sqrt{m_2}$ as $t \rightarrow \infty$, it means that the wave packet spreads well,  while if $P \sim O(1)$ at all times, then it means that the wave packet exhibits localization due to self-trapping or Lifshits.   
It is also convenient to use the compactness index $C\equiv P^2/m_2$ to measure the spreading of the wave packet and one can easily see that for a spreading wave packet $C \rightarrow O(1) $ as $t\rightarrow \infty$. On the other hand, if the wave is trapped due to high energy or condensed in several sites due to cold temperature and low density, $C \rightarrow 0$ as $t\rightarrow \infty$. 

All data, shown in the following figures of $m_2$, $P$, and $C$ are averaged over $N_r$ number of disorder realizations. For the $\alpha$ plots, we smoothed the averaged data of $m_2$ with the Hodrick-Prescott (HP) filter and calculated its finite-difference derivatives. The HP filter is used minimally so that all of our outputs had a standard deviation of less than 2\%. % The details of it can be found in \cite{Kati2020}.

%\begin{itemize}
%\item We analyze the normalized distributions using second moment which quantifies the wave packets degree of spreading and the participation number which measures the number of the strongly excited sites: 
%\begin{equation*}
%m_2 =\sum_{l=1}^N \frac{(l-\overline{l})^2 |\psi_l(t)|^2}{A}, \quad \quad 
%\overline{l}=\sum_{l=1}^N \frac{l|\psi_l(t)|^2}{A}, \quad \quad
%P\equiv \left(\frac{\sum_l |\psi_l|^4} {{A}^2} \right)^{-1}.
%\end{equation*}
%\item $P$ and $m_2$ can be used to quantify the sparseness of a wave packet through the compactness index:
%$C \equiv P^2/m_2 \rightarrow 0 \text{ as } t\rightarrow \infty.$
%\item Theoretical expectations for the subdiffusive growth of the average energy ($m_2 \approx t^\alpha$) are $\alpha\approx \frac{1}{2}$ in the strong chaos regime, and $\alpha\approx \frac{1}{3}$ in the asymptotic weak chaos regime.
%\item In the previous studies, the presence of two relevant densities $(x,y)$ have been ignored. Hence, in this study, we use properly defined initial states, and analyze the relation between energy and norm densities on the dynamics of the wave packet spreading.
%\end{itemize}

%PRL 102, 024101 (2009)

\section{Strong and weak chaos}\label{sec:weakandstrongchaos}

\begin{figure}[hbt!] 
    \centering
    \includegraphics[width=0.8\columnwidth]{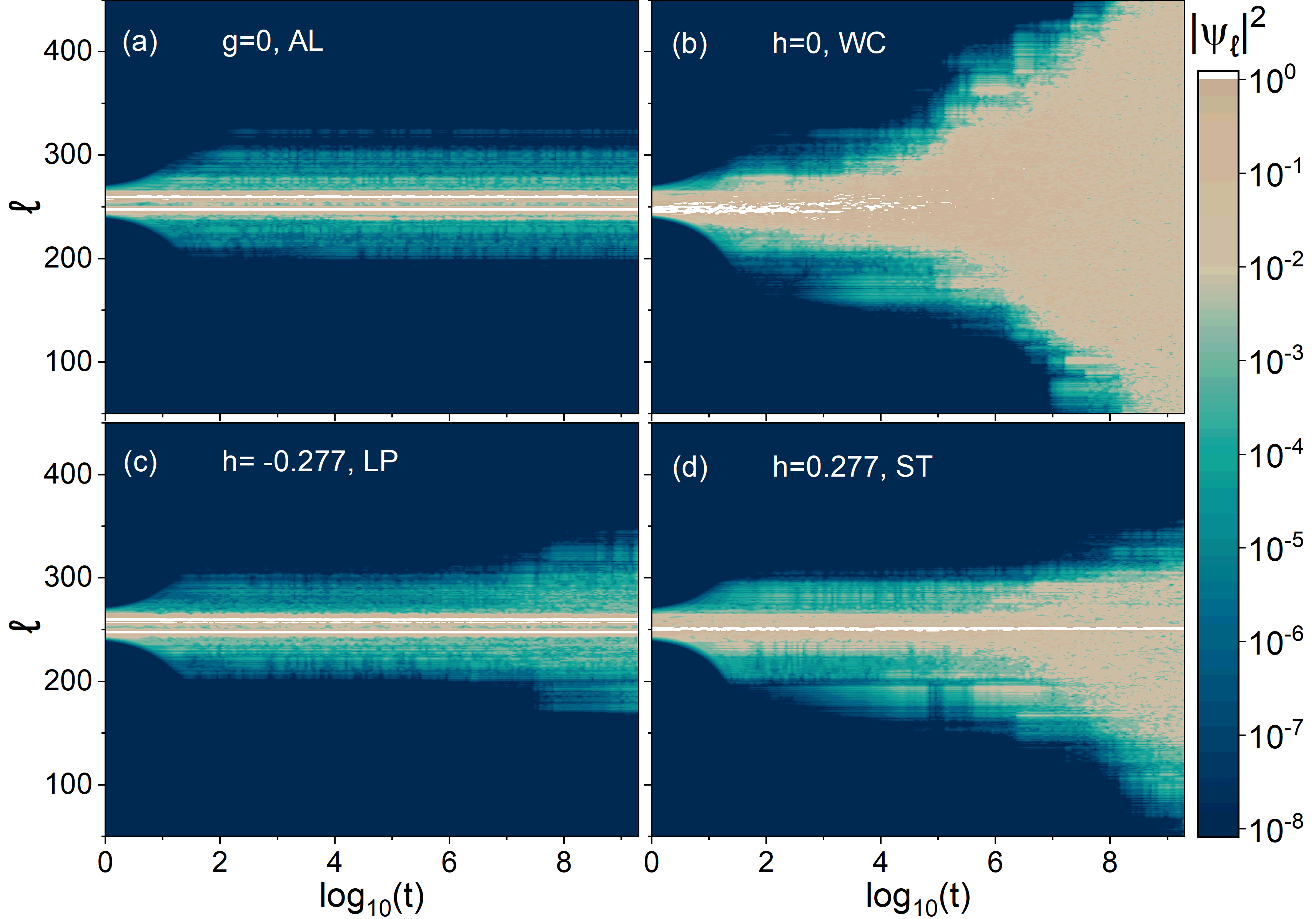}%pdf}
    \caption[Pictures of spreading waves in different regimes]{Different regimes of density resolved wave packet spreading.
The evolution of the norm density ${|\psi_\ell|}^2$ is plotted versus $\log_{10}t$.
(a) Anderson localization (AL): $g=0, h=-0.277$.
(b) weak chaos (WC): $g=1, h=0$.  
(c) Lifshits phase (LP): $g=1, h=-0.277$. 
(d) self-trapping (ST): $g=1, h=0.277$ (here $d t=0.05$).
For all cases  $a=0.1, W=4$, and one and the same disorder realization are used here for all regimes.}
    \label{f:BGSTALNM_R}
    \end{figure}

In the absence of nonlinear interactions $g=0$, a wave packet will evolve in time without appreciable spreading. We plot the evolution of
its norm density versus space and time in Fig. \ref{f:BGSTALNM_R}(a). After some short initial dynamics during which the field
established exponentially localized tails in space, the wave packet evolution essentially halts, signaling Anderson localization. 

When the nonlinearity is introduced, the Anderson localization is destroyed, and the wave packet starts to spread subdiffusively. Yet classical diffusion is not fully recovered \cite{2021arXiv210200210C}.  %The second moment $m_2$ of a spreading wave increases with time as $t^\alpha$, where $\alpha$ is found mathematically as $\sim 1/2$ for strong chaos, and $\sim 1/3$ for weak chaos \cite{Flach102}.
For a D-dimensional lattice with $\sigma$-body interactions, the complete dephasing of normal modes yields the second moment $m_2\sim t^\alpha$ of a spreading wave with $\alpha=2/(2+D\sigma)$, which is derived in \cite{Flach102}. This relation for strong chaos subdiffusive spreading yields $\alpha=1/2$ for 1D GP lattice with two-body interactions. The strong chaos regime occurs when disorder is weak enough s.t. all the normal modes of the linear problem have a large localization length, % the localization length of normal modes of the linear problem is large. 
so the nonlinearity can couple each normal mode to any other mode. All normal modes are expected to dephase completely if the frequency shift $\delta \sim g a$ is larger than the average spacing of eigenvalues -$d$- which is determined by $W$.

\begin{figure}[htp]%hbt!] 
\centering
\includegraphics[width=0.8\columnwidth]{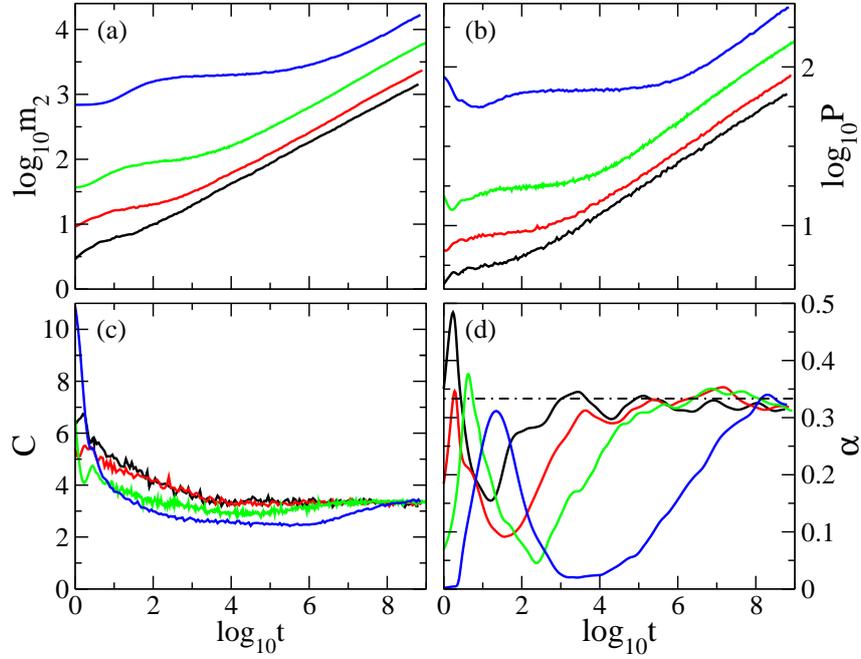}%WC.pdf}
\caption[Wave packet spreading in weak chaos regime]{Wave packet spreading in weak chaos regime in log time with $a<d$, and $h=0$. 
(a) Log of second moment. 
(b) Log of participation ratio.
(c) Compactness index.
(d) Derivative for the smoothed $m_2$ data. $a=0.0025$ and $W=2$ for blue, $a=0.05$ and $W=4$ for green, $a=0.245$ and $W=6$ for red, 
$a=0.9$ and $W=8$ for black line. The dashed line: $\alpha=1/3$. $N=2^{10}, N_r=200$.}
\label{f:wc}
\end{figure}

On the other hand, for a system with strong enough disorder, Flach et al. \cite{Flach102} predicted that a wave packet will evolve with the second moment $m_2\sim t^\alpha$ where $\alpha=2/(2+D(\sigma+2))$ assuming all normal modes are spatially localized.
This relation for weak chaos subdiffusive spreading yields $\alpha=1/3$ for 1D GP lattice with two-body interactions. 
%This behavior  with $\alpha=1/3$ for 1D GP lattice with two-body interactions is known as weak chaos wave spreading. 
For stronger disorder, the normal modes are more localized and the average spacing between their eigenvalues -$d$- is larger. The normal modes are not expected to dephase, and the interactions between them will be weak. More precisely, if the frequency shift, $\delta \propto g a$, is less than the average spacing of eigenvalues, modes  interact weakly, and the wave spreads with a weak chaos $m_2 \sim t^{1/3}$, observed for different $W$ in Fig. \ref{f:wc}(d).

\begin{figure}[htp]%hbt!] 
\centering
\includegraphics[width=0.8\columnwidth]{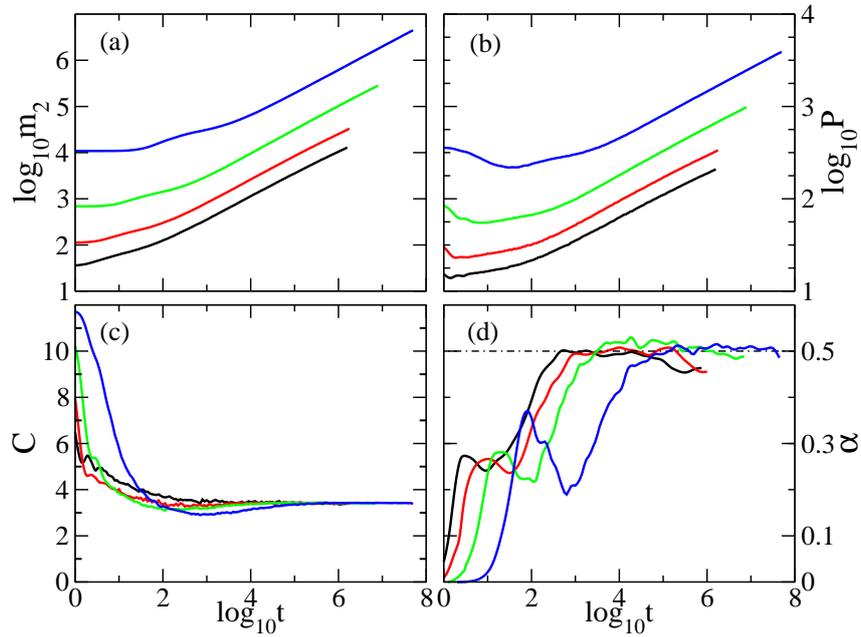}%SC.pdf}
\caption[Wave packet spreading in strong chaos regime]{Wave packet spreading in strong chaos regime in log time with $a>d$, and $h=0$. (a) Log of second moment. 
(b) Log of participation ratio.
(c) Compactness index.
(d) Derivative for the smoothed $m_2$ data. $a=0.047$ and $W=1$ for blue, $a=0.19$ and $W=2$ for green, $a=0.4$ and $W=3$ for red, 
$a=0.79$ and $W=4$ for black line. The dashed line: $\alpha=1/2$. $N=2^{13}$, $N_r=200$ for $W=1$, 500 for $W=2$, 750 for $W=3$, 1000 for $W=4$.}
\label{f:sc}
\end{figure}

Yet, if the frequency shift is greater than $d$, then all the modes will resonantly interact, and as a result, the sub-diffusion speeds up to $m_2 \sim t^{1/2}$. This strong chaos subdiffusive spreading is clearly observed for more than two decades in Fig. \ref{f:sc}(d). 
Strong chaos is a temporary regime since the relations $a(t)=\mathcal{A}/L(t)$ and $L(t\rightarrow\infty)\rightarrow\infty$ suggest that the frequency shift $g a(t)$ will eventually be less than $d$, i.e., the coupling of modes will get weaker, while $W$ is constant. Fig. \ref{f:sc}(d) clearly captures the decay of the exponent $\alpha$ to the values less than 1/2 signifying the wave packet crosses over from SC to WC. Nevertheless, it is important to note that
although the strong chaos regime is transient, it can be observed even for as long as the age of the universe if the disorder of the system is weak enough. We see that the time interval of SC spreading increases as we decrease the disorder strength in Fig. \ref{f:sc}(d).

%In the course of spreading, the frequency shift $g a$ reduces in time while $W$ is constant. Therefore, the strong chaos regime is transient, nevertheless it can be observed even for as long as the age of the universe if the disorder of the system is weak enough.

After long enough time passed, we presume $\alpha$ to reach and stay around $1/3$, noting that a similar expectation on the dynamical crossover from strong to weak chaos is done for Klein--Gordon chain in \cite{Laptyeva2010}. 
The regions of weak and strong chaos are shown in the densities phase diagram Fig. \ref{f:phased}.

\begin{figure}[hbt!] 
\centering
\includegraphics[width=0.8\columnwidth]{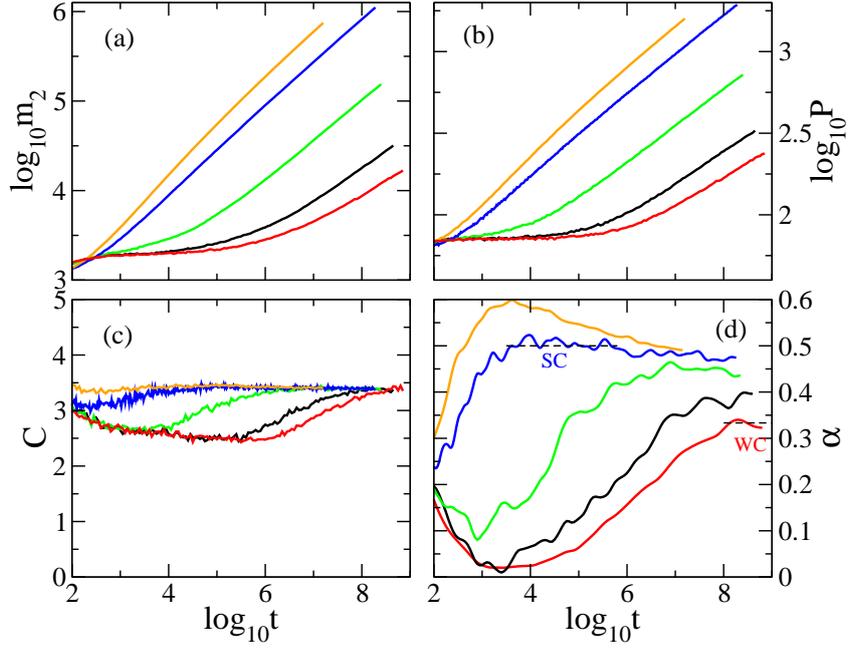}
\caption[Density resolved wave packet spreading at different norm densities]{Density resolved spreading at different norm densities, with fixed $h=0$, $W=2$ in log time.
(a) Log of second moment. 
(b) Log of participation ratio. 
(c) Compactness index.
(d) The derivative of smoothed $m_2$ data. As a reference, $\alpha=1/2$, and $\alpha=1/3$, are shown as black dashed lines.
Bottom to top: $a=0.0025,0.005,0.025,0.183,0.35$, shown respectively as red, black, green, blue, and orange lines. $N=2^{10}, N_r=200$.}
\label{f:crossover}
\end{figure}

%to overshooting dynamics
With increasing the nonlinear shift, we observe a transit from weak chaos to strong chaos regime dynamics (see Fig. \ref{f:crossover}). 
For a wave packet with an initial norm density much larger than $d$, the spreading has an initial acceleration due to the large particle density so that the wave abruptly spreads and the average norm density of the spreading wave does not satisfy $a>d$ for a long time. When the average norm density $a$ quickly reaches near to the value of $d$, the wave enters the crossover to the weak chaos regime, before observing strong chaos spreading for a long enough time. The examples of this situation are shown as a black dashed line in Fig. \ref{f:m2_x2y3_5_7}, and an orange line in Fig. \ref{f:crossover}. 
The crossover between SC and WC when $a \sim d$ is smooth which is shown in Fig. \ref{f:crossover}(d). Therefore, SC and WC are not separated with a solid line in the phase diagram \ref{f:phased}.

As stated above, strong chaos is confirmed ($m_2 \sim t^{1/2}$) with different disorder strengths for a long enough time ($>10^7$ for $W=1$). With weaker disorder strengths, it is possible to observe strong chaos for longer times as shown in Fig. \ref{f:sc}(d). However, we could not run simulations with $W<1$, since one has to study with a larger wave packet for lower $W$ due to its larger localization volume ($L$). Moreover, the time to reach a complete resonance of the wave takes a longer time, because it possesses a large size $L$. This means the exponent $\alpha$ reaches $1/2$ or $1/3$ more quickly for an initial wave packet with a smaller $L$, which implies higher disorder. Thus, we could observe $\alpha \sim 1/3$ for longer times, $10^3-10^9$, for $W=8$ as exhibited in Fig. \ref{f:wc}(d).

For better observation of strong and weak chaos, we had selected the initial energy density at $h=0$, since the spreading of a wave packet is faster %faster: the number of interacting modes increases in a shorter time, and the exponent alpha reaches higher values
in that range. One may see that the spreading is well in Fig. \ref{f:wc} and Fig. \ref{f:sc}; since the compactness index $C \rightarrow O(1)$.

In Fig. \ref{f:wc} and Fig. \ref{f:sc}, the first peak in the $\alpha$ plots disappear if the initial excited sites have a homogeneous energy density distribution. It is just a small interaction until the wave packet has homogeneity on energy.

\section{Self-trapping}\label{sec:selftrapping}

\begin{figure}[hbt!] 
\centering
\includegraphics[width=0.8\columnwidth]{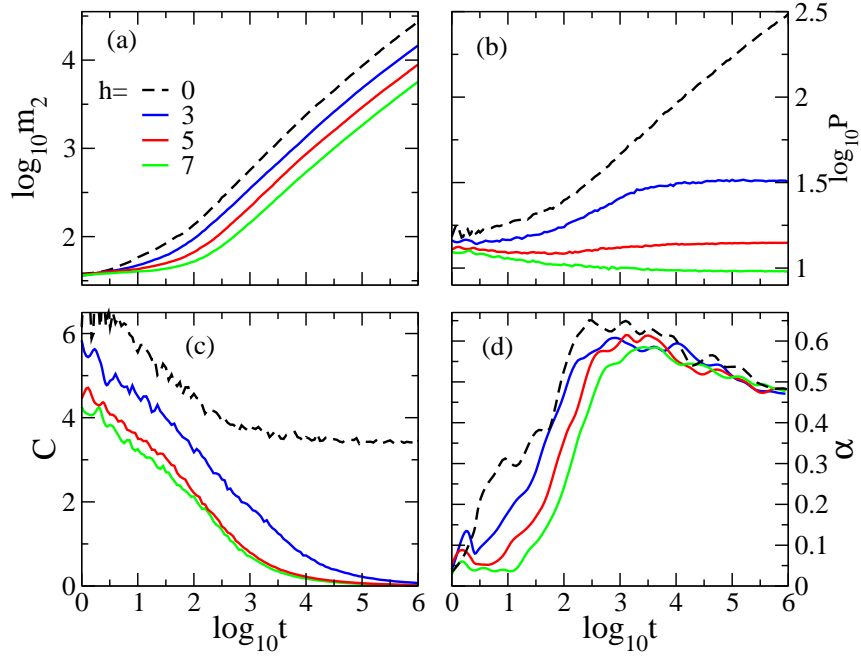}%x2y0_3_5_7.pdf}
\caption[Wave packet spreading at different energies, with fixed $a>d$]{Wave packet spreading at different energies, with fixed $a=2>d$ and $W=4$. Upper plots are log-log scaled $m_2(t)$ (left) and $P$ (right) vs. $t$. Lower plots are lin-log scaled $C$ (left) and $\alpha$ (right) vs $t$. $h=3,5,7$ are shown as blue, red. and green lines. $h=0$ is the black dashed line for the reference. $N=2^{13}$, $N_r=200$.} 
\label{f:m2_x2y3_5_7}%m2_x0763variousy
\end{figure}

    In the non-Gibbs regime, strong localization of states in real space may occur separately and hold the spreading. When a wave packet is self-trapped, the total number of strongly excited sites stays constant, and so the acceleration of the second moment slows down although the spreading from the tails of the wave packet continues. This means one must expect to see that $P \rightarrow O(1)$, while $m_2$ is still increasing with a power law, that signifies $C\rightarrow 0$ (refer to Fig. \ref{f:m2_x2y3_5_7}(c), and Fig. \ref{f:LPSTWCm2}(c)).

In Fig. \ref{f:m2_x2y3_5_7}, we showed the effect of high energy densities on spreading waves. Noting that the fastest spreading is at $h=0$, the signature of ST is observed on the participation ratio of wave packets in time. 
It either decreases or becomes constant while the spreading of the wave continues.

\section{Lifshits phase}\label{sec:lifshits}

\begin{figure}[hbt!] 
\centering
\includegraphics[width=0.8\columnwidth]{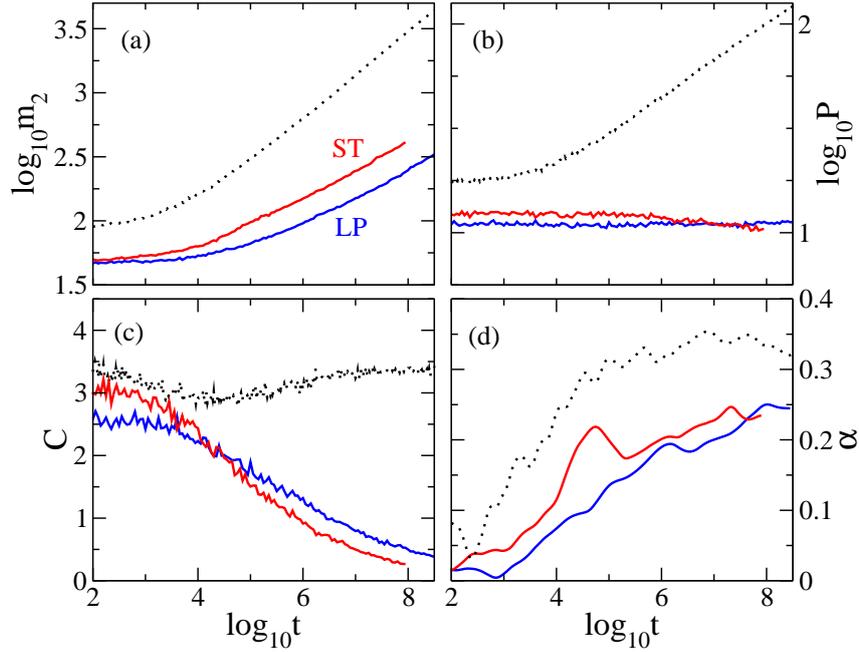}
\caption[Spreading waves in self-trapping and Lifshits]{ Wave packet spreading in self-trapping (ST) and Lifshits phase (LP) with $a=0.05$, $W=4$, $N=2^{10}, N_r=100$.  Upper plots are log-log scaled $m_2(t)$ (left) and $P$ (right) vs. $t$. Lower plots are lin-log scaled $C$ (left) and $\alpha$ (right) vs $t$. Lifshits phase: $h=-0.12$ (blue), and self-trapping $h=0.12$ (red). The weak chaos with $h=0$ (black dots) is a reference.} 
\label{f:LPSTWCm2}
\end{figure}

In the Lifshits regime, all the dynamics $P$, and $m_2$ slow down, which is similar to self-trapping. Nevertheless, the speed of spreading ($m_2$) is slower for Lifshits (see Fig. \ref{f:LPSTWCm2}, Fig. \ref{f:BGSTALNM_R}), although their acceleration in time is not much different. 
 
   We observed numerically that $C\rightarrow 0$ in the LP region indicated in Fig. \ref{f:phased}. For both regimes, in Fig. \ref{f:LPSTWCm2} we found the exponent $\alpha$ is smaller than the case for weak chaos, exhibiting that the evolution of wave is really slow. In Fig. \ref{f:BGSTALNM_R}, one can clearly see the difference between a normal spreading (weak chaos), and Lifshits. LP is much similar to Anderson localization. Although $m_2$ is still evolving due to the untrapped part of the wave, we assume that the trapped part will not delocalize for a very long time. Similar expectation stands for self-trapping due to its nonincreasing participation ratio.

\begin{figure}[hbt!] 
\centering
\includegraphics[width=0.8\columnwidth]{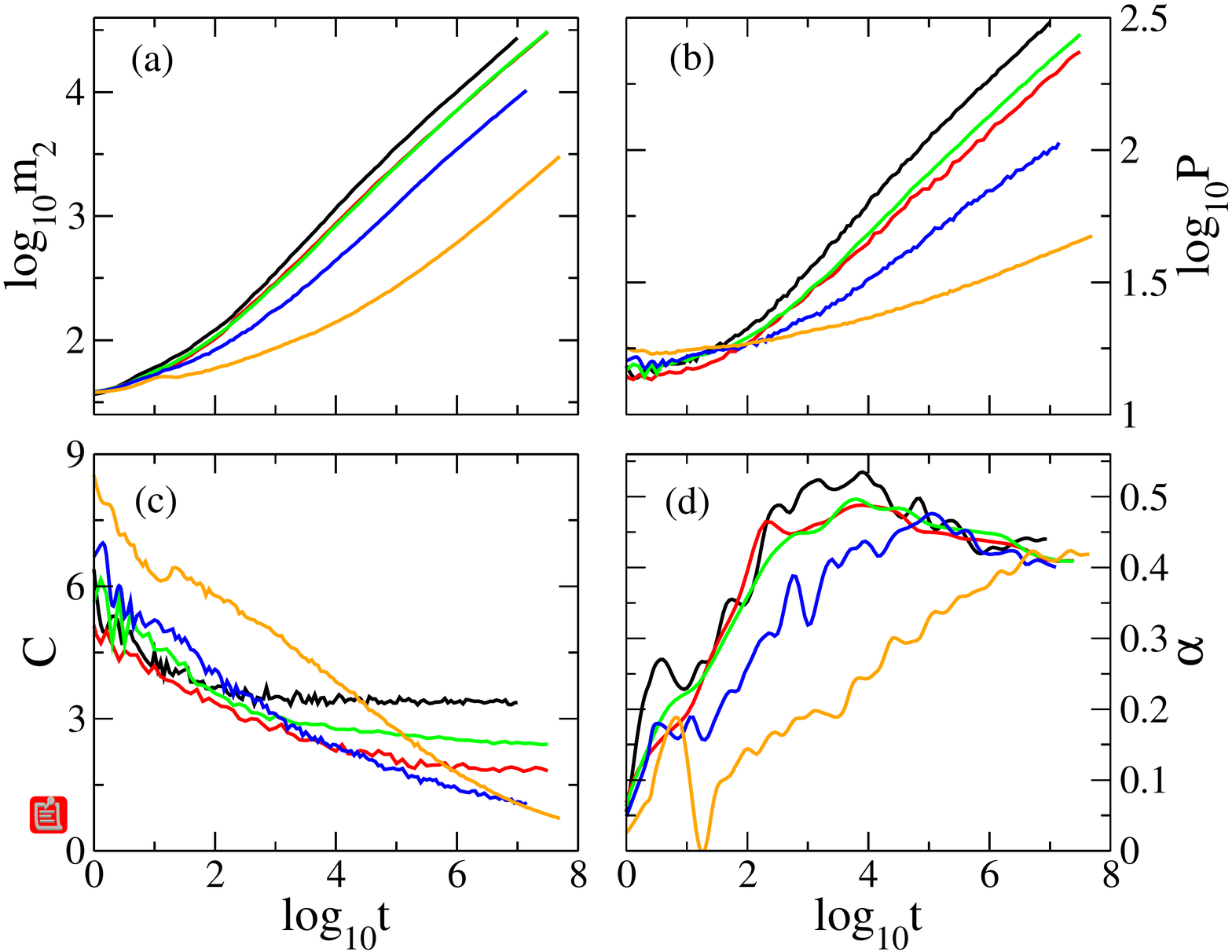}%m2_x0763variousy.pdf}
\caption[Wave packet spreading at different energies, with fixed $a>d$]{The effect of initial energy $h$ on wave packet spreading for fixed $a=0.763>d$, $W=4$ in log time. (a) Log of second moment. 
(b) Log of participation ratio. 
(c) Compactness index.
(d) The derivative of smoothed $m_2$ data.
$h=0,0.7,-0.7,-1,23,-1.85$ are shown as black, red, green, blue, and orange lines, which are averaged over $N_r=100,100,290,100,280$, respectively. $N=2^{12}$ for all.} 
\label{f:m2_x0763variousy}
\end{figure}

\section{Conclusion}

This chapter we found out the non-equilibrium dynamics of GP lattice in disordered media. %, considering all phases of density space. 
%\begin{itemize}
%\item We studied fully density resolved compact wavefront spreading in disordered DNLS.
%\item Strong chaos is observed for the disordered DNLS system for the first time.
%\item The wave spreads faster at $y=0$  than $|y|>0$. %The symmetric behavior around $y=0$ comes from.
%\item As also suggested in \cite{Laptyeva}, the boundaries between three different subdiffusive regimes (Weak chaos, strong chaos, self trapping) are not solid.
%\item  In the course of spreading the wave packets norm density will become smaller. Therefore, the effective coupling strength  between normal modes decreases as well. At the same time the number of excited normal modes grows.
%\item Above the self-trapping line, $y=2x$, there is no real spreading observed.
%\end{itemize}
We have analyzed fully density resolved compact wavefront spreading under a uniform disorder, realized in a Gross--Pitaevskii lattice. We observed strong chaos clearly for the first time and its crossover to weak chaos in the Gibbs regime. 
Further, we detected the fastest spreading at zero energy due to the continuous increase in temperature: $a(h=0)\rightarrow0, \beta\rightarrow0$. Keeping the norm constant, increasing or decreasing the energy density of the initial wave packet both reduces its speed of spreading gradually until it enters the ST or LP (see Fig. \ref{f:m2_x0763variousy}). There is no solid crossover from ST dynamics to SC to LP. 
The wave packet spreading in both the Lifshits phase and the self-trapping regime is characterized by a substantial slowing down from the subdiffusive spreading. At the same time, the self-trapping enforces highly localized almost single site excitations to be formed, at variance to the Lifshits phase dynamics where the wave packet structure resembles the Anderson localization case. 
On the other hand, LP, also known as the Lifshits glass regime for the Bose--Einstein condensate, exists near the ground state $h<a^2/2-2a$ when the norm density is small, i.e., $a<d$. In ST, the wave separates into two main parts: one is localized, and the rest is spreading in time such that its participation ratio of modes does not increase. Differently, the wave spreads quite slowly from its AL state in the LP without any separation, and the growth of its participation ratio of modes is suppressed but not held constant.

Positive energy densities are doomed to bring the wave packet into a non-Gibbs regime with potential fragmentation of the packets into a self-trapped condensate part and an infinite temperature background capable of spreading infinitely. One of the intriguing quantities is the ratio of the norm in the two field components and its asymptotic time dependence. Will the self-trapped component take over the entire wave packet norm at large enough times, or will some finite remain in the infinite temperature background? Zero energy densities keep the wave packet in the Gibbs regime and may lead to the entire packet heating up to infinite temperatures upon infinite spreading. Negative energy densities in the Gibbs regime can bring the wave packet closer to the ground state, and therefore zero temperatures upon spreading. Finally, initializing the wave packet in the Lifshits regime shows strong suppression of subdiffusive spreading. But even in this case, we notice a speedup of the spreading process with some potential fragmentation of the wave packet. It appears that there are a number of interesting and hard open problems to be addressed. In the future, it will be also interesting to study the possibility of LP and ST regimes in other nonlinear disordered systems and to understand their statistical behaviors. Our density resolved approach allows one to investigate the spreading of waves with multiple interacting excitations or to study localization in nonlinear disordered lattices.

%this part has to be integrated
%The wave packet spreading in both the Lifshits phase and the self-trapping regime is characterized by a substantial slowing down from the subdiffusive spreading. At the same time, the self-trapping enforces highly localized almost single site excitations to be formed, at variance to the Lifshits phase dynamics where the wave packet structure resembles the Anderson localization case.
 %this part has to be integrated ends

% as observed for weak and strong chaos

\cedp
%\include{Chapter6/chapter6}
%\cedp
%\include{Chapter7/chapter7}
%\cedp
%\include{Chapter8/chapter8}
%\cedp
\chapter{Final remarks}\label{chapter6}

Nonlinear Hamiltonian lattice systems are the fundamental class of models in mathematics, statistical and condensed matter physics. They have a wide range of applications from quantum computation to ultracold atomic gases forming Bose--Einstein condensates, to superconducting grains of Josephson junction networks, among others. Gross--Pitaevskii is one of the most successful nonlinear Hamiltonian models due to its availability to study integrability, chaos, and disorder in a fair manner with two integrals of motion. %%This model is applied to a wide range of physical settings from Fermi gases to Bose--Einstein condensates. %THIS WAS SAID TWO SENTENCE AGO
Its studies have a strong impact on the other nonintegrable Hamiltonian systems and have a connection with the subject of many-body localization in quantum systems.
%essential
There are many unsettling  aspects of the equilibrium and non-equilibrium dynamics of nonlinear systems. Particularly, the interaction and localization matters in the presence of disorder are remained to be explored. Our work  presented in this thesis just put a piece on the table of this infinitely large puzzle, and has a mission to help find its neighboring future segments.

In this thesis, we have studied the statistical mechanics of nonlinear Gross--Pitaevskii lattice with and without disorder, concerning both equilibrium and non-equilibrium dynamics. The prominent attributes of our work can be emphasized in the following points.

In Chapter \ref{chapter3}, we have initially concentrated on the equilibrium dynamics of interacting nonlinear ordered GP lattice and then investigate the out-of-equilibrium dynamics while approaching the integrable limit. We answered a former question on whether the transition from ergodic to non-ergodic regime occurs on the Gibbs infinite temperature line. Noting that we found the nonergodicity only exists in the non-Gibbs regime, we also discovered an ergodic subregion in the density space undefined by Gibbs thermodynamics. This, as a result, highlighted the idea that the Gibbs definition of partition function needs to be modified since a well-defined distribution function is needed for an ergodic regime.  Moreover, we successfully observed chaotic discrete breathers in thermal equilibrium and related their density and lifetime to the breaking of ergodicity. For future work, our method suggests a way to reveal and analyze the phenomena of ergodicity breaking in many-body interacting systems with quantitative and systematic assessment of non-ergodic regimes in their respective parameter space. We expect our method might be also useful to assess several topics in condensed matter physics such as anomalous heat conduction, interacting disordered systems, and spin glasses. In the future, weakly non-ergodic dynamics can be discovered close to the integrable limits of a broad class of nonlinear systems  from superconducting Josephson junction networks to disordered Bose--Einstein condensates in higher dimensions.

In Chapter \ref{chapter4}, we found out the equilibrium dynamics and statistics of our GP lattice in the presence of uncorrelated disorder. This fundamental problem of disordered GP lattice was not resolved before, yet there had been several experimental and theoretical studies on Bose gas in random potential. Expectedly, we found an analogy between most of our results with the ones in the literature for Bose gas in a random potential. For example, at the ground state, we measured the chemical potential and participation ratio both of which have produced similar results with the Bose gas, considering the weak and strong interaction regimes. We established new analytical results to confirm our numerics as well. In the second part of this study, we perturbed the ground state infinitesimally and derived its BdG equations. We calculated the extend of BdG modes on its spectrum of different energies via numerical and analytical techniques. The exact BdG modes showed an anomalous unexpected extension of their sizes at finite energy, which we later confirmed by TMM results using an approximation to the GS. Moreover, we analytically found that the side peak originates from a singularity at the wavenumber $q\sim \pi/2$. %starts from editor letter
Finally, we derived effective strong interaction field equations for the excitations and generalized them to higher dimensions.  This can constitute a substantial advance in the study of localization phenomena of interacting quantum many-body systems. It also advances the understanding and control of transport properties of excitations of disordered BEC in interacting quantum systems and explores the transport physics of these excitations in the lattice regime of finite momenta. %ends
The localization length of the modes has been used to separate the Bose-glass and superfluid regimes of Bose gas and is crucial to determine if the system is superfluid or fragmented. Using a lattice system brings several advantages. %starts from editor letter
Our study highlights  an  innovative  computational procedure to observe the resonant delocalization of quantum excitations of a disordered BEC. %ends
Hence we presume our results will contribute to the general understanding of superconductivity and superfluidity. 
It also represents a significant step for the bridge between classical and quantum systems, as it is feasible to map our results from the semiclassical GP lattice to the ones for ultracold Bose gas.
%Our  work  highlights  an  innovative  computational procedure to observe the resonant delocalization of quantum excitations of a disordered BEC

%All of the previous studies on BEC focus on the long-wavelength or small momentum regime, where the underlying lattice is less relevant.  Our work not only explains and vastly generalizes the observations reported in these papers but also and most importantly explores the transport physics of these excitations in the lattice regime of finite momenta. 

After we found the equilibrium dynamics of disordered GP lattice  in Chapter \ref{chapter4}, we addressed its non-equilibrium dynamics in Chapter \ref{chapter5}. %the nonequilibrium dynamics of disordered GP lattice.
We examined how controlling more than one integral of motion of a spreading wave help to discover its subdiffusive regimes. This was the first density-resolved wave spreading study, which allowed us to observe the transient strong chaos regime for a long enough period to confirm its theoretically expected value. Moreover, we found the suppression of spreading dynamics in the disorder-induced new phase of the densities space, called Lifshits. The self-trapping region was observed this time quite cleanly: along with a partial slowing down of spreading, a nonincreasing  participation ratio is found which must be the new signature of the entrance into a self-trapping region. Overall, this study answered and clarified several remained open problems in the wave spreading study of GP lattice. However, there are still unsolved questions for future studies, e.g., how to qualitatively and quantitatively distinguish the localized and delocalized part in the high energy -self-trapping- regime. Will the spreading keep its slowness in the Lifshits and self-trapped cases, or speed up, or even stop later times ($t>10^9$)? 
We showed that the theoretical scheme that has been used previously to distinguish the spreading regimes of strong \& weak chaos, and self-trapping is not valid. Introducing the Lifshits phase, our study brings out the need for a new mathematical relation to separate these four different subdiffusive regimes, assessing not only their norm but also their energy densities. It is still an open problem how to establish such an extensive formula. While we clarified several aspects from old studies on wave spreading, our study brought a lot of intriguing and difficult questions to answer in this field as well. %  These questions  % Can we build a formula which now includes to separate these different subdiffusive regimes more clearly?

Our work, which is summarized in Chapter \ref{chapter3}, \ref{chapter4}, \ref{chapter5}, has given some insight into the physics of the nonlinear Hamiltonian dynamics through the interplay of disorder and nonlinearity. Yet, it has also raised several challenging theoretical questions for future studies. %mentioned above.  %For the future, it would be also interesting to study the corresponding quantum many body systems

%For the future, it would be interesting to study the corresponding quantum many body systems and the way our result from the classical wave equations will transform into the quantum world.

%next line adds the Bibliography to the contents page
\phantomsection
\addcontentsline{toc}{chapter}{References}
%uncomment next line to change bibliography name to references
\renewcommand{\bibname}{References}
\bibliography{references}        %use a bibtex bibliography file refs.bib
\bibliographystyle{unsrt}  %use the plain bibliography style

\clearpage

%now enable appendix numbering format and include any appendices
\appendix
%\chapter*{Appendix}
\phantomsection
\addappheadtotoc
\appendixpage

%\chapter{Supplementary materials} 

\chapter{Symplectic integrators}\label{sec:symp}

Since we have a nonlinear differential equation, we used split-step integration via symplectic integrators to solve it. First, we will split the Hamiltonian to two solvable parts A and B as the following:

\begin{equation}\label{eq:splitHamiltonian}
    \mathcal{H}=A + \epsilon B,
\end{equation}
where $A=-\sum_{\ell=1}^N \left ( \psi_{\ell+1} \psi_{\ell}^* + \psi_{\ell+1} \psi_{\ell} \right )$, and $B=\sum_{\ell=1}^N \epsilon_{\ell} |\psi_{\ell}|^2 + \frac{g}{2} |\psi_{\ell}|^4$ with $N$ is the total number of the lattice sites, and $\epsilon$ is a parameter which equals to 1 in our case. We used the $\text{SBAB}_2$ symplectic method in order to integrate the GP lattice over long periods keeping total energy and total norm conserved.

%change now!!

%Now, we will explain the $\text{SBAB}_2$ method whose definition can also be found at \cite{Skokos:2009}.

Our Hamiltonian system can be defined as $\mathcal{H}(p,u)$ with $p=(p_1,...,p_N)$, and $u=(u_1,...,u_N)$ where $u_\ell$ and $p_\ell$ with $\ell=1,...,N$ are generalized coordinates and momenta, respectively. They can be defined by a vector $x(t)=x_1(t),...,x_{2N}(t)$ with $x_\ell=p_\ell$ and $x_{\ell+N}=u_\ell$. Thus, the equations of motion of the Hamiltonian can be represented as 

\begin{equation}
    \frac{d p_\ell}{dt}=-\frac{\partial \mathcal{H}}{\partial u_\ell}, \qquad \frac{d u_\ell}{dt}=\frac{\partial \mathcal{H}}{\partial p_\ell},
\end{equation}
where $\ell=1,...,N$, and $t$ is the time. Then, we can define the Poisson bracket of two functions $f(\vec{p},\vec{u})$, and $g(\vec{p},\vec{u})$ as

\begin{equation}
    \{f,g\}=\sum_{\ell=1}^N \left(\frac{\partial f}{\partial p_\ell} \frac{\partial g}{\partial u_\ell} - \frac{\partial f}{\partial u_\ell} \frac{\partial g}{\partial p_\ell}\right).
\end{equation}
Thus, the equations of motion of Hamiltonian can be written compactly as
\begin{equation}\label{eq:18}
    \frac{d \vec{x}}{dt} = \{ \mathcal{H},\vec{x} \} = L_\mathcal{H} \vec{x},
\end{equation}
where $L_\mathcal{H}=L_A +L_{\epsilon B}$ is the differential operator. The solution of Eq. (\ref{eq:18}) is
\begin{equation}
    \vec{x}(t)=\sum_{\ell \geq 0}\frac{t^n}{n!} {L_\mathcal{H}}^n \vec{x_0} = e^{t L_H} \vec{x_0}.
\end{equation}
We integrate Eq. (\ref{eq:18}) from time $t$ to $t+\tau$ where $\tau$ is the time step of the symplectic integration. The symplectic scheme consists of approximating the operator $\exp(\tau L_\mathcal{H})$ by an integrator of j steps involves products of $\exp(L_i \tau L_A)$ and $\exp(d_i \tau L_{\epsilon B})$ with $i=1,2,...,j$, that are the integrations over times $c_i \tau$ and $d_i \tau$. So, the $\text{SBAB}_2$ integrator is
\begin{equation}
    {\text{SBAB}}_2=e^{d_1 \tau L_{\epsilon B}} e^{c_2 \tau L_A} e^{d_2 \tau L_{\epsilon B}} e^{c_2 \tau L_A} e^{d_1 \tau L_{\epsilon B}},
\end{equation}
where $c_2 = \frac{1}{2}$, $d_1=\frac{1}{6}$, and $d_2=\frac{2}{3}$.  With $\text{SBAB}_2$, we are able to solve the Hamiltonian with an error of the order $\tau^4 \epsilon +\tau^2 \epsilon^2$ \cite{Skokos:2009}.

To integrate the part A of the Hamiltonian in Eq. (\ref{eq:splitHamiltonian}), we apply the action operator $\exp(\tau L_A)$ at time $t$ on the wave function $\psi_{\ell}$ where $\ell$ stands for the site number. So $\psi_\ell{'}$ at $t+\tau$ can be computed in three steps: 

\begin{enumerate}[label=(\roman*)]
  \setlength\itemsep{0em}

    \item The wave function is transformed from real space to Fourier space. Here, we use a fast Fourier transform method (FFT) which is also checked against the slow version of Fourier transform.
    \begin{equation}\label{eq:partAitem1}
        \phi_q = \sum_{\ell=1}^N \psi_\ell e^{2 \pi i (q-1) l /N},
    \end{equation}
where $\phi_q$ is the wave function defined in Fourier space.

    \item We do a rotation of $\phi_q$ : 
        \begin{equation}\label{eq:partAitem2}
    \phi_q {'} = \phi_q e^{2i \cos(2 \pi (q-1)/N) \tau }.
    \end{equation}
    \item Finally, we take the inverse Fourier transform of $\phi_q{'}$ : 
    
    \begin{equation}\label{eq:partAitem3}
        \psi_{\ell}{'} =\frac{1}{N} \sum_{q=1}^N \phi_q{'} e^{2 \pi i m (q-1)/N}.
    \end{equation}
\end{enumerate}
The integration of part $B$ of the Hamiltonian has only one step which is a simple rotation:
\begin{equation}\label{eq:partB}
    \psi_{\ell}{'} = \psi_{\ell} e^{-i (\epsilon_{\ell} + g |\psi_{\ell}|^2)\tau}.
\end{equation}

\chapter{The relation of $(\beta, \mu)$ and $(a,h)$ for linear lattice}%The relation of $(\beta, \mu)$ and $(a,h)$ for $g=0$}
\label{ap:betamu_a_h}

%Some of the results here are also shown in \cite{Rumpf2004}, and \cite{Rasmussen:2000}.

The Hamiltonian, and total norm of the linear GP lattice are defined as

\begin{equation}
    \mathcal{H}=\sum_{\ell} - (\psi_{\ell}^*\psi_{\ell+1}+\psi_{\ell}\psi_{\ell+1}^*), \quad\quad \mathcal{A}=\sum_\ell \psi_\ell {\psi_\ell}^*
\end{equation}
%\begin{equation}
%    i \dot{\psi_n} =\psi_{n+1} + \psi_{n-1} +|\psi_n|^2 \psi_n
%\end{equation}
with the energies $-2\mathcal{A} \leq \mathcal{H} \leq 2\mathcal{A}$. The low amplitude initial conditions $\mathcal{A}/N \ll 1$ can also approximate the results shown in this appendix. The partition function is
\begin{equation}
    \mathcal{Z}=\int \dots \int \rho \prod_{\ell}^N d\psi_{\ell} d\psi_{\ell}^*,\quad \text{where}\quad \rho=\exp[{-\beta (\mathcal{H}+\mu \mathcal{A})}].
\end{equation}
To reduce $\mathcal{Z}$ to Gaussian integrals, let us define $\rho$ as 
\begin{equation}\label{eq:ro}
    \rho=\prod_n^N e^{{-x_{\ell}}^2} e^{{-x_{\ell}'}^2}, \quad \text{where} \quad   x_{\ell}=\lambda_1 \psi_{\ell} -\lambda_2 \psi_{\ell+1}
\end{equation}
with
\begin{equation}\label{eq:lambda12}
    \lambda_{1,2}=(\sqrt{\beta(\mu+2)}\pm \sqrt{\beta(\mu-2)})/2, 
\end{equation}
as $\beta>0$, $\beta\mu \geq 0, \mu \geq 2$.
From Eq. (\ref{eq:lambda12}), one can directly find
\begin{equation}\label{eq:lambdasquare}
    \lambda_1^2=\frac{\beta}{2}(\mu+\sqrt{\mu^2-4}),\quad
    \lambda_2^2=\frac{\beta}{2}(\mu-\sqrt{\mu^2-4}),\quad
    \lambda_1\lambda_2=\beta
\end{equation}
Now, we will find the components of $\rho$ defined in Eq. (\ref{eq:ro}) explicitly.

\begin{equation*}
\begin{split}
&   \prod_{\ell}^N e^{-{x_{\ell}}^2}=\exp[-( \lambda_1\psi_1-\lambda_2\psi_{2})^2] \exp[-( \lambda_1\psi_2-\lambda_2\psi_{3})^2 ]\dots \exp[-( \lambda_1\psi_N-\lambda_2\psi_{1})^2]\\ &=\exp[-(\dots+{\lambda_1}^2|\psi_{\ell}|^2+{\lambda_1}^2|\psi_{\ell+1}|^2+{\lambda_2}^2|\psi_{\ell+1}|^2 + {\lambda_2}^2|\psi_{\ell}|^2]\times\\ &\exp[-\lambda_1\lambda_2(\psi_{\ell}^*\psi_{\ell+1}+\psi_{\ell}\psi_{\ell+1}^*)-\lambda_1\lambda_2(\psi_{\ell-1}^*\psi_{\ell}+\psi_{\ell-1}\psi_{\ell}^*) \dots )]\\
&=\dots \exp[-\beta ( \frac{\mu+\sqrt{\mu^2-4}}{2}|\psi_{\ell}|^2 +\frac{\mu+\sqrt{\mu^2-4}}{2}|\psi_{\ell+1}|^2] \times \\
&\exp[ \frac{\mu-\sqrt{\mu^2-4}}{2}|\psi_{\ell+1}|^2
    +\frac{\mu-\sqrt{\mu^2-4}}{2}|\psi_{\ell}|^2 - \psi_{\ell}^* \psi_{\ell+1}-\psi_{\ell} \psi_{\ell+1}^* )] \dots \\
   &= \prod_{\ell}^N \exp[-\beta\left(\mu|\psi_{\ell}|^2 - \psi_{\ell}^* \psi_{\ell+1}-\psi_{\ell} \psi_{\ell+1}^*  \right)] \equiv \prod_{\ell}^N \exp[-\beta \left( \mu A -H \right)]. 
\end{split}    
\end{equation*}
%\begin{align*}
% &   \prod_{\ell}^N e^{-{x_{\ell}}^2}=\exp[-( \lambda_1\psi_1-\lambda_2\psi_{2})^2] \exp[-( \lambda_1\psi_2-\lambda_2\psi_{3})^2 ]\dots \exp[-( \lambda_1\psi_N-\lambda_2\psi_{1})^2]\\ &=\exp[-(\dots+{\lambda_1}^2|\psi_{\ell}|^2+{\lambda_1}^2|\psi_{\ell+1}|^2+{\lambda_2}^2|\psi_{\ell+1}|^2 + {\lambda_2}^2|\psi_{\ell}|^2 -\lambda_1\lambda_2(\psi_{\ell}^*\psi_{\ell+1}+\psi_{\ell}{\psi_{\ell+1}}^*)-\lambda_1\lambda_2(\psi_{\ell-1}^*\psi_{\ell}+\psi_{\ell-1}{\psi_{\ell}}^*) \dots )]\\
%    &=\dots \exp[-\beta ( \frac{\mu+\sqrt{\mu^2-4}}{2}|\psi_{\ell}|^2 +\frac{\mu+\sqrt{\mu^2-4}}{2}|\psi_{\ell+1}|^2  + \frac{\mu-\sqrt{\mu^2-4}}{2}|\psi_{\ell+1}|^2
%    +\frac{\mu-\sqrt{\mu^2-4}}{2}|\psi_{\ell}|^2 - {\psi_{\ell}}^* \psi_{\ell+1}-\psi_{\ell} \psi_{\ell+1}^* )] \dots \\
%   &= \prod_{\ell}^N \exp[-\beta\left(\mu|\psi_{\ell}|^2 - {\psi_{\ell}}^* \psi_{\ell+1}-\psi_{\ell} \psi_{\ell+1}^*  \right)]\\
%   &\equiv \prod_{\ell}^N \exp[-\beta \left( \mu A -H \right)] 
%    \end{align*}

\begin{equation}
    \prod_{\ell}^N x_{\ell} =\dots (\lambda_1\psi_{\ell-1}-\lambda_2\psi_{\ell}) (\lambda_1\psi_{\ell}-\lambda_2\psi_{\ell+1})\dots
\end{equation}

\begin{equation}
    \det{\left(\frac{\partial x_{\ell}}{\partial \psi_{\ell}}\right)}=\lambda_1^N - \lambda_2^N
\end{equation}

\begin{equation}
    \det{\left(\frac{\partial \psi_{\ell}}{\partial x_{\ell}}\right)}\approx 1/\lambda_1^N, \quad \text{since} \quad \lambda_1>\lambda_2
\end{equation}
Thus, the partition function can be written as
  \begin{align}
        \mathcal{Z}&=\int \int \rho \prod_{\ell}^N d\psi_{\ell} d\psi_{\ell}^* = \left( e^{-x^2} \det\left(\frac{\partial \psi_{\ell}}{\partial x_{\ell}}\right) \prod_{\ell}^N dx_{\ell} \right)^2\\
        &=\left( \left({\sqrt{\frac{\pi}{2}}}\right)^N \frac{1}{{\lambda_1}^N} \right)^2 = \left(\frac{\pi}{2 {\lambda_1}^2}\right)^N.
  \end{align}
The thermodynamic value of the total energy, and total norm are
\begin{equation}
    \mathcal{H}=\left(\frac{\mu}{\beta}\frac{\partial}{\partial\mu}-\frac{\partial}{\partial\beta}\right)\ln{\mathcal{Z}}, \quad \mathcal{A}=-\frac{1}{\beta}\frac{\partial}{\partial\mu}\ln{\mathcal{Z}}.
\end{equation}

\begin{equation}\label{eq:turevmu}
 \frac{\partial}{\partial\mu}\ln{\mathcal{Z}}= N  \frac{\partial}{\partial\mu}\ln{ \left(\frac{\pi}{2 {\lambda_1}^2}\right)}=N \frac{\partial}{\partial\mu} \left ( \ln(\pi)-\ln{\left[ \beta (\mu+\sqrt{\mu^2-4}) \right]} \right)=-\frac{N}{\sqrt{\mu^2-4}},
\end{equation}
where Eq. (\ref{eq:lambdasquare}) is used.

\begin{equation}\label{eq:turevbeta}
  \frac{\partial}{\partial\beta}\ln{\mathcal{Z}}= N  \frac{\partial}{\partial\beta}\ln{ \left(\frac{\pi}{2 {\lambda_1}^2}\right)}= N \frac{\partial}{\partial\beta} \left ( \ln(\pi)-\ln{\left[ \beta (\mu+\sqrt{\mu^2-4}) \right]} \right)=-N/\beta
\end{equation}
From Eq. (\ref{eq:turevmu}), and (\ref{eq:turevbeta}), the energy density is
\begin{equation}\label{eq:h_}
h=\frac{\mathcal{H}}{N}=\frac{1}{N}\left[\frac{\mu}{\beta} \frac{\partial}{\partial\mu}\ln{\mathcal{Z}} -\frac{\partial}{\partial\beta}\ln{\mathcal{Z}}\right]= 
- \frac{\mu}{\beta} \frac{1}{\sqrt{\mu^2-4}} + \frac{1}{\beta},
\end{equation}
and the norm density
\begin{equation}\label{eq:a_}
    a=\frac{\mathcal{A}}{N}=\frac{1}{N}\left[-\frac{1}{\beta}\frac{\partial}{\partial\mu}\ln{\mathcal{Z}}\right]=\frac{1}{N\beta}\frac{N}{\sqrt{\mu^2-4}}=\frac{1}{\beta\sqrt{\mu^2-4}}. 
\end{equation}
Let $h=-ca$. Then, by using Eq. (\ref{eq:h_}), and Eq. (\ref{eq:a_}) we can write $\beta$ and $\mu$ as 
\begin{equation}\label{eq:betamu}
    \beta=\frac{-2h}{4a^2-h^2}=\frac{2c}{a(4-c^2)}, \quad \mu=\frac{4a^2+h^2}{-2ah}=\frac{4+c^2}{2c}. 
\end{equation}
The explicit derivation of Eq. (\ref{eq:betamu}) is as follows.
\begin{align}
    \beta&=\frac{-2h}{4a^2-h^2}=(-2)\left[ \frac{\sqrt{\mu^2-4}-\mu}{\beta\sqrt{\mu^2-4}}\left(\frac{4}{\beta^2(\mu^2-4)}-\frac{(\sqrt{\mu^2-4}-\mu)^2}{\beta^2(\mu^2-4)} \right)^{-1} \right]\\
    &=\frac{(-2)(\sqrt{\mu^2-4}-\mu)\beta\sqrt{\mu^2-4}}{4-(\sqrt{\mu^2-4}-\mu)^2}=\frac{(-2)\beta(\mu^2-4-\mu\sqrt{\mu^2-4})}{(-2)(\mu^2-4-\mu\sqrt{\mu^2-4})}=\beta
\end{align}

\begin{align}
    \mu&=\frac{4a^2+h^2}{-2ah}=-\frac{2a}{h}-\frac{h}{2a}=\frac{-2}{\beta\sqrt{\mu^2-4}} \frac{\beta\sqrt{\mu^2-4}}{\sqrt{\mu^2-4}-\mu}-\frac{\sqrt{\mu^2-4}-\mu}{2\beta\sqrt{\mu^2-4}}\beta\sqrt{\mu^2-4}\\
    &=\frac{-2}{\sqrt{\mu^2-4}-\mu}-\frac{(\sqrt{\mu^2-4}-\mu)^2}{2(\sqrt{\mu^2-4}-\mu)}=\frac{-4-(2\mu^2-4-2\mu\sqrt{\mu^2-4})}{2(\sqrt{\mu^2-4}-\mu)}\\
    &=\frac{-\mu^2+\mu\sqrt{\mu^2-4}}{\sqrt{\mu^2-4}-\mu}=\frac{\mu(\sqrt{\mu^2-4}-\mu)}{\sqrt{\mu^2-4}-\mu}=\mu. 
\end{align}

%\section{DNLS thermodynamics} \label{gng}

%\section{Gibbs distribution}\label{sec:gibbsdnls}

\chapter{The relation of $(\beta, \mu)$ and $(a,h)$ for nonlinear lattice} %$g\neq 0$}
\label{ap:temperature_mu_a_h_g}
%Temperature and chemical potential}% in Gibbs regime}
%, by neglecting the effect of disorder

\begin{figure}[hbt!] 
    \centering
    \includegraphics[width=0.75\textwidth]{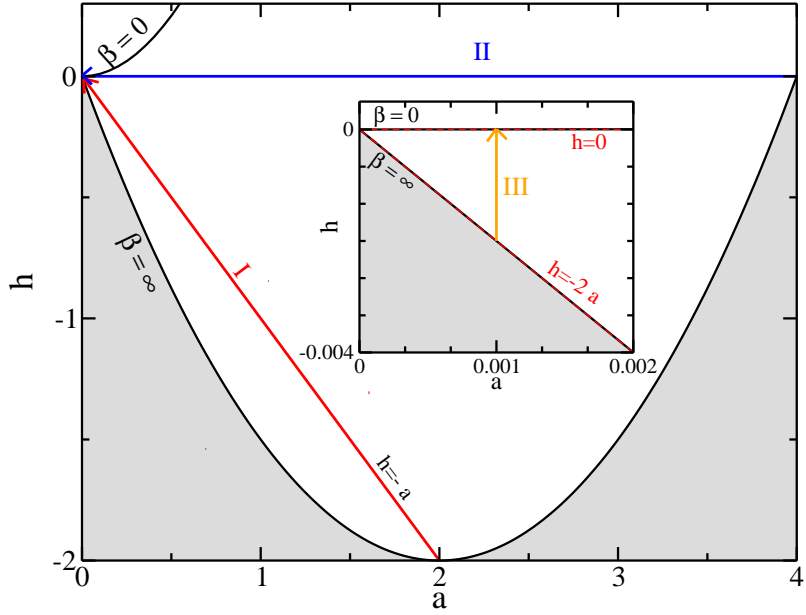}
    \caption[The phase diagram of GPL with three angles of approach to study temperature and chemical potential]{The phase diagram of GPL, showing our study on how temperature and chemical potential changes in three directions. I (red arrow): on $h=-a$ line from $a=2$ to $a=0$. II (blue arrow): on $h=0$ line from $a=4$ to $a=0$. III (inset, orange arrow): on fixed $a=10^{-3}$ from $h\approx -2a$ to $h=0$. }
     \label{fig:phasezoom}
\end{figure}

We can define a point in the phase diagram of GP lattice by two pairs  ($\beta,\mu$) or correspondingly $(a, h)$, via transfer integral operator method, described in \ref{sec:tim}. Hereby, we show how inverse temperature ($\beta$) and chemical potential ($\mu$) change as we approach the origin of the phase diagram from different angles. To define the angle, we use the fact that near to the origin the nonlinear term of Hamiltonian can be neglected s.t. we can write the density relation of energy and norm linearly with $h=-c a$. Here, $c$ is representing the angle of approach varying from 0 to 2 in the Gibbs regime. The inset of Fig. \ref{fig:phasezoom} exhibits  the $\beta=0$ and $\beta=\infty$ black lines almost match with the $h=0$, and $h=-2a$ red dashed lines near to the origin, respectively.

  \begin{figure}[hbt!] 
    \centering
    \includegraphics[width=0.5\textwidth]{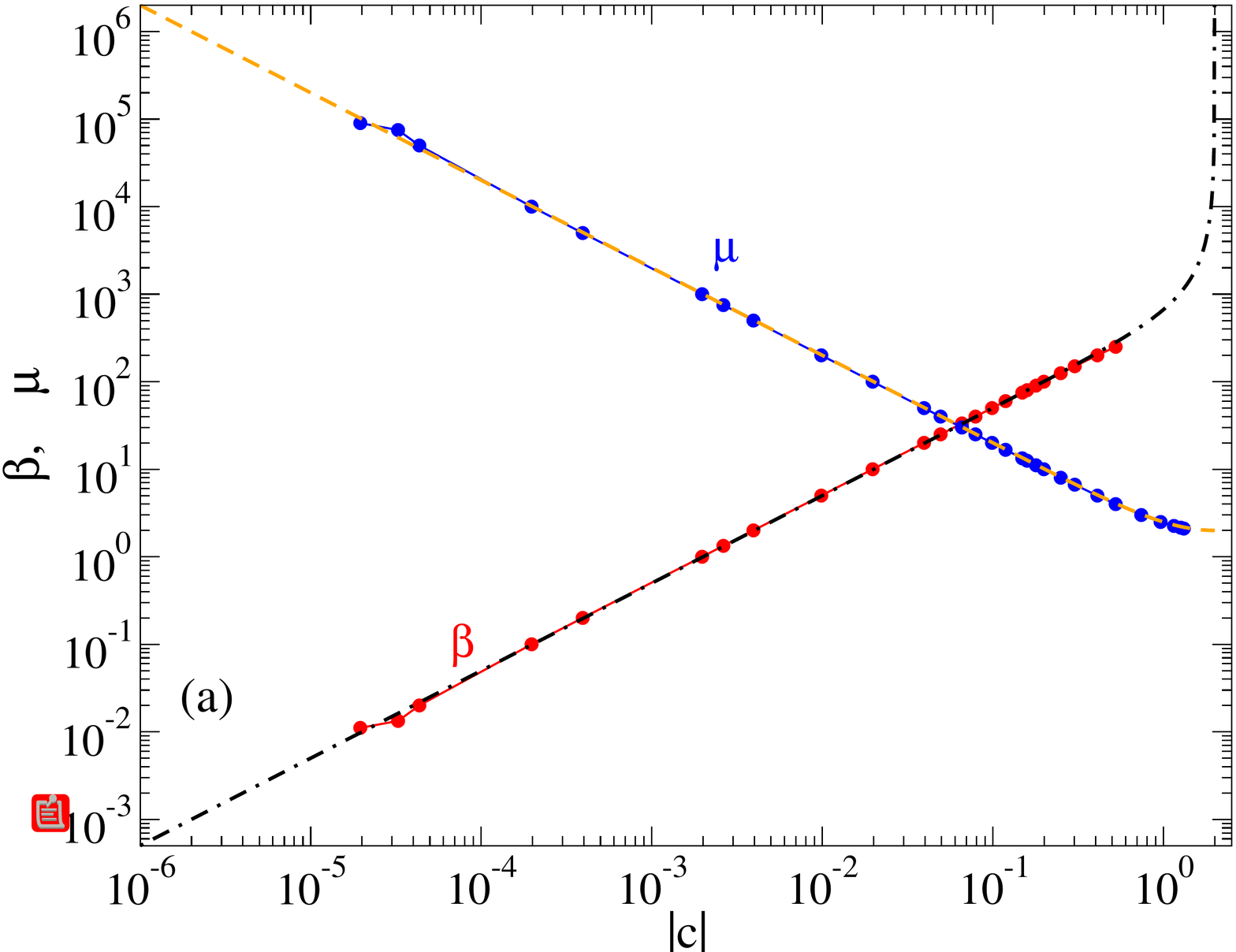}%c_beta_mu_new.pdf}
     \includegraphics[width=0.5\textwidth]{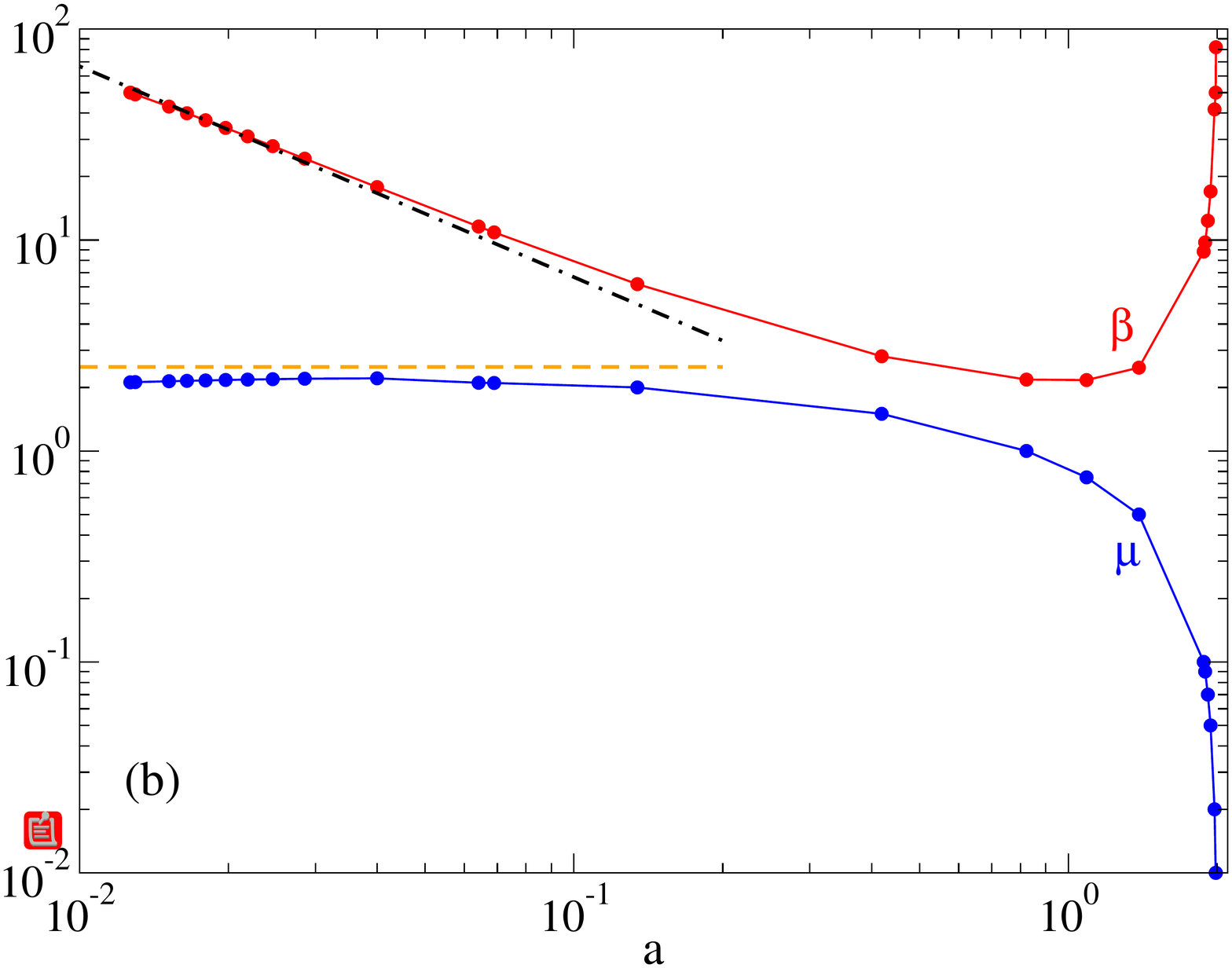}    
    \caption[Temperature and chemical potential vs. (a) $h/a$ on  $a=10^{-3}$ line, and vs. (b) $a$ on $h=-a$ line]{Inverse temperature $\beta$ and chemical potential $\mu$ are plotted  (a) versus $|c|=h/a$, with fixed $a=10^{-3}$ (b) versus $a$ with fixed $c=1$ ($h=-a$). The data by TIO method ($g=1$), $\beta_{TIO}$ and $\mu_{TIO}$ are shown as line connected circles, red and blue, respectively. The analytical approaches from Eq. (\ref{eq:betamuc}) are shown as black dot-dashed line for $\beta=\frac{2c}{a(4-c^2)}$, and orange dashed line for $\mu=(4+c^2)/2c$.}
     \label{fig:c_beta_mu_new}
\end{figure}

\begin{figure}[hbt!] 
     \centering
     \includegraphics[width=0.6\textwidth]{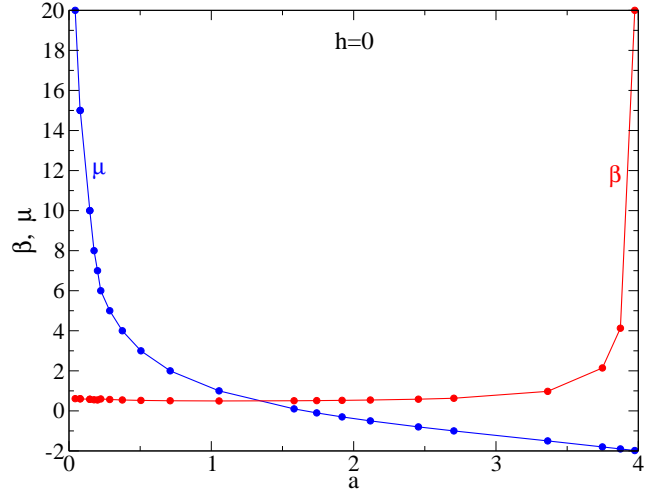}
     \caption[Temperature and chemical potential vs. $a$ on $h=0$ line]{Inverse temperature $\beta$ and chemical potential $\mu$ are plotted versus $a$ on $h=0$ line ($g=1$, $c=0$).  The numerical data  by TIO method, shown as line connected circles in red for $\beta$, and in blue for $\mu$.}
     \label{fig:yequalszero_mu_beta_vs_a}
\end{figure}

Let us recall Eq. (\ref{eq:betamu}) which is valid for zero nonlinearity. Now, the definitions of $\beta$ and $\mu$ in terms of $a$ and $h$ %, which are derived in Appendix (Eq. \ref{eq:betamu}) 
are considered for the limit of vanishing nonlinearity, near to the origin. 
\begin{equation}\label{eq:betamuc}
    \beta (ga \rightarrow 0)\approx\frac{-2h}{4a^2-h^2}=\frac{2c}{a(4-c^2)}, \quad \mu(ga \rightarrow 0)\approx\frac{4a^2+h^2}{-2ah}=\frac{4+c^2}{2c}.
\end{equation}

Fig. \ref{fig:c_beta_mu_new} and Fig. \ref{fig:yequalszero_mu_beta_vs_a} show how $\beta$ and $\mu$ varies as some pair of finite densities $(a,h)$ approaches to the origin of the phase diagram from different directions.

Fig. \ref{fig:c_beta_mu_new}(a) exhibits how $\beta$ and $\mu$ varies as one changes the angle of direction of approach towards the origin. %$a=10^{-3}$ is hold constant. 
We keep $a=10^{-3}$ near to the origin while $g=1$, since the nonlinear term is expected to be negligible in the small norm density limit. Then we increase the energy density $h$ and find how $\beta$ and $\mu$ changes by TIO method. The direction is shown as III (orange line) in the inset of Fig. \ref{fig:phasezoom}, starting from near to the ground state  $h\approx-2a$ ($\beta\rightarrow \infty$) to $h=0$ ($\beta\rightarrow 0$).  The TIO numeric results of $\beta$, and $\mu$ matches their analytic approximations in Eq. (\ref{eq:betamuc}), very well.   %$\beta=\frac{2c}{a(4-c^2)}$ and $\mu=(4+c^2)/2c$, nicely. 

In Fig. \ref{fig:c_beta_mu_new}(b), we show how $\beta$ and $\mu$ change on the $h=- a$ line. We fix $c=-h/a =1$, in order to stay on the $h=-a$ line. We use the TIO method with $g=1$ and start from $a\approx 2$ which is the ground state, and approach to the origin in the direction I, as shown in Fig. \ref{fig:phasezoom} with a red line. The analytics from Eq. (\ref{eq:betamuc}), and the numerics obtained by TIO method, both show that while approaching the origin, $\mu$ goes to a constant value $\mu=2.5$, as the temperature goes to infinity. On the other hand, as $a$ approaches to 2, which corresponds to the ground state, $\beta$ diverges to infinity while $\mu$ goes to zero, found by TIO method. The reason for that is while $\beta\rightarrow0$, $\beta\mu$ has to stay constant, and thus $\mu\rightarrow\infty$. The analytical approximations  in Eq. (\ref{eq:betamuc}) roughly match the TIO results in the small $a$ limit.

In Fig. \ref{fig:yequalszero_mu_beta_vs_a} we  found by TIO method how the temperature and chemical potential varies as we approach from $a=4$ to the origin keeping $h=0$. The direction of approach is shown as II in Fig. \ref{fig:phasezoom} with a blue line. Since the ground state is $h=g a^2/2 -2 a$, %(see Appendix \ref{ap:zerotemperature}), 
the TIO data for $a=4$ is at the ground state. We expect that the analytic approaches of $\beta$ and $\mu$ in Eq. (\ref{eq:betamuc}) will be valid in the small norm density limit, which give $\beta\rightarrow 0$, and $\mu\rightarrow \infty$. % in the $a\rightarrow 0$ limit. 

This study is inspired by the wave packet spreading (see more details in Chapter \ref{chapter5}), since both the absolute energy and norm densities of wave packets decrease in time as the wave spreads: $\lim_{t\rightarrow\infty} (a,h)\rightarrow (0,0)$.

%decrease $a$ from the and see how $\beta$ and $\mu$ changes from $T=0$ to infinity. The analytic result obtained by $g=0$ assumption is plotted with the numeric result by transfer integral method (TIM).
\chapter{Lyapunov characteristic exponent calculation}\label{ap:lce}

We introduce a perturbation $\delta_\ell$ to the equilibrium wave function $\psi_\ell=\sqrt{a}\exp(i\phi_\ell)$ for $\ell=1,\dots,N$. We insert the perturbed version of the wave function into the equations of motion of GP lattice in Eq. (\ref{eq:eqsofmotion}). After neglecting the second order perturbation terms, we obtain its equations of motion as Eq. (\ref{eq:DNLSlce}):

\begin{equation}\label{eq:DNLSlce2}
i\dot{\delta}_\ell=-(\delta_{\ell+1}+\delta_{\ell-1})+2|\psi_\ell|^2\delta_\ell+\psi_\ell^2\delta_\ell^{\ast}. 
\end{equation}
Eq. (\ref{eq:DNLSlce2}) is solved by using symplectic $SBAB_2$ integrator scheme. The Hamiltonian, 
  \begin{equation}
\small \mathcal{H} =\sum_{\ell}\big[-(\delta_{\ell}^{\ast}\delta_{\ell+1}+\delta_{\ell} \delta_{\ell+1}^{\ast})+\frac{g}{2}(4|\psi_{\ell}|^2|\delta_{\ell}|^2+\psi_{\ell}^2\delta_{\ell}^{\ast 2} )\big]
\label{eq:b1}
\end{equation}
corresponds to Eq. (\ref{eq:DNLSlce}) is split as 
  \begin{eqnarray}
    A =\sum_{\ell}\big[-(\delta_{\ell}^{\ast}\delta_{\ell+1}+\delta_{\ell} \delta_{\ell+1}^{\ast})\big],\\
   B =\frac{g}{2}\sum_{\ell}\big[(4|\psi_{\ell}|^2|\delta_{\ell}|^2+\psi_{\ell}^2\delta_{\ell}^{\ast 2})\big].
  \label{eq:b2}
  \end{eqnarray}
B can be written as $B=P+Q$, where
\begin{equation}
\small P=2g \sum_{\ell}|\psi_{\ell}|^2|\delta_{\ell}|^2,~~Q=\frac{g}{2}\sum_{\ell}\psi_{\ell}^2\delta_{\ell}^{\ast 2}
\label{eq:b3}
\end{equation}
\subsubsection*{Action of the operator $\exp(\tau L_A)$ on $\delta_{\ell}$}
Fast Fourier transform (FFT) is used for Hamiltonian $A$.
  \begin{eqnarray}
    \delta_q=\sum_{\ell=1}^N \delta_\ell e^{2 \pi i (q-1) (l-1)/N},\\
   \delta_q{'}=\delta_q e^{2 i \cos(2\pi(q-1)/N)\tau},\\
   \delta_{\ell}{'}=\sum_{q=1}^N {\delta_q}' e^{-2 \pi i (\ell-1) (q-1)/N},
  \label{eq:b4}
  \end{eqnarray}
    where $\delta_{\ell}{'}$ is $\delta_{\ell}$ at $t+\tau$.
\subsubsection*{Action of the operator $\exp(\tau L_P)$ on $\delta_{\ell}$}
 It can be solved exactly as 
\begin{equation}
{\delta_{\ell}}{'}=\delta_{\ell} \exp{{-2 i g {{|{\psi_{\ell}}|}^2} \tau}}
\label{eq:b5}
\end{equation}

\subsubsection*{Action of the operator $\exp(\tau L_Q)$ on $\delta_{\ell}$}
\begin{equation}
\delta_{\ell}{'}=(\delta_{a\ell}+i \delta_{b\ell}),
\label{eq:b6}
\end{equation}
 \begin{eqnarray}
  \delta_{a\ell}= c_1 \cosh(At)+\frac{1}{A}\sinh(At)\big[bc_1-a c_2\big],\\
  \delta_{b\ell}= c_2 \cosh(At)-\frac{1}{A}\sinh(At)\big[bc_2+a c_1\big],
\label{eq:b7}
 \end{eqnarray}
  where $A=\sqrt{a^2+b^2}$, $a=Re(g \psi_{\ell}^2)$, $b=Im(g \psi_{\ell}^2)$, $c_1=Re(\delta_{\ell})$ and $c_2=Im(\delta_{\ell})$.  Equations are integrated by using the $SBAB_2$ scheme given in Sec. \ref{sec:symp} \cite{Bodyfelt:2011}.

\chapter{Infinite temperature of ordered \& disordered GP lattice}\label{ap:betazero}
%Gibbsian - NonGibbsian Crossover

For ordered GP lattice, the relation between energy and norm densities at the infinite temperature  is shown in \cite{Rasmussen:2000} as $h=g a^2$. Hereby, we present a derivation for the disordered case, while considering $g=1$. The relations for ordered GP lattice can also be easily subtracted from the following derivation. A similar derivation of disordered case can be also found in \cite{Johansson:2004}.

As the inverse temperature $\beta\rightarrow 0$; $\beta\mu$ stays constant, thus $\mu\rightarrow\infty$, and the modified Bessel function term in Eq. (\ref{eq:Zbessel}) goes to 1. Moreover, all kinetic energy terms disappear. Hence, the infinite temperature case corresponds to uncoupled sites. The results are valid then for any lattice dimension. We can rewrite the partition function in Eq. (\ref{eq:Zbessel}) as

\begin{equation}
\mathcal{Z}=(2\pi)^N \int_0^\infty \prod d\mathcal{A}_\ell \exp[-\beta (\mathcal{H}_0+\mu \mathcal{A})],
\end{equation}
where $\mathcal{H}_0=\sum_\ell \epsilon_\ell \mathcal{A}_\ell +\frac{1}{2}{\mathcal{A}_\ell}^2$. With the definitions
\begin{equation}
\mu_\ell=\mu+\epsilon_\ell, \qquad y(\beta,\mu_\ell)=\frac{1}{\beta \mu_\ell}-\frac{\beta}{(\beta\mu_\ell)^3},    
\end{equation}
we can write $\mathcal{Z}$ as
\begin{equation}
\mathcal{Z}=(2\pi)^N \prod_{\ell=1}^N y(\beta,\mu_\ell)=(2 \pi)^N \prod_{\ell=1}^N  \frac{1}{\beta\mu_\ell} \prod_{\ell=1}^N \left( 1-\frac{1}{\beta {\mu_\ell}^2}\right)   . 
\end{equation}
The natural logarithm of $\mathcal{Z}$ gives
\begin{equation}\label{eq:lnZbeta0}
\begin{split}
\ln \mathcal{Z}&=N\left[ \ln 2\pi -\ln \beta \right]-\sum_{\ell=1}^N \ln\mu_\ell + \sum_{\ell=1}^N \ln \left(1-\frac{1}{\beta \mu_\ell^2}\right)\\
&=N\ln 2\pi - \sum_{\ell=1}^N \ln(\beta \mu_\ell) - \sum_{\ell=1}^N \frac{\beta}{\beta^2 \mu_\ell^2}. 
\end{split}
\end{equation}
By this definition in (\ref{eq:lnZbeta0}), we can find 
\begin{equation}\label{eq:dlnZdmu}
\frac{\partial}{\partial \mu}\ln\mathcal{Z}=-\sum_{\ell=1}^N \frac{\beta}{\beta\mu_\ell}+\sum_{\ell=1}^N \frac{2\beta}{\beta^2\mu_\ell^3},
\end{equation}
and 
\begin{equation}\label{eq:dlnbetadmu}
-\frac{\partial}{\partial\beta}\ln\mathcal{Z}=\sum_{\ell=1}^N\frac{\mu_\ell}{\beta\mu_\ell}-\sum_{\ell=1}^N \frac{1}{\beta^2 \mu_\ell^2}.    
\end{equation}
Via Eq. (\ref{eq:dlnZdmu}) and Eq. (\ref{eq:dlnbetadmu}), we obtain 
% into the definitions of $\langle A\rangle$ and $\langle H \rangle$
\begin{equation}\label{eq:Abeta0}
\langle \mathcal{A}\rangle = -\frac{1}{\beta}\frac{\partial}{\partial\mu}\ln\mathcal{Z} =\sum_{\ell=1}^N \frac{1}{\beta \mu_\ell}=\frac{N}{\beta\mu},    
\end{equation}
and
\begin{equation}
\langle \mathcal{H}\rangle = \sum_{\ell=1}^N \frac{\varepsilon_\ell}{\beta \mu_\ell}-\sum_{\ell=1}^N \frac{1}{\beta^3 \mu_\ell^3}(-2\beta\mu+\beta\mu_\ell)
=\frac{N}{\beta^2 \mu^2} + \sum_{\ell=1}^N \frac{\varepsilon_\ell}{\beta \mu_\ell}.  
\end{equation}
%\approx \frac{N}{\beta^2 \mu^2}
The last term is zero for ordered case which exactly gives the relation $h= a^2$. It is, however, negligible for a disordered system in the thermodynamic limit $N\rightarrow\infty$:
\begin{equation}\label{eq:Happroxbeta0}
\langle \mathcal{H} \rangle= \frac{N}{\beta^2\mu^2}+ \sum_{\ell=1}^N  \frac{\varepsilon_\ell}{\beta (\mu+\varepsilon_\ell)}=\underbrace{\frac{N}{\beta^2\mu^2}}_{\sim N}+ \sum_{\ell=1}^N  \frac{1}{\beta \mu} \bigg[ \underbrace{\sum_{\ell=1}^N \varepsilon_\ell}_{\sim \sqrt{N}} -\underbrace{\frac{1}{\mu}\sum_{\ell=1}^N \varepsilon_\ell^2}_{\sim N/\mu} \bigg]
\end{equation}
In Eq. (\ref{eq:Happroxbeta0}) the first term is in the order of $N$, as the second term is in the order of $\sqrt{N}$, and the third is in the order of $N/\mu$, while $\mu\rightarrow \infty$ when $\beta=0$. Therefore we obtain
\begin{equation}\label{eq:Hbeta0}
 \langle \mathcal{H} \rangle\approx   \frac{N}{\beta^2\mu^2}.
\end{equation} 
From Eq. (\ref{eq:Abeta0}) and (\ref{eq:Hbeta0}), the energy density $h=1/(\beta^2 \mu^2)$ and the norm density $a=1/\beta \mu$ gives  $h=a^2$ (originally $h=g a^2$) while nonlinearity $g=1$.

\chapter{Numerical ground state of disordered GP lattice}
\label{ap:zerotemperaturedisorder}

From Eq. (\ref{eq:psilt}), total norm, and energy are
\begin{equation}
\mathcal{A}=\sum_{\ell} a_{\ell}, \quad H=\sum_{\ell} \epsilon_{\ell} a_{\ell} + \frac{g}{2} a_{\ell}^2 -2 J \sqrt{a_{\ell} a_{\ell+1}} \cos{(\phi_{\ell}-\phi_{\ell+1})}.     
\end{equation}
The minimum energy is obtained while  $\phi_{\ell}=\phi_{\ell+1}$ for all sites. 
Since all the phases rotate with the same angle in time according to the relation $G_\ell(t)\propto e^{i\mu t}$, we can choose the initial time $t=0$ when the ground state is completely real. Hence, we can compute it efficiently. 

Numerically, imposing the constraint on total norm, we obtain the ground state via the Nelder-Mead simplex algorithm \cite{lagarias}, using the convergent solution after thousands of iterations for a fixed disorder realization, explained in Sec. \ref{sec:GroundStateRenormalization}.

\chapter{Analytic localization length of BdG modes near zero energy} \label{ap:analyticlocalizationlength} %spectrum properties

This appendix presents an earlier approximation of the analytic localization length. Here we neglected the nondecaying correlation between $\zeta_\ell$ and $\epsilon_\ell$ (see Eq. \ref{eq:corr}) which is responsible for the side peak evolution. Nevertheless, this approximation produces Eq. (\ref{eq:analyticlocalizationlength}) for $q\rightarrow 0$ which is the same relation found in Eq. (\ref{eq:analyticlocalizationlengthk0}). 

While $g a \gg J$, the fluctuations $\delta\zeta_\ell$ become negligible, and $\bar{\zeta_\ell} \approx 2$ (see Fig. \ref{fig:zeta_a_corr}(a)). With these approximations, in the strongly interacting regime, we can rewrite Eq. (\ref{eq:chipi}) using Eq. (\ref{eq:muavg}). 

\begin{align}\label{eq:chipi3}
\lambda \chi_{\ell} &=(-\Tilde{\epsilon_{\ell}}+4J)\chi_{\ell}+(\Tilde{\epsilon_{\ell}}-2J){\Pi_{\ell}} -J(\chi_{\ell+1} +\chi_{\ell-1}) \nonumber\\
\lambda \Pi_{\ell} &=  (\Tilde{\epsilon_{\ell}}-4J){\Pi_{\ell}}+ (-\Tilde{\epsilon_{\ell}}+2J)\chi_{\ell} +J(\Pi_{\ell+1} +\Pi_{\ell-1}), 
\end{align}
where $\tilde{\epsilon}_\ell= {\epsilon}_\ell -\mu \approx \epsilon_\ell-g a+2J$, in which we used Eq. (\ref{eq:muavg}) and the approximation $\bar{\zeta}=2$. Let us rewrite Eq. (\ref{eq:chipi3}) with $S_\ell=\chi_\ell+\Pi_\ell$, and $D_\ell=\Pi_\ell-\chi_\ell $ as

\begin{align}
\lambda S_\ell+2(ga+J-\epsilon_\ell)D_\ell-J(D_{\ell+1}+D_{\ell-1}) &=0 \label{s_l}\\  
\lambda D_\ell+2J S_\ell -J(S_{\ell+1}+S_{\ell-1})&=0.\label{d_l} 
\end{align}
At $\lambda=0$, $D_\ell=0$, and $S_\ell$ is constant, which confirms the delocalized mode with diverging localization length in Fig. \ref{fig:PlaN20000log}. We then insert the expression for $D_\ell$ into $S_\ell$, and obtain 
\begin{equation}\label{sldl}
({\lambda}^2/J -4g a-6J +4\epsilon_\ell) S_\ell -J(S_{\ell+2}+S_{\ell-2}) +(S_{\ell+1}+S_{\ell-1})(2ga+4J-2\epsilon_\ell)=0.    
\end{equation}
The spectrum width of BdG equations with disorder can be found by setting $\epsilon_\ell=-W/2$ for all sites so that we can substitute $S_\ell \approx e^{iq\ell}\tilde{S}$, which gives 
\begin{equation}
{\lambda_{\text{max}}}=\sqrt{(ga+W/2+4J)^2-(ga+W/2)^2}    
\end{equation}
 while $q=\pi$ for the maximum eigenvalue.

 Assuming $\xi(\lambda)\gg 1$, we substitute

\begin{equation}\label{chiell}
 S_\ell=e^{iq \ell} \tilde{S_\ell}, \quad \text{with} \quad {\tilde{S}}_{\ell \pm 1}\approx \tilde{S_\ell} \pm \frac{d \tilde{S_\ell}}{d\ell} 
\end{equation}
in the continuous limit assuming $\tilde{S_\ell}$ varies weakly as $q$ is not close to $\pi$. 
Then, we can write 
\begin{align}\label{variationXI}
S_{\ell+1}+S_{\ell-1}&=\left\{\cos{(q)} \tilde{S}_\ell+i\sin{(q)} \frac{d \tilde{S_\ell}}{d\ell} \right\} 2 e^{iq\ell},\nonumber\\
S_{\ell+2}+S_{\ell-2}&=\left\{\cos{(2 q)} \tilde{S}_\ell+2i\sin{(2 q)} \frac{d \tilde{S_\ell}}{d\ell} \right\} 2 e^{i q\ell},    
\end{align}
where $q$ is determined by the condition that $S$ does not depend on the coordinate $\ell$ in the absence of disorder. Thus, substituting Eq. \ref{chiell} in Eq. \ref{sldl}, we find the spectrum of the Bogoliubov Hamiltonian 
\begin{equation}\label{k0_1}
\lambda^2-[ga +4J\sin^2{{q}/2}]^2+(ga)^2=0,
\end{equation}
which corresponds to the well-known dispersion relation

\begin{equation}\label{eq:soundvelocity}
\lambda =\pm \sqrt{ 8J g a\sin^2{{q}/2} +16J^2 \sin^4{{q}/2} } \approx \pm \sqrt{2g a J}q   
\end{equation}
in the limit of small $q$ \cite{Pippan,Menotti}. 
We observe a  side-peak at $q=\pm \pi/2$ from the solution of Eq. \ref{sldl} by transfer matrix method, which gives

\begin{equation}\label{eq:sidepeak}
\lambda_{\text{sp}}= 2 \sqrt{J^2+ g a J}   
\end{equation}
in the weak disorder limit. Eq. (\ref{eq:sidepeak}) explains the energy shift of the sidepeak $\lambda_{sp}$ as we increase norm density $a$ in Fig. \ref{fig:PlaN20000log}. % with $g\equiv J\equiv 1$. 

We insert Eq. \ref{chiell}, and Eq. \ref{variationXI} into Eq. \ref{sldl}, then use $q$ definition in Eq. \ref{k0_1}, and get
 \begin{equation}\label{RandomEquation}
\tilde{S_\ell}  \epsilon_\ell (1- \cos{q})=i \frac{d  \tilde{S_\ell}}{d\ell} [J\sin{2q}-\sin{q}(ga+2J-\epsilon_\ell)]. 
 \end{equation}
After using trigonometric relations we can write it as

\begin{equation}\label{eq:slprimeoversl}
 \frac{1}{\tilde{S_\ell}}\frac{d\tilde{S_\ell}}{d\ell}=\frac{i\epsilon_\ell \tan{(q/2)}}{\sqrt{\lambda^2+g^2a^2}-\epsilon_\ell}.     
\end{equation}
In order to find the localization length, let us use the following relations in the limit of $q (\lambda)\rightarrow 0$:
\begin{equation}
\frac{d{\tilde{S_\ell}}}{d\ell} \approx \tilde{S}_{\ell+1}-\tilde{S_\ell}, \qquad \tilde{S_\ell} \sim e^{\ell/\xi}, \quad \frac{\tilde{S}_{\ell+1}}{\tilde{S_\ell}} = e^{1/\xi}.
\end{equation}
Now, we can define the inverse localization length as
\begin{align}\label{eq:analyticloclength1}
\frac{1}{\xi}&=\ln{\left|\frac{\tilde{S}_{\ell+1}}{\tilde{S_\ell}}\right|}\approx  \ln{\left|\frac{1}{\tilde{S_\ell}}\frac{d\tilde{S_\ell}}{d\ell}+1\right|}
\approx\frac{1}{\tilde{S_\ell}}\frac{d\tilde{S_\ell}}{d\ell}-\frac{1}{2}\left(\frac{1}{\tilde{S_\ell}}\frac{d\tilde{S_\ell}}{d\ell}\right)^2= -\frac{1}{2}\left(\frac{1}{\tilde{S_\ell}}\frac{d\tilde{S_\ell}}{d\ell}\right)^2\nonumber\\
&= \frac{\tan^2{(q/2)}W^2}{24(\lambda^2+ g^2 a^2)+ 2 W^2} = \frac{W^2 \left(\sqrt{\lambda^2+g^2a^2}-ga\right)}{24\left(ga+4J-\sqrt{\lambda^2+g^2a^2}\right)\left(\lambda^2+g^2a^2+W^2/12\right)} 
\end{align}
in which we used Taylor expansion, $e^{1/\xi}\approx 1+ 1/\xi$ and $\ln(x+1)\approx x-x^2/2$ 
where the term $x$ is eliminated because $\langle \epsilon_\ell \rangle=0$ in Eq. (\ref{eq:slprimeoversl}), yet $x^2/2$ not since ${\langle \epsilon_\ell \rangle}^2=W^2/12$.

%as red dashed lines which has almost matching results with TMM in the large norm densities. The inset of Fig.\ref{fig:analyticloclength} shows that the small side peak disappears for the analytic result due to the extra approximations explained earlier. The analytic $\xi(\lambda\rightarrow 0)$ found by Eq.\ref{eq:analyticlocalizationlength} matches the TMM and the analytic $\xi$ in  (\ref{eq:analyticloclength1}) in the low energies shown as cyan dashed line in Fig.\ref{fig:PlaN20000log}.

  \begin{figure}[hbt!] 
     \centering
     \includegraphics[width=0.6\textwidth]{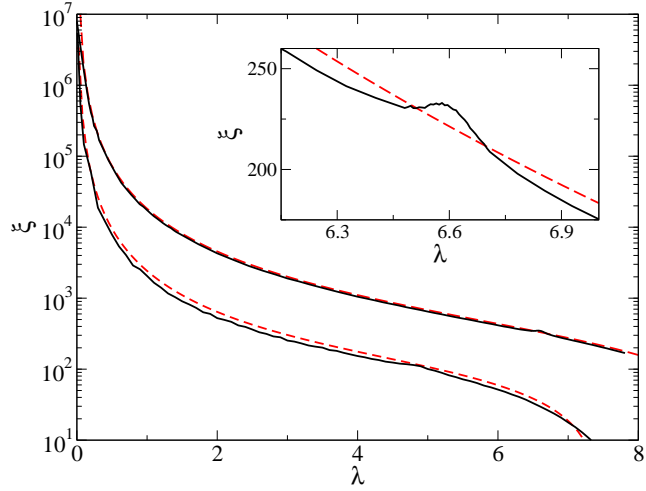}%
     \caption[The localization length of BdG modes vs. eigenvalues]{The localization length of BdG modes versus the eigenvalue $\lambda$ with $a=5,10$ from bottom to top.  Black lines - transfer matrix calculation results with $10^8$ number of iterations. Red dashed line - analytic result (\ref{eq:analyticloclength1}) in the strong interaction and large localization length approximation. Inset: The result for $a=10$ is zoomed.}
     \label{fig:analyticloclength}
     \end{figure}

In the $q\rightarrow 0$ limit, $\lambda \rightarrow 0$ and  $\tan^2(q/2)\approx \sin^2(q/2) \approx  \lambda^2/8Jg a $ via Eq. (\ref{k0_1}) in which the term with $\sin^4(q/2)$ can be neglected. Therefore, the localization length

\begin{equation}\label{eq:analyticlocalizationlength}
\xi(\lambda\rightarrow 0)\approx \frac{192 J(ga)^3}{W^2 \lambda^2},
\end{equation}
and the spectrum shows a linear behavior in $q$, $\lambda \approx \pm \sqrt{2g a J} q$ with sound velocity $\sqrt{2g a J}$ that is found by Eq. (\ref{eq:soundvelocity}).

We solve Eq. (\ref{eq:chipi3}) and Eq. (\ref{sldl}) by transfer matrix method and find the localization length $\xi$. Normalization is performed with QR decomposition after each 5 iteration steps with a total number of iterations $10^8$ (see Appendix \ref{ap:tmm} for the calculation details).

Eq. (\ref{sldl}) and Eq. (\ref{eq:analyticloclength1}) both assumes that $\zeta_\ell$ and $\epsilon_\ell$ are uncorrelated, since all $\zeta_\ell$ terms are taken as 2. Although $\bar{\zeta_\ell}\approx2$ and $\delta\zeta_\ell\sim 0$ in the strong interaction limit (see Fig. \ref{fig:zeta_a_corr}(a)), this assumption neglects the correlation between $\zeta_\ell$ and $\epsilon_\ell$, which in fact stays constant as $a\rightarrow\infty$ (see Fig. \ref{fig:zeta_a_corr}(b)). Neglecting the nondecaying correlation between GS field and disorder causes the disappearance of the side peaks (see  Fig. \ref{fig:analyticloclength}). This result may indicate that the correlation between $\zeta_\ell$ and $\epsilon_\ell$ is responsible for the singularity in Eq.\ref{invloc_fin_a}, which generates the side peaks. On the other hand, the figure \ref{fig:analyticloclength} gives a good approximation for $\xi(\lambda \sim 0)$, and shows that the analytic results almost match the TMM (\ref{4by4eski}) results with $\zeta_\ell\approx 2$  in the large norm densities. Nevertheless, in spite of a tiny enhancement of the localization length at $q\sim \pi/2$ in the inset, we no longer observe a side peak. %, which gives a good approximation of $\xi(\lambda \sim 0)$.
%To conclude, the approximation $\zeta_\ell=2$ in Eq. (\ref{eq:analyticloclength1}) and Eq. (\ref{sldl}) can produce valid results only for small energies ($\lambda$). 

%The TMM results are valid  in the strong interaction regime which can be reached by weakening the disorder strength $W$ or increasing the norm density $a$. 

\begin{figure}[htp]
     \centering
     \includegraphics[width=0.6\textwidth]{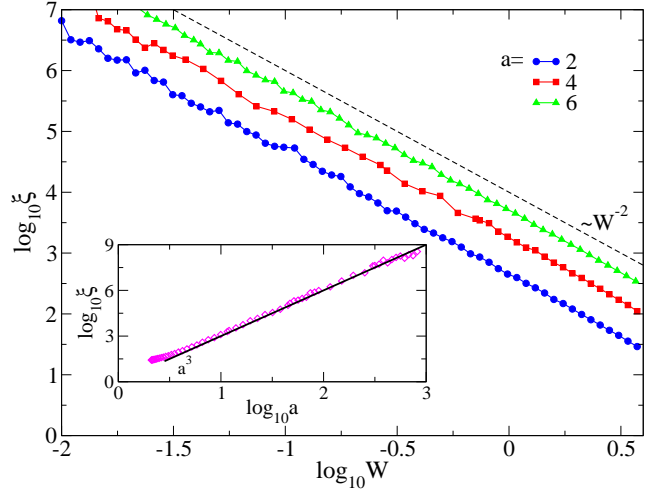}%{loclength_vs_W_a_la3tez.pdf}%loclength_vs_W_a_la3_.pdf}%loclength_vs_W_la3.pdf}%loclength_W.pdf}
     \caption[Localization length of BdG modes vs. disorder strength and norm density in the strongly interacting lattice]{Localization length of BdG modes $\xi$ vs. disorder strength $W$ and norm density $a$ in the strong interaction limit, calculated by TMM (Appendix \ref{ap:tmm}) plotted in log-log scale, with $\lambda=3$, and $N=10^9$. The main figure shows $a=2, 4, 6$ results, with the slope $\sim W^{-2}$ shown as black dashed lines (see Eq. \ref{eq:analyticlocalizationlength}). Inset: $\xi$ versus $a$, for fixed $W=4$ shown as magenta diamonds. The slope with $\sim a^{3}$ is shown as black line (see Eq. \ref{eq:analyticlocalizationlength}).}
     \label{fig:loclength_vs_W_la3}
\end{figure}

The localization length, which is found by Eq. (\ref{sldl}) with TMM, is plotted in Fig. \ref{fig:loclength_vs_W_la3}. In the weak disorder strength limit, it possess $\xi \sim 1/W^2$ relation for a range of fixed norm densities $a=2,4,6$, which is confirmed by the analytic result in Eq. (\ref{eq:analyticlocalizationlength}).  In the inset of Fig. \ref{fig:loclength_vs_W_la3}, correspondingly we plot $\xi$ for the large norm density limit which gives $\xi \sim a^3$. This result both confirms the numerical observation $P_{sp}{\lambda_{sp}}^2 \sim a^3$ in Fig. \ref{fig:plasquare_sidepeak}, and the analytic result in Eq. (\ref{eq:analyticlocalizationlengthk0}). %Moreover, the enhancement of localization lengths with the increasing $a$ in Fig. \ref{fig:loclength_vs_W_la3} also approves the exponential increase in $P$ in Fig. \ref{fig:PlaN20000log} with growing $a$.

\chapter{Transfer matrix method for the BdG modes near zero energy} \label{ap:tmm}

Eq.(\ref{eq:chipi3}) can be solved by transfer matrix method as follows:

\begin{equation}\label{eq:tmm1}
\begin{bmatrix} 
\chi_{\ell+1}  \\
\Pi_{\ell+1} \\
\chi_\ell \\
\Pi_\ell 
\end{bmatrix} =
\begin{bmatrix} 
{(-\lambda-\tilde{\epsilon}_\ell)}/{J}+4 & \tilde{\epsilon}_\ell/J-2 & -1 & 0 \\
\tilde{\epsilon}_\ell/J-2  & ({\lambda-\tilde{\epsilon}_\ell})/{J}+4 & 0 & -1 \\
1  & 0 & 0 & 0 \\
0 & 1 & 0 & 0
\end{bmatrix}
\begin{bmatrix} 
\chi_\ell  \\
\Pi_\ell \\
\chi_{\ell-1} \\
\Pi_{\ell-1}
\end{bmatrix},
\end{equation}
where the $4\times4$ matrix is known as the transfer matrix $T_\ell (\ell=1,\dots,N)$. Following \cite{Slevin_2014}, we start the transfer matrix multiplication with 
%We can define the initial state orthogonally as
\begin{equation}
Q_1=
\begin{bmatrix} 
1 & 0  \\
0 & 1 \\
0 & 0 \\
0 & 0 
\end{bmatrix},
\end{equation}
%\begin{bmatrix} 
%\chi_1  \\
%\Pi_1 \\
%\chi_0 \\
%\Pi_0 
%\end{bmatrix}=
matrix  with orthogonal columns, and multiply it by the transfer matrices $T_1 T_2 \dots T_N$. To control the round-off error, after each $q=5$ iterations (number of multiplications by the transfer matrix $T$), we apply QR decomposition which gives two vectors: a normalized $2 \times 4$  matrix $Q_j$, and a $2\times2$ upper triangular matrix $R_j$: 
\begin{equation}
Q_{j} R_{j} =T_{({j-1})q+1}  \dots T_{jq}  Q_{j-1}  \quad (j=1, \dots, N/q).  
\end{equation}
%The smallest positive Lyapunov exponent can be estimated by
%\begin{equation}
%\tilde{\gamma}_2= \frac{1}{N} \sum_{j=1}^{N/q} \ln{R_j(2,2)}
%\end{equation}
%in the limit $N\rightarrow \infty$. 
%Here, in each QR factorization step, we store the second diagonal element of $R_j$ which has the smaller positive Lyapunov exponent. Practically, we need to use large but finite number of iterations: $N=10^8$.
%Then we find the localization length as
%\begin{equation}
%\xi= 1/\gamma_2.
%\end{equation}
%%where
%Then, we multiply it by the transfer matrices $T_1 T_2 \dots T_N$ to find all $\{\chi_\ell,\Pi_\ell\}$.  After each 5 iteration (multiplications by the transfer matrix $\hat{T}$), we apply QR decomposition \cite{Slevin_2014} which gives two vectors: the normalized $2 \times 4$ matrix $Q_\ell$ which needs to be replaced by the previous one, and 
%%\begin{equation}
%%R_j=
%%\begin{bmatrix} 
%%\exp(\gamma_{2j-1}) & 0  \\
%%0 & \exp(\gamma_{2j}) \\
%%\end{bmatrix}
%%\end{equation}
%%with $\gamma_{2j-1}>\gamma_{2j}$. %, in which we store the smallest positive Lyapunov exponent in each iteration $j$ that is $\gamma_{2j}$. %, where $j$ denotes the iteration number.  %The Lyapunov exponents are related to the diagonal elements of $R$ by
%\begin{equation}
%\gamma_i= \lim_{N\rightarrow \infty} \frac{\ln{R({i,i})}}{N} \approx \frac{\ln{R({i,i})}}{N}, 
%\end{equation}
%for large $N$. 

The smallest positive Lyapunov exponent can be estimated by
\begin{equation}
\tilde{\gamma}_2= \frac{1}{N} \sum_{j=1}^{N/q} \ln{R_j(2,2)}
\end{equation}
in the limit $N\rightarrow \infty$. Here, in each QR factorization step, we store the second diagonal element of $R_j$ which has the smaller positive Lyapunov exponent.
Practically, we need to use large but finite number of iterations: $N=10^8$.
\begin{equation}
\xi= 1/\gamma_2.
\end{equation}
%%The smallest positive Lyapunov exponent can be estimated in the limit $N\rightarrow\infty$. Here, in each QR factorization step, we store the second diagonal element of $R_j$ which has the smaller positive Lyapunov exponent $\gamma_{2j}$. Practically, we need to use large but finite number of iterations. %: $N=10^8$.
%%Thus, after $N\geq 10^8$ iterations, we find the localization length as
%%\begin{equation}
%%\xi = \frac{N}{\sum_{j=1}^{N/q} \gamma_{2j}}.    
%%\end{equation}%= \frac{1}{\min{(\gamma_i)}}
We use the same transfer matrix method with QR decomposition to solve Eq.(\ref{eq:chipi3}) and find the localization length. %We can solve Eq.(\ref{eq:chipi3}) more simply
Eq. (\ref{sldl}) can be written in matrix form:
\begin{equation}\label{4by4eski}
\begin{bmatrix} 
S_{\ell+2}  \\
S_{\ell+1} \\
S_\ell \\
S_{\ell-1} 
\end{bmatrix} =
\begin{bmatrix} 
\frac{2(ga-\epsilon_\ell)}{J}+4 & \frac{E+4\epsilon_\ell}{J} & \frac{2(ga-\epsilon_\ell)}{J}+4 & -1 \\
 1 & 0 & 0 & 0 \\
0  & 1 & 0 & 0 \\
0 & 0 & 1 & 0
\end{bmatrix}
\begin{bmatrix} 
S_{\ell+1}  \\
S_{\ell}\\
S_{\ell-1} \\
S_{\ell-2},
\end{bmatrix}
\end{equation}
with gauge 
\begin{equation}\label{gauge}
E={\lambda}^2/J -4g a-6J. 
\end{equation}
%%which is shown as black dashed lines in Fig. \ref{fig:PlaN20000log}. 
It gives the exact same solution with Eq.(\ref{eq:tmm1}).  
As Eq. (\ref{sldl}) is the simpler version of Eq. (\ref{eq:chipi3}), the TMM procedure to solve Eq. (\ref{sldl}) is more efficient due to less CPU time. 

\chapter{Localization length : Transfer matrix method}

\label{LLTMM}

We solve Eq.~(\ref{onlyS}) with the approximate GS field (\ref{gglsquare}) with the transfer matrix method:

%The equation (\ref{sldl}) can be solved by transfer matrix method (TMM) by rewriting it in the matrix form:

\begin{equation}\label{4by4new}
\begin{bmatrix}
S_{\ell+2}  \\
S_{\ell+1} \\
S_\ell \\
S_{\ell-1}
\end{bmatrix} =
T_\ell
\begin{bmatrix}
S_{\ell+1}  \\
S_{\ell}\\
S_{\ell-1} \\
S_{\ell-2},
\end{bmatrix}
\end{equation}

where the transfer matrix
\begin{equation}\label{tmm1new}
T_\ell=
\begin{bmatrix}
u_\ell+\zeta_{\ell+1} & {\lambda^2}/{J^2}-2-u\zeta_\ell & u_\ell+\zeta_{\ell-1} & -1 \\
 1 & 0 & 0 & 0 \\
0  & 1 & 0 & 0 \\
0 & 0 & 1 & 0
\end{bmatrix},
\end{equation}
where $u_\ell=2g G_\ell^2/J +\zeta_\ell$.
%\frac{2(ga-\epsilon_\ell)}{J}+4 & \frac{{\lambda}^2/J -4g a-6J+4\epsilon_\ell}{J} & \frac{2(ga-\epsilon_\ell)}{J}+4 & -1 \\
% 1 & 0 & 0 & 0 \\
%0  & 1 & 0 & 0 \\
%0 & 0 & 1 & 0
Following \cite{Slevin_2014}, we start the transfer matrix multiplication with
\begin{equation}
Q_0=
\begin{bmatrix}
1 & 0  \\
0 & 1 \\
0 & 0 \\
0 & 0
\end{bmatrix}
\end{equation}
matrix  with orthogonal columns, and multiply it by the transfer matrices $T_1 T_2 \dots T_N$.
To control the round-off error, after each $q=5$ iterations (number of multiplications by the transfer matrix $\hat{T}$), we apply QR decomposition which gives two vectors: a normalized $2 \times 4$  matrix $Q_j$, and a $2\times2$ upper triangular matrix $R_j$:
\begin{equation}
Q_{j} R_{j} =T_{({j-1})q+1}  \dots T_{jq}  Q_{j-1}  \quad (j=1, \dots, N/q)
\end{equation}
The smallest positive Lyapunov exponent can be estimated by
\begin{equation}
\tilde{\gamma}_2= \frac{1}{N} \sum_{j=1}^{N/q} \ln{R_j(2,2)}
\end{equation}
in the limit $N\rightarrow \infty$. Here, in each QR factorization step, we store the second diagonal element of $R_j$ which has the smaller positive Lyapunov exponent.
Practically, we need to use large but finite number of iterations: $N=10^8$.
\begin{equation}
\xi= 1/\gamma_2.
\end{equation}

In addition, Eq.~(\ref{Slap}) can be solved by transfer matrix method:

\begin{equation}\label{2by2}
\begin{bmatrix}
S_{\ell+1} \\
S_\ell
\end{bmatrix} =
T_\ell
\begin{bmatrix}
S_{\ell}\\
S_{\ell-1}
\end{bmatrix},
\end{equation}

where
\begin{equation}
T_\ell=
\begin{bmatrix}\label{tmm2new}
\zeta_\ell -\lambda^2/2 J g G_\ell^2 & -1 \\
 1 & 0
\end{bmatrix}.
\end{equation}

We start the transfer matrix multiplication with
\begin{equation}
V_0=
\begin{bmatrix}
1   \\
0
\end{bmatrix}
\end{equation}
vector, and multiply it by the transfer matrices: $V_j = T_{j}  V_{j-1}  \quad (j=1, \dots, N)$. %$T_1 T_2 \dots T_N$ such that
%The $2\times2$ matrix produces two Lyapunov exponents $\gamma_1$ and $-\gamma_1$. %Therefore, we do not need to use QR method.
To deal with the round-off error, after every $5$ iterations, we normalize the vector $V_j$ and estimate the smallest positive Lyapunov exponent by
\begin{equation}
\tilde{\gamma}_1= \frac{1}{N} \sum_{j=1}^{N/q} \ln{\Vert{V_j}\Vert}, \qquad
\end{equation}

After $N=10^8$ iterations, we find the localization length $\xi$ as
\begin{equation}
\xi= 1/\tilde{\gamma_1}.
\end{equation}

\chapter{Localization length calculation in the strong interaction regime}
\label{LLCD}

In this appendix, we outline the computation of the localization length in the strong interaction regime as given by Eq.~(\ref{invloc_fin}). We start from Eqs.~(\ref{stand},\ref{kapp}).
In leading order $1/(ga)$ we obtain $\tilde{E}=2-E$ and the onsite disorder potential
%is expanded to leading order in $1/(ga)$:
\begin{align}
\varkappa_l = \frac{\epsilon_l(2-2E)-\epsilon_{l-1} -\epsilon_{l+1}}{2ga}\;.
\end{align}
We define the onsite disorder correlation function
\begin{align}
K(n-m) =K(m-n) =\langle\varkappa_n\varkappa_m\rangle \;.
\end{align}
The range of the correlations is finite because $K(\ell)$ takes non-zero values only for $\ell=0,\pm 1,\pm 2$:
\begin{eqnarray}
K(0)&= & \langle \varkappa_l^2 \rangle = \frac{W^2}{48g^2a^2}\left( (2-2E)^2 +2 \right)\;,
\\
K(1)&= & \langle \varkappa_l \varkappa_{l+1} \rangle = \frac{W^2}{24g^2a^2}(2E-2)\;,
\\
K(2) &= & \langle \varkappa_l \varkappa_{l+2} \rangle = \frac{W^2}{48 g^2 a^2}\;.
\end{eqnarray}
The Fourier transformed correlation function is then readily obtained:
%\begin{widetext}
\begin{align}%{eqnarray}
\label{kq}
K(q)&=\sum_{\ell} K(\ell)e^{i q \ell} =K(0)+2\sum_{\ell=1}^{\infty} K(\ell)\cos q \ell\\ 
&= \frac{W^2}{12 g^2a^2}\left( (1-E)^2 + \frac{1}{2} -2(1-E)\cos q + \frac{1}{2}\cos 2q \right).\nonumber
\end{align}
%\end{widetext}
Next we compute the above expression at double argument value $K(2q)$ and then use the dispersion relationship of the homogeneous Eq. (\ref{stand})
\begin{equation}
\label{Edef}
E=2(1-\cos(q))
\end{equation}
to replace $q$ by $E$. After some additional simple algebra, the result reads
\begin{align}
\label{k2q}
K(2q) = \frac{W^2}{48 g^2 a^2} E^2 (2-E)^2 \;.
\end{align}
Anderson localization with correlated disorder was studied in many publications (see Ref.~\cite{Lifshits88:book} for continuum models and Refs.~\cite{Griniasty88,Luck89,Izrailev99,Titov05} for lattice models). The inverse correlation length in the model (\ref{stand}) is given by (see Sec.~5.2.1 of the review \cite{Izrailev12})
\begin{align}\label{invloc_gen_a}
  \frac{1}{\xi}=\frac{K(2q)}{8\sin^{2}{q}}.
\end{align}
 Substituting $K(2q)$ from (\ref{k2q}) into Eq.~(\ref{invloc_gen_a}) and using Eq.~(\ref{Edef}) yield
%\begin{widetext}
\begin{align}\label{invloc_fin_a}
\xi=\frac{96g^2a^2}{W^2}\frac{4-E}{E(2-E)^2}
%\equiv
%\frac{24g^2a^2}{W^2}\cos^{-2}(q)\tan^{-2}({q}/{2}) \;.
\end{align}
%\end{widetext}

\cedp
%\include{Appendix1/appendix1}
%\cedp
%\include{Appendix2/appendix2}
%\cedp
%\include{Appendix3/appendix3}
%\cedp

\end{document}